 \documentclass[aps,showpacs,amsmath,dvips,showkeys,preprint]{Tawrevtex4}

\usepackage{graphicx}
\usepackage{color}
\usepackage{natbib}

\def \be  {\begin{equation}}
\def \ee  {\end{equation}}
\def \ee  {\end{equation}}
\def \bea {\begin{eqnarray}}
\def \eea {\end{eqnarray}}

\def \Tr  {\bf{Tr}}

\newcommand{\nn}{\nonumber}
\newcommand{\de}{\partial}
\renewcommand{\figurename}{{\bf Fig.}}
\renewcommand{\tablename}{{\bf Tab.}}

\begin{document}

\preprint{ECTP-2013-09b\hspace*{0.5cm}and\hspace*{0.5cm}WLCAPP-2013-06b}

\title{Generalized Uncertainty Principle: Approaches and Applications}
\author{Abdel Nasser TAWFIK\footnote{http://atawfik.net/}}
\affiliation{Egyptian Center for Theoretical Physics (ECTP), Modern University for Technology and Information (MTI), 11571 Cairo, Egypt}
\affiliation{World Laboratory for Cosmology And Particle Physics (WLCAPP), Cairo, Egypt}

\author{Abdel~Magied~DIAB}
\affiliation{World Laboratory for Cosmology And Particle Physics (WLCAPP), Cairo, Egypt}
\date{\today}

\begin{abstract}
We review highlights from string theory, black hole physics and doubly special relativity and some "thought" experiments which were suggested to probe the shortest distance and/or the maximum momentum at the Planck scale. The models which are designed to implement the minimal length scale and/or the maximum momentum in different physical systems are analysed entered the literature as the Generalized Uncertainty Principle (GUP). We compare between them. The existence of a minimal length and a maximum momentum accuracy is preferred by various physical observations. Furthermore, assuming modified dispersion relation allows for a wide range of applications in estimating, for example, the inflationary parameters, Lorentz invariance violation, black hole thermodynamics, Saleker-Wigner inequalities, entropic nature of the gravitational laws, Friedmann equations, minimal time measurement and thermodynamics of the high-energy collisions. One of the higher-order GUP approaches gives predictions for the minimal length uncertainty. Another one predicts a maximum momentum and a minimal length uncertainty, simultaneously. An extensive comparison between the different GUP approaches is summarized. We also discuss the GUP impacts on the equivalence principles including the universality of the gravitational redshift and the free fall and law of reciprocal action and on the kinetic energy of composite system. The concern about the compatibility with the equivalence principles, the universality of gravitational redshift and the free fall and law of reciprocal action should be addressed. We conclude that the value of the GUP parameters remain a puzzle to be verified.

\end{abstract}

\pacs{04.20.Dw,04.70.Dy, 04.60.-m}
\keywords{Generalized uncertainty principle, black hole thermodynamics, quantum gravity} 

\maketitle

\tableofcontents
\makeatletter
\let\toc@pre\relax
\let\toc@post\relax
\makeatother 

 \newpage

\section{A very short history}
\label{history}

An idea about the existence of a minimal length and/or time was speculated in the ancient time. In modern Physics, the {\it chronon}, which is the hypothetical fundamental or the indivisible interval of time with the value of the ratio between the diameter of the electron and the velocity of light and proposed by Robert Levi  \cite{Levi} in 1927, would be the first minimum measurable time interval proposed, $\sim 10^{-24}~$s. Within this time interval, the Special Relativity (SR) and Quantum Mechanics (QM) are conjectured to unify in framework of the Quantum Field Theory (QFT). In light of this, the possible existence of a minimal length scale rose the awareness of the physicists. For example, the Planck time is given as $(G\,\hbar/c^5)^{1/2}$ with the dimensions of {\it ''time''} which is to be formed from $G$, where $c$ is speed of light, $\hbar$ is the Planck constant and $G$ stands for the Newtonian gravitational constant. 

Recently,  various scenarios for the minimal length scale have been reviewed \cite{TD,Sabine}. Accordingly, the main developments for defining the minimal length have been guided by: 
\begin{itemize} 
\item Singularities in fundamental theories, like Fermi theory of $\beta$-decay. This leads to concrete cut-off  and even QM obtains a minimal length scale.
\item Distasteful arbitrary procedure of the cut-off. This leads to modification of the canonical commutation relations of position and momentum operators.
\item Redefining role of the gravity in testing physics at short distance and in various {\it ''gedanken''} (thought) experiments. This leads to an approach that the minimal length scale is connected with some gravitational aspects. 
\item The trans-Planckian problem or the black-hole thermodynamic properties. This leads to a modification in the dispersion relation. This an an essential milestone in implementing modified dispersion relation as an alternative to generalized uncertainty principle (GUP). 
\item QM at minimal length scale and QFT. This leads to modifications of the canonical commutation relations in order to accommodate a minimal length scale.
\item The string theory, which leads to GUP based on string scattering in the {\it super-Planckian regime}.
\end{itemize}

Defining a fundamental length was necessary to overcome singularities in fundamental theories. First, regularizations like \underline{\it cut-off} in some dimensional quantities was implemented. But since the cut-off would not be independent of the frame of reference, problems with the Lorentz invariance principle did appear \cite{Sabine}. It dates back to 1930's \cite{Heisenberg2,Pauli}, where it was found that the effect of regularizations with respect to the cut-off should be the same as that of a fundamentally discrete space-time \cite{Sabine}. At that time, neither a fundamental finite-length nor a maximum frequency was known \cite{exmp5,exmp3,exmp6,exmp7}. Thus, the fundamental length was thought as a realm of subatomic physics, $10^{-15}~$m \cite{Sabine}.

According to Ref. \cite{Sabine}, the fundamental minimal length was refined by Heisenberg \cite{Heisenberg}, who pointed out that the Fermi theory of $\beta$-decay \cite{frmbta1,frmbta2} should be non-normalizable. At high energy, the theory (four-fermion coupling) should break. In electroweak interactions, it was suggested that theory should be replaced by an exchange of a gauge boson \cite{Heisenberg2} and accordingly, QM with a minimal length scale defines discrete mass-spectrum. This leads to a better understanding of the QFT \cite{Sabine}. On the other hand, the space and time discrete approaches were not really inviting \cite{Sabine}.

At that time, both weak and strong forces were not known, but an idea about gravity as a non-fundamental force was introduced by Bronstein, the so-called to Schwarzshild singularity \cite{Gorelik}. This is to be interpreted as the gravity does not allow an arbitrarily concentration of mass in infinitesimal space-time.. This makes the gravity basically different from the electrodynamics, for instance. An upper bound for the mass density can be estimated as $\rho\, \lesssim c^{2}/G\, V^{2/3}$  \cite{cr25,cr27}. Accordingly, uncertainties in measuring the Christoffel symbols were studied \cite{mead20}. This supports the conclusion proposed by Bronstein in 1930's \cite{Bronstein}. 

Through a {\it ''distasteful arbitrary procedure''} \cite{Snyder}, it was proposed that the momentum space cut-off is very likely. Thus, a modification in the \underline{\it canonical commutation relations} was proposed instead of the cut-off. Accordingly, the non-commutative space-time is conjectured to affect the Heisenberg uncertainty in a way that a tiny space-resolution should be taken into consideration. This proposed resolution follows the Lorentz invariance principle \cite{Yang}. A peculiar role of gravity at short distances was proposed by Mead \cite{mead22}. 

The idea of utilizing fundamental limits governing mass and size of the physical systems in measuring the time dates back to nearly six decades. Salecker and Wigner suggested the use of a quantum clock \cite{wigner57,wigner58} in measuring {\it distances} between events in space-time. Measuring rods are entirely avoided, as they are likely macroscopic objects \cite{wigner58}. The Wigner second constrain requires that only one single simultaneous measurement of both energy and time can be accurate. Therefore, it is believed to be more severe that the Heisenberg uncertainty principle (HUP). 

In calculating the emission rate of photons approaching a black-hole horizon, the trans-Planckian modes should be taken into account \cite{entr3}. Later on, modified dispersion relation (MDR) was suggested by Unruh to solve this problem \cite{unruh25}. Accordingly, the minimal wavelength is defined. It is the scale, which is able to solve the {\it trans-Planckian problem}. Also, generalizing Poincare algebra to the Hopf algebra was proposed \cite{MR28}. This generalization is based on an assumption of modification in the space-time coordinates commutators \cite{MR28}. To the mathematical basis of QM, a minimal length was proposed \cite{14,16,kmpf32} resulting in modifications in the canonical commutation relations, including the dispersion relation and a generalization of the Poincare algebra. A finite extension was observed in the string theory \cite{st33,st34,st35,st36}, where the string scattering in {\it super-Planckian} scale leads to generalized uncertainty principle (GUP). It is apparent that GUP prevents the existence of a scale smaller than the string extension.

In supporting the phenomena that uncertainty principle would be affected by QG, various examples can be mentioned. In the context of polymer quantization, the commutation relations are given in terms of the polymer mass scale \cite{polymer}. Also, the standard commutation relations in QM are conjectured to be generalized (changed) at the length scales of the order of the Planck length \cite{garay1,Scardigli,gupps2,16}. Such modifications are supposed to play an essential role in the quantum gravitational corrections  \cite{qgc}.  Accordingly, the standard uncertainty relation of QM is replaced by a gravitational uncertainty relation having a minimal observable length (of the order of Planck length) \cite{gupp3a,gupp3b,gupp3c,gupp3d}.

We also review argumentation against the GUP approaches in section \ref{sec:ev}. We first start with the {\it equivalence principle}, which is one of the five principles forming the basis of GR, where the motion of the gravitational test-particle in a gravitational field should be independent on the mass and composition of the particle \cite{Ray}. On the other hand, when taking into consideration the Strong (SEP) \cite{7} and Weak Equivalence Principle (WEP) \cite{7}, the gravitational field  should couple to everything \cite{Ray}.

\newpage

\section{Introduction}
\label{intro}

As introduced in section \ref{history}, the existence of a minimal length was a great prediction deduced from different approaches (related to QG) such as the black hole physics \cite{3,4} and the string theory \cite{guppapers,5}. The mean idea is simply that the string is conjectured not to interact at distances smaller than its size, which is determined by its tension. For completeness, we highlight that the information about the string interactions would be included in the Polyakov-loop action \cite{daxson}. The existence of a minimal length leads to generalized (Heisenberg) uncertainty principle (GUP) \cite{guppapers}. At Planck (energy) scale, the corresponding Schwarzschild radius becomes comparable to the Compton wavelength. The high energies (Planck energy) seem to result in further decrease in the Schwarzschild radius $\Delta x$ in the presence of gravitational effects. In light of this, $\Delta x \approx \ell_{Pl}^2\Delta p/\hbar$. This observation and the ones deduced from various {\it gedanken} experiments suggest that the GUP approach should be essential, especially at some concrete scales. 

The Heisenberg uncertainty principle (HUP) represents one of the fundamental properties of quantum system. Accordingly, there should be a fundamental limit for the measurement accuracy, with which certain pairs of physical observables, such as the position and momentum and energy and time, can not be measured, simultaneously. In other words, the more precisely one observable is measured, the less precise the other one shall be detected. 
In QM, the physical observables are described by operators in Hilbert space. Given an observable $A$, we define an operator as a standard deviation of $A$, $\Delta A=A-\langle A \rangle$, where the expectation value is given by $\langle (\Delta A)^2 \rangle = \langle {A^2} \rangle - {\langle A \rangle}^2$.
Using Schwartz inequality \cite{Schwarz} $\langle \alpha | \alpha\rangle \langle \beta | \beta \rangle \geq | \langle \alpha | \beta \rangle |^2$, 
which is valid for any ket and bra state $|\alpha\rangle = \Delta A | \alpha {'} \rangle$ and $|\beta \rangle = \Delta B | \beta {'} \rangle$, respectively.
By using Dirac algebra, we get the Cauchy-Schwartz inequality $(\Delta A)^{2}\;  (\Delta B) ^{2}\geq \frac{1}{4} | \langle \Delta A\; \Delta B \rangle |^{2}$,  
\bea
\Delta A  \Delta B &\geq& \frac{1}{2} | \langle \Delta A \Delta B \rangle |. 
\eea

In Heisenberg algebra, the position $\hat{x}$ and momentum operator $\hat{p}$ satisfy the canonical commutation relation $[\hat{x} ,\hat{p}]= \hat{x} \hat{p} -\hat{p} \hat{x}= i \hbar$.  As a consequence, for position and momentum uncertainties, $ \Delta x $ and $ \Delta p$, respectively, of a given state, the Heisenberg uncertainty relation reads (in natural units)
\be 
\Delta x\; \Delta p \geq \frac{\hbar}{2}.
\ee 
  
In order to detect an arbitrarily small length scale, one has to utilize tools of sufficiently high energy (high momentum) and thus very short wavelength. This is the principle, on which colliders/accelerators, such as the Relativistic Heavy-Ion Collider (RHIC) \cite{rhicSite}, Large Hadron Collider (LHC) \cite{lhcSite}, FermiLab \cite{fermilabSite}, etc., are based. On the other hand, there are reasons to believe that at high energies, the gravity becomes important. In light of this, the former conclusion would be no longer true. In other words, the linear relation between energy and wavelength would be violated, as well.

The detectability of quantum space-time foam with gravitational wave interferometers has been discussed \cite{disct1}. But in relation with the limited measurability of the smallest quantum distances, this was criticized \cite{disct1,disct2}. Four decades later, Barrow applied Wigner inequalities \cite{wigner57,wigner58} in describing the quantum constrains on the black-hole lifetime \cite{barrow96}. The black-hole running time is proportional to the Hawking lifetime. The latter is to be calculated under the assumption that the black hole is a black body and therefore the Stefan-Boltzmann law can be utilized. Also, it is found that the Schwarzschild radius is correspondent to the constrains on Wigner size. Furthermore, the power of information processing of a black hole has been estimated through the emission of the Hawking radiation \cite{swLitr}.

\subsection{Generalized (gravitational) uncertainty principle}
\label{sec:ggup}

Based on the various GUP approaches, the existence of a minimal length suggests that the space in the Hilbert space representation \cite{16} describes a non-commutative geometry, which can also arise as a momentum over curved spaces \cite{19}. From various {\it gedanken} experiments, which have been designed to measure the area of the apparent horizon of a black hole in QG \cite{20}, the uncertainty relation was preformed \cite{3}. The modified Heisenberg algebra introduces a relation between QG and Poincare algebra \cite{20}. In an $n$-dimensional space and under the effects of GUP, it is found that even the gravitational constant $G$ \cite{Extra} and the Newtonian law of gravity \cite{7} are subject of modifications. The interpretation of QM through a quantization in $8$-dimensional manifold implies the existence of an upper limit in the accelerated particles \cite{21}. Nevertheless, the quadratic and linear GUP approaches \cite{3,16,12} assume that the momenta approach the maximum value at very high energy (Planck scale)  \cite{12}. 

A new GUP approach fits well with the string theory and the black hole physics (with quadratic term of momenta) and with the Doubly Special Relativity (DSR) (with linear term of momenta)  \cite{advplb}. This approach predicts a minimal measurable length and a maximum measurable momentum, simultaneously and suggests that the space should be quantized and/or discritized. But, it has severe difficulties discussed in Ref. \cite{pedram}. Therefore, 
\begin{itemize}
\item another GUP approach is conjectured to absolve an extensive comparison with Kempf, Mangano and Mann (KMM) \cite{16} and
\item a new GUP approach was introduced to characterize a minimal length uncertainty and a maximal momentum, simultaneously, \cite{pedram}.
\end{itemize}
The latter has been performed in Hilbert space \cite{Nouicer}. Here, a novel idea of minimal length modelled in terms of the quantized space-time was implemented. Thus, this new approach agrees well with QFT and Heisenberg algebra, especially in context of non-commutative coherent states representation. The resulting GUP approach can be studied at UV finiteness of Feynman propagator \cite{Nouicer}.

\subsection{Physics of generalized (gravitational) uncertainty principle}
\label{sec:gupphys}

There are various observations showing that the GUP approaches offer a valuable possibility to study the influences of the minimal length on the properties of a wide range of physical systems, especially at the quantum scale  \cite{3,7,Scardigli}. The effects of the linear GUP approach have been studied on 
\begin{itemize}
\item compact stars \cite{Ali:2013ii}, 
\item Newtonian law of gravity \cite{Ali:2013ma}, 
\item inflationary parameters and thermodynamics of the early Universe \cite{Tawfik:2012he}, 
\item Lorentz invariance violation \cite{Tawfik:2012hz} and 
\item measurable maximum energy and minimum time interval \cite{DahabTaw}. 
\end{itemize}
 Regardless some applicability constrains, the effects of QG on the quark-gluon plasma (QGP) are introduced, as well  \cite{Elmashad:2012mq}.
 It was found that the GUP can potentially explain the small observed violations of the weak equivalence principle in neutron interferometry experiments \cite{exp}, section \ref{sec:ev}, and also predicts a modified invariant phase space which is relevant to the Lorentz transformation. It is suggested \cite{nature2012} that GUP can be measured directly in Quantum Optics Lab  \cite{Das1,afa2}. Furthermore, deformed commutation relations would cause new difficulties in quantum as well as in classical mechanics. We give a list of some of these problems as follows. 
\begin{itemize}

\item 1-dimensional harmonic oscillator with minimal uncertainty in position \cite{16} and minimal uncertainty in position and momentum \cite{sss} and $d$-dimensional harmonic oscillator with position minimal uncertainty  \cite{ddd,www},
\item problem of $3$-dimensional Dirac oscillator \cite{w}  and the solution of ($1+1$)-dimensional Dirac oscillator within  Lorentz covariant algebra \cite{ww}, 
\item $1$- and $3$-dimensional Coulomb problem within deformed Heisenberg algebra in perturbation theory  \cite{22,23,24,25,26},
\item scattering problem in deformed space with minimal length \cite{29},
\item ultra-cold neutrons in gravitational field with minimal length \cite{30,31,32},
\item influence of minimal length on Lamb shift, Landau levels, and tunnelling current in scanning tunnelling microscope \cite{Das,afa2}
\item Casimir effect in a space with minimal length \cite{35},
\item effect of non-commutativity and the existence of a minimal length on the phase space of cosmological model \cite{36},
\item various physical consequences of non-commutative Snyder space-time geometry \cite{37}, and 
\item classical mechanics in a space with deformed Poisson brackets  \cite{38,39,40}.
\end{itemize}
In sections \ref{implquadratic} and \ref{implLinear}, we review different implications of quadratic and linear GUP approaches, respectively, on 
\begin{itemize}
\item physics of early Universe, 
\item inflation parameters, 
\item Lorentz invariance violation, 
\item black hole thermodynamics, 
\item compact stellar objects, 
\item Saleker-Wigner inequalities, 
\item entropic nature of gravitational force, 
\item time measurement and 
\item thermodynamics of high-energy collisions.
\end{itemize}

The present review article is organized as follows.  In section \ref{sec:ggup}, definition for the generalized (extended) uncertainty principle (GEUP) is introduced. The relationship between the minimal length and maximum momentum is also presented. As introduced in previous sections, there are various approaches to GUP proposing the existence of nonvanishing minimal length that leads to non-commutative geometry. The physics of GUP approaches is reviewed in section \ref{sec:gupphys}.

In section \ref{sec:gup1}, we summarize the behavior of some well-known expressions for GUP. These expressions contain quadratic term of momenta with a minimal uncertainty on position. In section \ref{ssec:1}, we shall investigate the modification of the uncertainty relation due to the high-energy fixed-angle scatterings at short length such as the string length. In section \ref{ssec:2}, the uncertainty relation through various {\it gedanken} experiments which are designed to measure the area of the apparent horizon of black hole is reviewed. These thought experiments assume QG due to recording the photons of the Hawking radiation, which are emitted from the apparent horizon. Due to quantized space-time of the QFT and the geometric approach to curvature of momentum space, an algebraic approach can be expressed in the coproducts and the description of the Hopf-algebra \cite{MR28} leading to modified commutation relation between position and momenta, section \ref{ssec:3}. In section \ref{ssec:4}, the modified de Broglie relation which leads to changing the commutation relation between position and momentum and the investigation of minimal length and/or nonvanishing minimal length. In section \ref{ssec:5}, a new commutation relation containing a linear term as an addition of the quadratic term of momenta and predicts of the maximum measurable of momenta, shall be investigated.

In section \ref{sec:min1}, the relations describing the minimal length uncertainty are outlined. Two proposals for the modification of the momentum operator are introduced. The proposal of a minimal length uncertainty with a further modification in the momentum shall be reviewed. The main features in Hilbert space representation of QM of the minimal length uncertainty will be studied. Furthermore, their difficulties are also listed out. We show how to overcome these difficulties, especially  in Hilbert space representation.

In section \ref{sec:max}, the GUP approaches relating to string theory and black hole physics (lead to a minimum length) and the ones relating to DSR (suggest a similar modification of commutators) shall be studied. The main features and difficulties in Hilbert space representation will be reviewed, as well, and we show how to overcome these difficulties.

Section \ref{implquadratic} is devoted to the applications of the quadratic GUP approach. We list out seven applications; physics of early Universe, inflation parameters, black hole thermodynamics, compact stellar objects, Saleker-Wigner inequalities, entropic nature of gravitational force and time measurement.

Section \ref{implLinear} is devoted to the applications of the linear GUP approach on the same problems as given in section \ref{implquadratic}. Additional two problems are also discussed, namely the Lorentz invariance violation and thermodynamics of high-energy collisions.

In section \ref{other}, other alternative approaches to GUP such as the one suggested by Nouicer, in which an exponential term of momentum and minimal length appears, shall be introduced. This approach agrees well with the GUP which is originated in the theories for QG. There is another approach coming up with higher orders of  the minimal length uncertainty and maximal observable momentum. Finally, we compare between these approaches.
 
The effects of GUP on the principles of GR are studied in section  \ref{sec:ev}. The results estimated in various thought experiments are compared with the possible effects of GUP. It is found that the GUP apparently changes the natural statement of the kinematic energy of the deformed system. Argumentation against the GUP approaches shall be reviewed in section \ref{sec:ev}. These can be divided into two groups; the equivalence principle and the kinetic energy of composite system. The first group includes the universality of the gravitational redshift and the free fall and the law of reciprocal action. 

\newpage

\section{Generalized (Gravitational) Uncertainty Principle }
 \label{sec:gup1}
 
The Quantum Gravity (QG) describes the quantum behaviour of gravitational field and unifies  the  Quantum Mechanics (QM) with the General Relativity (GR). As we discussed in previous sections, there are different approaches such as string theory, black hole physics and double special relativity, in which likely the Heisenberg uncertainty principle (HUP) is conjectured to be violated. Accordingly, various quantum mechanical systems would be subjects of modification.

The consistent unification of the classical description of GR with QM still an open problem. One attempt assumes that the two theories can be used as a guiding principle to the search of a fundamental theory of QG. Another one gives several arguments ranging from theoretical analysis in string theory to more sophisticated or even {\it gedanken} experiments in order to measure the minimal length. Accordingly, a new contribution to the quantum uncertainty with a gravitational origin leading to a length scale as a Planck length  in the determination of space-time coordinates can be concluded. 

On one hand, these approaches provide essential predictions. We have listed out some of these in section \ref{intro}. Other applications shall be elaborated in sections \ref{implquadratic} and \ref{implLinear}. DSR suggests a possibility to relate the transition from the quantum behavior at the microscopic level to the classical behavior at the macroscopic level with the modification of QM induced by a modification of the relativity principles. Thus, the laboratory tests should be able to judge about these theories. On the other hand, the predictions remain uncertain due to the limitations of the current technologies. Nevertheless, the minimal length has been observed in condensed matter and atomic physics experiments, such as Lamb shift \cite{Das,Das1}, Landau levels \cite{Das,Das1}, and the Scanning Tunnelling Microscope (STM) \cite{Das1}.

\subsection{Introduction}
\label{ssec:0}

As discussed, it seems that HUP likely breaks down at energies close to the Planck scale. Taking into account the gravitational effects, an emergence of a minimal measurable distance seems to be inevitable. More generally, the generalized (gravitational) uncertainty principle (GUP)  can be expressed as \cite{16}
\be 
\Delta x  \Delta p \geq \frac{\hbar}{2} \left(1+\alpha (\Delta x)^2 + \beta (\Delta p)^{2} +\zeta \right),
\ee
where both $\beta$ and $\zeta$ are positive and independent variables. The uncertainties in position $ \Delta x $ and momentum $\Delta p$ may depend on the expectation values of the operators $\textbf{x}$ and $\textbf{p}$, respectively; $\zeta= \alpha \langle x \rangle^2 +\beta \langle p \rangle^2$. 

\begin{figure}[htb]
\centering{
\includegraphics[scale={0.5}]{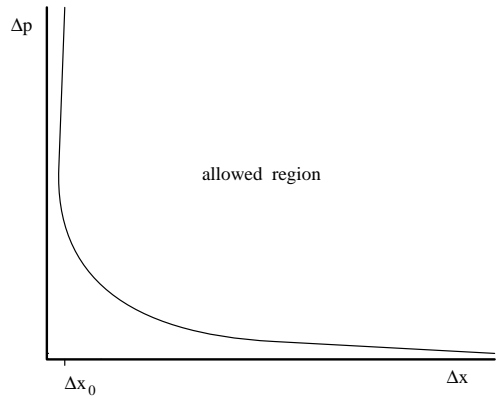}
\caption{The momentum uncertainty is given in dependence on the position uncertainty. This implies the existence of a minimal length of uncertainty $\Delta x_{0}$. The graph taken from Ref. \cite{16}. }
\label{fig1}
}
\end{figure}

In Fig. \ref{fig1}, the minimal momentum uncertainty $\Delta p$ is given as function of the position uncertainty $\Delta x$. It is apparent that the position minimal uncertainty $\Delta x_0 \ne 0$ (finite) and $\Delta x_{min} \propto \Delta p_{max}$ (proportional) \cite{16}. Therefore, $\left[\textbf{x},\textbf{p}\right] = i \hbar \left( 1+\alpha \textbf{x}^{2}+\beta \textbf{p}^{2}\right)$ describes the resulting commutation relation.  
In QM, both \textbf{x} and \textbf{p} can be represented as operators acting on position-  $\phi(x)=\langle x| \phi (x) \rangle$ and momentum-space wavefunctions $\phi(p)=\langle p| \phi (p) \rangle$, where $|x\rangle$ and $|p\rangle$ are the position and momentum eigenstates, respectively. Both operators \textbf{x} and \textbf{p} are essentially self-adjoint. Their eigenstates can be approximated to an arbitrary precision by sequences of the physical states $|\phi _n \rangle$ of the increasing localization in position- $\lim_{n\rightarrow\infty} \Delta x_{|\phi _n \rangle}=0$ or momentum-space $\lim_{n\rightarrow\infty} \Delta p_{|\phi _n \rangle}=0$.

As pointed out in Refs. \cite{14,15}, with the inclusion of minimal uncertainties $\Delta x_0 >0$ and/or  $\Delta p_0 >0$, this situation would change drastically. For example, a non-vanishing minimal uncertainty in position is given as $(\Delta x)_{|\phi _n \rangle}^{2} = \langle \phi | \left(\textbf{x}-\langle \phi | x |\phi \rangle \right)^{2}| \phi \rangle \geq \Delta x_{0}, \longrightarrow |\phi  \rangle$,  implying that no physical state would exist with such a position eigenstate \cite{16}. This is because an eigenstate would - of course - have vanishing position uncertainties. It is apparent that a minimal position uncertainty means that the position operator is no longer essentially self-adjoint but symmetric. The preservation of symmetry assures that all expectation values should be real. When self-adjointness is abandoned, the introduction of minimal uncertainties is likely \cite{16}. 

The Heisenberg algebra will no longer find a Hilbert space representation on the position wavefunctions $\langle x| \phi (x) \rangle$ \cite{16}, because of the absence of position eigenstates $|x\rangle$ in representation of the Heisenberg algebra. In light of this, the discussion should be restricted to $\Delta x_{0} \ne 0$ and therefore $\alpha=0$, where there is no minimal momentum uncertainty. Similarly, a minimal momentum uncertainty is conjectured to abandon the momentum space wavefunctions \cite{16}. This allows to work with the convenient representation of the commutation relations on the momentum space wavefunctions. Thus
\bea
\Delta x  \Delta p &\geq& \frac{\hbar}{2} \left(1+ \beta (\Delta p)^{2} +\zeta \right),
\eea
where the constant $\zeta$ is positive and related to the expectation value of the momentum, $\zeta = \beta \langle p \rangle^{2}$.

\subsubsection{Space non-commutativity} 

The HUP has a strong relationship to the canonical commutation or the commutative phase-space structures. When HUP should be broken down due to GUP, an operational form of the non-commutative (NC) phase-space structures shall be observed. The generic expressions were introduced in Ref. \cite{Subir2}
\bea
\left[x_{i},\, p_{j}\right] &=& i \hbar\, \left[  \delta_{i j}\, \left(1+\beta f_{1} \left(\textbf{p}^{2}\right) \right)+f_{2}\, \left(\textbf{p}^{2}\right)\; p_{i}\, p_{j}\right],\\
\left[x_{i},\, x_{j} \right] &=& i \hbar f_{i j}(p) \ne 0.
\eea
The presence of a minimum length  or a maximum momentum or both of them apparently leads to GUP originated in  the NC algebras. Accordingly, Kempf \cite{17} proposed the following algebraic relations: 
\bea
\left[x_{i}, p_{j}\right] &=& i\, \hbar\, \left[   \delta_{i j}\, \left( 1+\beta\, p^{2} \right)+ \beta^{'}\, p_{i}\, p_{j}\right],\\
\left[x_{i}, x_{j}\right] &=& i\, \hbar\, \left(\beta ^{'} - 2\, \beta \right) \left(x_{i}\, p_{j} - x_{j}\, p_{i} \right),\\
\left[p_{i}, p_{j}\right] &=& 0.
\eea
Other algebraic relations were introduced in Ref. \cite{16}
\bea
\left[x_{i}, p_{j} \right] &=& i\, \hbar\,  \delta_{i j}\, \left(1 + \beta\, p^{2}\right),\\
\left[x_{i}, x_{j} \right] &=& - 2\, i\, \hbar\, \beta\, \left(x_{i}\, p_{j} - x_{j}\, p_{i} \right),\\
\left[p_{i}, p_{j} \right] &=& 0.
\eea
Recent algebraic relations have been presented \cite{chang2q}
\bea
\left[x_{i}, p_{j} \right] &=& i\, \hbar\,  \left[   \delta_{i j}\, \left( 1+\beta\, p^{2} \right)+ \beta^{'} p_{i} p_{j}+O(\beta^{'2}, \beta^{2}) \right] ,\\
\left[x_{i}, x_{j} \right] &=& i\, \hbar\, \frac{\left(2\, \beta - \beta^{'} \right)+\left(2\, \beta + \beta^{'} \right)\,\beta\, p^{2}}{1+\beta\, p^{2}} \, \left( x_{i}\, p_{j} - x_{j}\, p_{i} \right),\\
\left[p_{i}, p_{j} \right] &=& 0.
\eea

The GUP approaches which are  consistent with the NC algebras offer the possibility for space discreteness and/or quantization. In other words, the physical states of space should be non-commute. Despite that the physical states can not be measured, simultaneously, the space discretization seems to be possible.

\subsection{String theory}
\label{ssec:1}

In order to check how the theory tackles in consistences of QG at the Planck scale \cite{guppapers}, a GUP approach was first proposed by Amati {\it et al.} \cite{guppapers}. The ultra high-energy scatterings of strings have been analysed. Some interesting effects are compared to those which were found in {\it usual} field theories, especially the ones originating from the soft short-distance behavior of the string theory \cite{guppapers}. The hard processes are studied at a short distance as in high-energy fixed-angle scatterings.  The latter are apparently not able to test distances shorter than the characteristic string length $\lambda_{s}=(\hbar \alpha)^{1/2}$, where $\alpha$ is the string tension. 

Another scale is dynamically generated: $D$-dimensional gravitational Schwarzschild radius $R(E) \sim (G_N E)^{1/(D-3)}$ is conjectured to approach the string length $\lambda_s$ \cite{guppapers}. This depends on whether $R(E)$ smaller or greater than $\lambda_s$. If $R(E)>\lambda_{s}$, then new contributions at distances of the order of $R(E)$ appear. This indicates a classical gravitational instability that can be attributed to the black hole formation. If the opposite should be the case ($R(E)<\lambda_{s}$), then their contributions are irrelevant. Obviously, there are no black holes with a radius smaller than the string length. In this case, the analysis of short distances can go on. It has been shown that the larger momentum transfers do not always correspond to shorter distances. Precisely, the analysis of the angle distance relationship suggests the existence of a scattering angle $\theta_M$. When the scattering should take place at $\theta<\theta_M$, then the relation between the interaction distance and the momentum transfer is the classical one, i.e. follows the Heisenberg relation with $q\sim\hbar/b$, where $b$ is the impact parameter. But when $\theta\gg\theta_M$, then the classical picture is no longer valid. An important new regime where $\langle q \rangle \sim b$ would be constructed. This suggests a modification of the uncertainty relation at the Planck scale \cite{guppapers}
\be 
\Delta x \sim \frac{\hbar}{\Delta p} +Y\, \alpha\, \Delta p,
\ee
where $Y$ is a suitable constant. Consequently, the existence of a minimal observable length of the order of String size $\lambda_{s}$ is likely.

\subsection{Black hole physics} 
\label{ssec:2}

Several works have been devoted to perform the uncertainty relations and their measurability bounds in QG \cite{3}. Thought experiments have been proposed to measure the area of the apparent horizon of a black hole \cite{3}. Accordingly, a generalization of the uncertainty principle has been concluded, which agrees well with the one deduced from the string theories \cite{guppapers,2,3}.  Also, in string theories, the tool of gedanken string collisions at the Planck energy was very useful \cite{guppapers,4}.  In addition to these, the renormalization group analysis has been applied to the string \cite{5}. A main physical ingredient was the Hawking radiation \cite{entr3}. The black hole approach to GUP, which is a rather model independent approach, agrees, especially in its functional form, with the one obtained in framework of the string theory. 

The thought experiment proceeds by observing the photons scattered by the studied black hole. The main physical hypothesis of the experiment is that the black hole emits Hawking radiation. Detecting the Hawking radiation, it turns to be possible to grab a black hole  ''image'' \cite{entr3}. Besides, measuring the direction of the propagating photons that are emitted at different angles and tracing them back, one can - in principle - locates the position of the black hole center \cite{entr3}. In such a way, the radius $R_{h}$ of the apparent horizon will be measured. Apparently, this measurement has two sources of uncertainty \cite{3}.  
\begin{itemize}
\item The first one is based on the fact that a photon with wavelength $\lambda$ can't carry information about a more detailed scale than $\lambda$ itself \cite{3}. As in the classical Heisenberg analysis, the resolving power of the microscope gives the minimum error $\Delta x^{(1)} \sim \lambda/\sin \theta$, 
where $\theta$ is the scattering angle. Then, the final momenta should have the uncertainty $\Delta p\sim h \sin(\theta)/\lambda$. During the emission process, the mass of the black hole varies from $M$ to $M-\Delta M$ \cite{3}, where $\Delta M=h/(c\, \lambda)$. The radius of the horizon changes, accordingly. The corresponding uncertainty is intrinsic to the measurement.

For example, the metric field of Reissner black hole \cite{Carroll} is given as 
\be 
ds^{2} = -\left(1 -\frac{2 G}{r}+\frac{G Q}{r^{2}}\right) d t^{2} + \left(1 -\frac{2 G}{r}+\frac{G Q}{r^{2}}\right) d r^{2} + r^{2} d \Omega^{2}.
\ee
Also, the apparent horizon is defined as the outer boundary of a region of closed trapped surfaces. In spherical topology and Boyer-Lindquist coordinates \cite{BL1967}, the apparent horizon is located at  $r = R_{h}$ 
\be 
R_{h} = G\, M\, \left[1+\left(1- \frac{Q^{2}}{G\, {M^{2}}}\right)^{1/2}\right].
\ee
The Boyer-Lindquist coordinates  are a generalization of the coordinates used for the metric of a Schwarzschild black hole. This can be used to express the metric of a Kerr black hole \cite{kerr1963}. Accordingly, the line element for a black hole with mass $M$, angular momentum $J$, and charge $Q$ reads
\bea
d s^2 &=& -\frac{\Delta}{\Sigma} \left(d t - K\, \sin^2(\theta)\, d \phi\right)^2 + \frac{\sin^2(\theta)}{\Sigma}\, \left(\left(R^2+K^2\right) d \phi - K\, d t\right)^2 +\frac{\Sigma}{\Delta}\, d R^2 + \Sigma\, d \theta^2, \hspace*{10mm}
\eea
where $\Delta = R^2 - 2 M R + K^2 + Q^2$, $\Sigma = R^2 + K^2 \cos^2(\theta)$ and $K=J/M$.
In Boyer-Lindquist coordinates, the Hamiltonian of a test particle is separable in Kerr space-time. From Hamilton-Jacobi theory, a fourth constant of the motion can be derived. This is known as Carter's constant \cite{CC1968}

\item The second source of uncertainty is the case, when $1 - 2 G/r+G Q/r^{2}$ vanishes. In 1D and for $M\,\gg\,\Delta M$ and $Q^2= G\, M^2$, the position uncertainty reads
\bea 
\Delta x ^{(2)} &=& G M \pm \sqrt{G^{2} (M+\Delta M)^{2}-G Q^{2}}, \\
\Delta x ^{(2)} &>& G \sqrt{2 M \Delta M} \geq \frac{2G}{c^2} \Delta M=\frac{2G}{c^3} \frac{h}{\lambda}.
\eea
By means of inequality $\lambda/\sin \theta \geq \lambda$, the uncertainty in $\Delta x^{(2)}$ and the quantity itself can be combined, linearly 
\bea
\Delta x & \gtrsim & \lambda +\kappa\, \frac{l_p^{2}}{\lambda} \label{eq:f}
\eea
where $\Delta x \gtrsim \frac{\hbar}{\Delta p} + c~G  \Delta p$ or $\Delta x  \gtrsim \frac{\hbar}{\Delta p} + \beta  \Delta p$, with $\kappa$ is a constant. The other numerical constant $\beta$ cannot be predicted by the model-independent arguments presented so far. It is natural to investigate whether the relation given in  Eq. (\ref{eq:f})  reproduces what was obtained considering only a very specific measurement. This principle would assure that the results should have a more general validity in QG.
\end{itemize}

In a {\it gedanken} experiment of a micro 4-dimensional black hole \cite{7}, another approach has been deduced. This approach is given as function of time and energy. When position with a precision $\Delta x$ is measured, the quantum fluctuations of the metric field around the measured position with energy amplitude can be expected as $\Delta E \sim  \frac{c\,  \hbar}{2\, \Delta x}$.
The Schwarzschild radius associated with the energy fluctuation $\Delta E$, $R_s = 2\,  G_N\,  \Delta E/c^4$.
The energy fluctuation $\Delta E$ would grow up and the corresponding the radius $R_s$ would become larger and larger, until it reaches the same size as $\Delta x$. As it is well known, the critical length is the Planck length, $R_s = \Delta x \equiv l_p$, where $l_p ^{2}=G_N  \hbar/c^3$ and the associated energy is the Planck energy $\epsilon_p=\hbar\, c/(2\, l_p) =  \sqrt{\hbar\, c^5/G_N}/2$.

When the discussion is limited to the Planck energy, the Schwarzschild radius $R_s$ is considerably enlarged. The situation can be summarized by the inequalities
$\Delta x \gtrsim  \frac{c  \hbar}{2 \Delta E}  \Longrightarrow   \Delta E \ll \epsilon_p$ or $\Delta x \gtrsim 2  G_N  \Delta E/c^4  \Longrightarrow   \Delta E \sim \epsilon_p$.
If these two inequalities are combined linearly, then
\bea
\Delta x &\gtrsim&  \frac{c  \hbar}{2 \Delta E} + \frac{2  G_N  \Delta E}{c^4}. 
\eea
This is a generalization of the uncertainty principle to the cases in which gravity gets very important, i.e. to energies of order of $\epsilon_p$. We have discussed this in connection with the various colliders and the indirect relation between energy and wavelength. We noticed that this relation might be violated at very high energy due to the dominant role of gravity at this energy scale. It is obvious that the minimum value of $\Delta x$ is reached for $\Delta E_{max} \sim \epsilon_P$, 
$\Delta x_{min} = 2\, l_p$.

\subsection{Snyder form}
\label{ssec:3} 

A relationship between a dual structure and the associated product rules fulfilling certain compatibility conditions is introduced by the Hopf algebra \cite{MR28}. An additional structure was found in this geometric approach. The curvature of momentum-space is expressed in terms of coproducts and antipodes of the Hopf algebra \cite{MR28}. In light of this, a theory for quantized space-time was proposed \cite{8,9}. In resolving the infinities problem in early days of QFT different possibilities are investigated.  A de-Sitter space with real coordinates $(\eta_{0},\eta_{1},\eta_{2},\eta_{3},\eta_{4})$ was taken into account. By choosing different parametrizations of the hypersurface than the ones proposed in Ref. \cite{MR28}, one can also use different coordinates in the momentum-space. One such parametrizations, coordinates $\pi_{\nu}$ are related to Snyder basis \cite{MR28}:
\bea
\eta_{0} &=& - m_{p}\, \sinh\left(\frac{\pi_{0}}{m_{p}}\right) - \frac{\vec{\pi}^{2}}{2\, m_{p}} \exp\left(\frac{\pi_{0}}{m_{p}}\right),\\
\eta_{i} &=& - \pi_{i}\, \exp\left(\frac{\pi_{0}}{m_{p}}\right),\\
\eta_{4} &=& - m_{p}\, \cosh\left(\frac{\pi_{0}}{m_{p}}\right) - \frac{\vec{\pi}^{2}}{2\, m_{p}} \exp\left(\frac{\pi_{0}}{m_{p}}\right),
\eea
where on the hypersurface $\eta_{4}$ is not constant and $\pi_{\nu}$ is the bicrossproduct basis of the Hopf algebra \cite{MR28}.

The position $X$ and time $T$ operators, which act on functions of variables $(\eta _{0},\eta _{1},\eta_{2},\eta _{3},\eta _{4})$, are defined as \cite{8,9} 
$X_{i} = i\, a\, \left(\eta_{4}\, \frac{\partial}{\partial\, \eta_{i}} - \eta_{i}\, \frac{\partial}{\partial\, \eta_{4}}\right)$ and 
$T = \frac{i\, a}{c}\left(\eta_{4}\, \frac{\partial}{\partial\, \eta_{i}} + \eta_{i}\, \frac{\partial}{\partial\, \eta_{4}}\right)$, respectively,  where $i=1,2,3$ and $a$ is a natural unit of length. Also, the energy  $P_{i} = (\hbar/a)\, \eta_{i}/\eta_{4}$, and momentum operators
$P_{T} = (\hbar/a)\, \eta_{0}/\eta_{4}$  \cite{8,9}.
Thus, the commutators between positions and momenta read
\bea
\left[X_i,\, P_j\right] &=& i\, \hbar\, \left[1+\left(\frac{a}{\hbar}\right)^{2}\, P^{2}\right],
\eea
where $P^{2}=\sum_{j}^3 P_j\, P_j$.

\subsection{Modified de Broglie relation}
\label{ssec:4}

The modified de Broglie relation has been investigated by Hossenfelder {\it et al.} \cite{11}. It is conjectured that the wave number $\kappa$ is linearly depending on   $p$ (especially for small values) and asymptotically approaching an upper limit, which is proportional to a minimal length $M_p \sim L_p^{-1}$ \cite{11}. Such a function reads $\kappa(p) = p -\sigma \frac{p_3}{m_p^{2}}$. By taking into consideration the commutation relation between $x$ and $\kappa(p)$, GUP can be deduced as $\Delta x\; \Delta p \geq  \frac{\hbar}{2} \left\langle \partial p/\partial \kappa \right\rangle$ \cite{11} or 
\bea
\Delta x\; \Delta p &\geq &  \frac{\hbar}{2} \left(1+ \sigma \frac{<p^2>}{M_p}\right).
\eea
This gives the commutation relation (de Broglie)
\bea
\left[ X_i\,, P_j  \right] &=& i\, \hbar\, \left(1+ \sigma \frac{p^2}{M_p}\right).
\eea
It is obvious that these algebraic relations match well with the generalized uncertainty commutation relation presented in Refs. \cite{3,7,16,19}.

\subsection{Doubly Special Relativity}
\label{ssec:5}

The doubly relativistic theories are group of transformations with two invariants \cite{12}, the constant speed of light and an invariant energy scale. Nevertheless, they remain Lorentzian. A non-linear realization of Lorentz transformations ($E$, $p$) parametrized by an invariant length $l$ was proposed in Ref. \cite{13}. The auxiliary linearly transforming variables $\epsilon$, and  $\pi$, respectively, read
\bea
\epsilon &=& E\, f\left(l\, E, l^{2}\, p^2 \right), \\
\pi_{i} &=& P_i\, g\left(l\, E, l^{2}\, p^2 \right).
\eea
With rotations realized as linearly depending on the dimensional scale \cite{12}, the two functions $f$ and $g$ parametrize non-linear realization of the Lorentz transformations. Corresponding to the choice of $f$ and $g$ \cite{Amelino,garay1,Scardigli}, the Lorentz transformations of the energy-momentum of a particle in different inertial frames should differ from the transformations, which recover a non-linear realization of the Lorentz transformation, when $l\, E \ll 1$ and $l^{2}\, p^{2}\ll 1$
\bea
f &=& \frac{1}{2} \left[\left(1+ l^2\, p^{2} \right) \frac{e^{l\, E}}{l\, E} -\frac{e^{-l\, E}}{l\, E} \right], \\
g &=& e^{l\, E}.
\eea
For a particle of mass $m$, the relation between the energy and momentum $\left(1- l^2\, p^{2} \right) e^{l\, E}+e^{-l\, E}=e^{l\, m}+e^{-l\, m}$ \cite{Amelino,garay1,Scardigli}. Accordingly, $\exp(l\, E) = (\cosh(l\, m)+\sqrt{\cosh^{2}(l\, m)-\left(1-l^{2} p^{2}\right)})/\left(1-l^{2} p^{2}\right)$.
Furthermore, the upper bound on the momentum is $p_{max}^{2}<1/l^{2}$. This suggests the existence of a minimal measurable length restricting the momentum to take any arbitrary value. Thus, an upper bound, $P_{max}$, is likely. At the Planck scale, this leads to a maximal momentum due to the fundamental structure of space-time \cite{12}.

Following commutation relation was suggested in Ref. \cite{12}
\bea
\left[X_i\,, P_j \right] &=& i\, \hbar\, \left[ e^{-l\, E} \delta_{i\, j} + \frac{l^2\,  p_{i} p_{j}}{\cosh {(l\, m)}} \right].
\eea
It is obvious that when the mass $m$ becomes much larger than the inverse of the length scale $l$, a classical phase-space is approached. This result obviously relates the transition from quantum to classical behavior with the modification of QM. The latter is induced by a modification of the relativity principle \cite{12}. 

If we consider massless particle, then $\exp(l\, E) =1/1- l\, |\textbf{p}|$ and the commutation relation should be modified \cite{12} 
\bea
\left[ X_i,\, P_j \right] &=& i\, \hbar\, \left[(1 - l\, |\textbf{p}|)\, \delta_{i j}+ l^2\,  p_{i}\, p_{j} \right]. \label{uuu}
\eea
When the momentum approaches its maximum value, a non-trivial limit for the canonical commutation relation shall be reached \cite{12}.  

\newpage 

\section{Minimal length uncertainty relation} 
\label{sec:min1}
 
Based on HUP, it exists no restriction on the measurement precision for the particle's position, $\Delta x$. This minimal position uncertainty can be made arbitrarily small even down to zero \cite{Scardigli}. The theoretical argumentation to avoid such a limit is reviewed in section \ref{history}. It is obvious that going down to such a limit is not essentially the case of the framework of GUP, because of the existence of a minimal length uncertainty, section \ref{history}, which obviously modifies the Hamiltonian of the physical system leading to modifications, especially at the Planck scale, in the energy spectrum of the quantum system, which in turn predict small corrections in the measurable quantities. As discussed in section \ref{sec:gup1}, this has been observed in condensed matter and atomic physics experiments, such as Lamb shift \cite{Das,Das1}, Landau levels \cite{Das,Das1}, and the Scanning Tunnelling Microscope (STM) \cite{Das1}. Thus, a hope arises that the quantum gravity effects may be observable in the laboratory.

We review two GUP approaches suggesting the existence of minimal length uncertainty. We summarize the mean features to each of them in Tab. \ref{tab:1}. In section \ref{sec:mm1}, we show the proposal of the minimal length uncertainty with momentum modification  \cite{Das,Das1}. In section \ref{sec:hs1}, we study the main features in Hilbert space representation of QM for the minimal length uncertainty \cite{16}.

\subsection{Momentum modification} 
\label{sec:mm1}

Via Jacobi identity, the GUP approach modifies the Heisenberg algebra 
\bea  \label{veg}
\left[x_{i},\, p_{j}\right]=i \, \hbar\, \left(\delta _{i j}(1+\beta\, p^{2}) + 2\, \beta\, p_{i}\, p_{j} \right),
\eea
This ensures \cite{Das,Das1} that $\left[x_{i},\, x_{j}\right]=\left[p_{i},\, p_{j}\right]=0$. Thus, both position and momentum operators read  
\bea 
X_{i} &=& x_{0 i}, \label{dass}\\  
P_{j} &=& p_{0 j}\, (1+\beta\, p_{0}^{2}). \label{dasss}
\eea
It is obvious that $p_{0}^{2} = \sum_{j}^3 p_{0 j} p_{0 j}$ satisfies the canonical commutation relations $\left[x_{0 i},\, p_{0 j}\right] = i\, \hbar\, \delta_{i j}$ and $p_{0 j}$ is defined as the momentum at low-energy scale; $p_{0 j}=-i\, \hbar(\de/\de\, x_{0 j})$, while $P_{j}$ is considered as the momentum at high-energy scale.

\subsubsection{Main difficulties with this proposal}

As discussed earlier, the introduction of a minimal length leads to modification in the canonical commutation relations, while the position space at the Planck scale must differ from the position in the canonical system, because the absence of zero-state in the position eigenstates. Thus, it is useful to modify the position space rather to allow for modification in momentum space. The latter leads to non-commutation of space $\left[x_{i},\, x_{j}\right] \ne 0$.

From the assumptions given in Eqs. (\ref{dass}) and (\ref{dasss}), it is impossible to utilize Hilbert representation for the position space, since no zero physical state exists. With  the definition of the modified  momentum at the highest energy scales, Eq.  (\ref{dasss}), the non-commutative values of the momentum states $\left[p_{i},\, p_{j}\right] \ne 0$. We conclude that {\bf this approach fails to be represented in the Hilbert space}.

\subsection{Hilbert space representation} 
\label{sec:hs1}

We discuss a generalized framework to implement the appearance of a non-zero minimal uncertainty in the position. The discussion can be confined to exploring the applications of such a minimal uncertainty in the context of non-relativistic QM. Various features of the Hilbert space representation of QM, especially at the Planck scale, are introduced in Ref. \cite{16}.
\bea 
\Delta x\, \Delta p &\geq & \frac{\hbar}{2}+ \beta_0\, l_p^{2}\, \frac{(\Delta p)^{2}}{\hbar ^{2}}. 
\eea
The extra term, $\beta_0\, l_p ^{2}\, (\Delta p)^{2}/\hbar ^{2}$, finds its origin in nature of the spacetime at the Planck energy $\epsilon_p $  (of $10^{39}$ GeV) \cite{Scardigli,16}. The simplest GUP approach implies the appearance of a non-zero minimal uncertainty $\Delta x_0$  
\bea 
\Delta x\, \Delta p \geq \frac{\hbar}{2} \left(1+ \beta\, (\Delta p)^{2} \right),  \label{eg}
\eea
where $\beta = \beta_0/(M_p\, c^2) =\beta_0\, l_p ^2/\hbar ^2$ is the GUP parameter.

As a non-trivial assumption,  the minimal observable length is conjectured to have a minimal but non-zero uncertainty. Therefore,  the Hilbert space representation on position space wavefunctions of ordinary QM \cite{16} is no longer possible, as no physical system with a vanishing position eigenstate $|x \rangle$ is allowed \cite{16}. In light of this, a new Hilbert space representation which should be compatible with the commutation relation in GUP, Eq. (\ref{eg}), must be constructed. This means working with the convenient representation of the commutation relations on momentum space wavefunctions \cite{16}. Accordingly, the Heisenberg algebra of GUP is given as \cite{16,gupps2,3,Inflation2q,14,kmpf32,15,Scardigli}
\bea 
\left[x,\, p\right] &=& i\, \hbar\, \left(1+\beta\, p^{2}\right). \label{KMM1}
\eea
The Heisenberg algebra can be represented in the momentum space wavefunctions $\phi(p)=\langle p| \phi (p) \rangle $ and $\partial_{p}=i \hbar (\partial/\partial x)$
\bea 
\textbf{P}\, \cdot \phi(p) &=& p\, \phi(p), \label{eq:app} \\ 
\textbf{X}\, \cdot \phi(p) &=& i\, \hbar\, \left(1+\beta\, p^{2}\right) \partial_{p} \phi(p), \label{eq:app1}
\eea
where $\textbf{X}$ and $\textbf{P}$ are symmetric operators on the dense domain $S_{\infty}$ with respect to the scalar product $\langle \phi | \psi \rangle = \int_{-\infty}^{\infty} \frac{dp}{1+\beta p^2} \phi^{*} (p) \psi (p)$, the identity operator $\int_{-\infty}^{\infty} \frac{dp}{1+\beta\, p^2} | p \rangle \langle p | = 1$ and the scalar product of the momentum eigenstates changes to $
\langle p | p^{'} \rangle = \left(1+ \beta\,  p^{2} \right) \delta \left( p - p^{'} \right)$.
While the momentum operator  essentially still self-adjoint, the functional analysis of the position operator as expected from the appearance of the minimal uncertainty in positions should be changed. For $ (\Delta\, p)^{2} = \langle p^{2}\rangle -\langle p\rangle^{2}$ \cite{16}
\bea 
\Delta x\, \Delta p &\geq & \frac{\hbar}{2}\left(1+\beta\, (\Delta p)^{2} +\beta\, \langle p \rangle ^{2} \right).
\eea
This relation can be rewritten as a second-order equation for $\Delta p$. Then, the solutions for $\Delta p$ are  \cite{16}
\bea
\Delta p &=& \left(\frac{\Delta x}{\hbar\, \beta}\right) \pm \sqrt{\left(\frac{\Delta\, x}{\hbar\,  \beta}\right)^{2}-\frac{1}{\beta}-\langle p \rangle^{2}}.
\eea
A minimum position uncertainty $\Delta x_{min} (\langle p \rangle) = \hbar\, \sqrt{\beta} \sqrt{1+\beta\, \langle p \rangle ^{2}}$.
Therefore, the absolutely smallest uncertainty in position, where $\langle p \rangle =0$, $\Delta x_{0}=\hbar\, \sqrt{\beta}$.
There is a non-vanishing minimal momentum uncertainty as accepted from Fig. \ref{fig1}.

For Hilbert space representations, one can't work on position space. One has to resort a generalized Bargmann-Fock representation \cite{fock,fock1}. Here, the situation with non-zero minimal position uncertainties should be specified. For $n$D, the generalised Heisenberg algebra, Eq. (\ref{eg}), reads \cite{16,gupps2,3,Inflation2q,14,kmpf32,15,Scardigli}
\be 
\left[x_{i},\, p_{j}\right]= i\, \hbar\,  \left(1+\beta\, \vec{p}^{2} \right). \label{egu1}
\ee
It requires that 
\be 
\left[p_{i},\, p_{j} \right] = 0,  \label{egu2}
\ee
in order to allow the generalization of the momentum space representation \cite{16}
\bea
\textbf{P}_{i} \cdot \phi(p) &=& p_{i}\, \phi(p),\\
\textbf{X}_{i} \cdot \phi(p) &=& i\, \hbar\, \left(1+\beta \vec{\textbf{p}}^{2} \right) \partial_{p_{i}} \phi(p),
\eea
and $\partial_{p_{i}}=i\, \hbar\, (\partial/\partial p_{i})$. It turns to be obvious that
\bea
\left[{\bf X}_{i},\, {\bf X}_{j} \right] &=& 2\, i\, \hbar\, \beta\, \left({\bf P}_{i}\, {\bf X}_{j} - {\bf P}_{j}\, {\bf X}_{i} \right), \label{egu3} 
\eea
leads to a non-commutative geometric generalization of the position space.

Furthermore, the commutation relations, Eqs. (\ref{egu1}), (\ref{egu2}) and (\ref{egu3}) do not violate the rotational symmetry \cite{16}. In fact, the rotation generators can be expressed in terms of position and momentum operators  \cite{16}
${\bf L}_{i j} = ({\bf X}_{i}\, {\bf P}_{j} -  {\bf X}_{j}\, {\bf P}_{i})/(1+ \beta\, \vec{\textbf{p}}^{2})$, where their representation in momentum wavefunctions ${\bf L}_{i j}\, \psi(p) = -\, \, i\, \hbar\, \left( p_{i}\, \partial_{p_{j}} - p_{j}\, \partial_{p_{i}}\right)\, \psi(p)$
are essentially the same as encountered in ordinary QM. However, the main change now appears in the relation
\be
\left[x_{i},\, x_{j} \right] = -\, 2\, i\, \hbar\, \beta\, \left( 1+ \beta\, \vec{\textbf{p}}^{2} \right)\; L_{i\, j}.
\ee
Once again, this relation reflects the noncommutative nature of the spacetime manifold at the Planck scale.

\subsubsection{Eigenstates of position operator in momentum space}

The position operators generating momentum-space eigenstates are given as \cite{16}
\bea 
\textbf{X} \, \phi_{\lambda} (p) &=& \lambda\, \phi_{\lambda}(p),\\
i\, \hbar\, \left(1+\beta\, p^{2}\right)\, \partial_{p}\, \phi_{\lambda}(p) &=& \lambda\, \phi_{\lambda}(p).
\eea
This differential equation can be solved to obtain formal position eigenvectors $\phi_{\lambda} (p) = C\, \exp(-i \lambda/(\hbar\, \sqrt{\beta})\, \tan^{-1}\, \sqrt{\beta}\, p)$ \cite{16}.  By applying the normalization condition, the formal position eigenvectors in momentum-space can be found $
\phi_{\lambda} (p) = \sqrt{\sqrt{\beta}/\pi} \exp(- i\, \lambda/(\hbar\, \sqrt{\beta}) \tan^{-1} \sqrt{\beta}\, p)$  \cite{16}. 
This is the generalized momentum-space eigenstate of the position operator in the presence of both a minimal length and a maximal momentum. To this end, we calculate the scalar product of the momentum space eigenstate of the position operator $|\phi_{\lambda}(p)\rangle$, $\langle \phi_{\lambda^{'}} |\phi_{\lambda} \rangle = \sqrt{\beta}/\pi\, \int_{-\infty}^{\infty} \frac{dp}{1+\beta p^2}\exp(-i (\lambda-\lambda^{'})/(\hbar \sqrt{\beta}) \tan^{-1} \sqrt{\beta} p)$  \cite{16}.
Thus,
\bea
\langle \phi_{\lambda^{'}} |\phi_{\lambda} \rangle &= \frac{2\, \hbar\, \sqrt{\beta}}{\pi\, \left(\lambda-\lambda^{'}\right)} \sin\left(\frac{(\lambda-\lambda^{'})}{2\, \hbar\, \sqrt{\beta}} \pi\right).
\eea

\begin{figure}[htb]
\vspace*{-2cm}
\centering{
\includegraphics[scale={.4}]{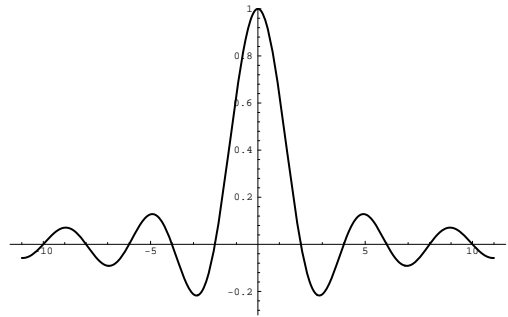}\vspace*{-2.5cm}
\caption{$\langle \phi_{\lambda^{'}} | \phi_{\lambda}\rangle$ is plotted as function of $\lambda - \lambda^{'}$, which is normalized to $\hbar\, \sqrt{\beta}$. The graph taken from Ref. \cite{16}. }
\label{fig:phi2}
}
\end{figure}

Fig. \ref{fig:phi2} compares the behavior of $\langle \phi_{\lambda^{'}} | \phi_{\lambda} \rangle$ as function of $\lambda - \lambda^{'}$ normalized to $\hbar\, \sqrt{\beta}$. The curve stands for the special case $\Delta p_{0}$, which is is set of eigenvectors parametrised by $\lambda\, \in\, [-1,1[$ \cite{16}. It apparent that the standard position eigenstates are generally no longer orthogonal, because the formal position eigenvectors $|\phi_{\lambda} \rangle$ are not physical states. They are not part of the domain of $\texttt{p}$. The latter physically means that they have infinite uncertainty in momentum and - in particular - infinite energy $\left\langle \phi_{\lambda} \left| \frac{\textbf{p}^{2}}{2\, m} \right| \phi_{\lambda} \right\rangle = \texttt{divergent}$. 
This is the main difficulty with this approach; the energy of the short wavelength modes seems to diverge \cite{16}.

\subsubsection{Maximal localization states} 
\label{Localization}

The maximum localization around $\zeta$ position states $|\phi_{\zeta}^{ml}\rangle$, $\left\langle \phi_{\zeta} ^{ml}\left| \hat{X} \right|\phi_{\zeta} ^{ml} \right\rangle = \zeta$, and $\Delta x_{min} = \Delta x_{0}$ depends on $\langle p \rangle$. These states satisfy the inequality \cite{16}
\bea  
\left|\left|\left((x-\langle x \rangle)+ (p-\langle p \rangle )\frac{\left[x, p \right]}{2 (\Delta p)^{2}}\right) | \phi \rangle \right|\right| \geq 0, \label{equaa} 
\eea 
which immediately implies 
\bea
\Delta x\, \Delta p &\geq & \frac{1}{2}\left| \langle \left[x, p \right]\rangle \right|.
\eea

For first-order GUP parameter, we can use the approximate relation \cite{16,Scardigli,Das,Das1}
\bea 
\left|\langle \left[x,\, p \right] \rangle\right| &\approx & i\, \hbar\, \left(1+\beta (\Delta p)^{2} +\beta\, \langle p \rangle ^{2} \right).
\eea
In the momentum space and from Eqs. (\ref{eq:app}) and (\ref{eq:app1}), this gives the differential equation \cite{16}
\bea 
\left\{\left[i\hbar(1+\beta p^{2} \right)\partial_{p}  - \langle x \rangle] + i\, \hbar\, \frac{\left(1+\beta (\Delta p)^{2} +\beta \langle p \rangle ^{2} \right)}{2 (\Delta p)^{2} } (p-\langle p \rangle )\right\} \phi (p) &\approx & 0,
\eea
which can be solved  as
\bea
\phi (p) &\approx & C (1+\beta p^{2})^ \frac{-\left(1+\beta (\Delta p)^{2} +\beta \langle p \rangle ^{2} \right)}{4\, \beta\, (\Delta p)^{2}}  \exp \left[\left(\frac{\langle x \rangle}{i\, \hbar\, \sqrt{\beta}} - \frac{\left(1+\beta\, (\Delta p)^{2} +\beta\, \langle p \rangle ^{2} \right) \langle p \rangle}{2\, \sqrt{\beta} (\Delta p)^{2}}\right)\; \tan^{-1}(\sqrt{\beta}\, p)\right].
\eea

At $\langle p \rangle =0$ and critical momentum uncertainty $(\Delta p)^{2}=1/\beta$, the absolutely maximal localization reads  $\phi ^{ml}_{\zeta} (p) \approx  C\, (1+\beta\, p^{2})^{-\frac{1}{2}}\; \exp\left(- i\, \frac{\langle x \rangle \tan^{-1}(\sqrt{\beta} p)}{ \hbar\, \sqrt{\beta}}\right)$  \cite{16}. The momentum space wavefunctions $|\phi_{\zeta} ^{ml} \rangle$ of a maximum localization around $\zeta$ reads
\bea
\phi ^{ml}_{\zeta} (p) &=& \sqrt{\frac{2 \sqrt{\beta}}{\pi}} \left(1+\beta p^{2}\right)^{-\frac{1}{2}} \; \exp\left(- i \frac{\zeta \tan^{-1}(\sqrt{\beta} p)}{ \hbar \sqrt{\beta}}\right).
\eea
These states generalize the plane waves in the momentum-space and describe maximal localization in the ordinary QM. This leads to proper physical states with finite energy \cite{16}
\bea 
\left\langle \phi_{\zeta} ^{ml}\left| \frac{\hat{\textbf{P}}^{2}}{2 m} \right|\phi_{\zeta} ^{ml} \right\rangle &=& \frac{2 \sqrt{\beta}}{\pi} \int_{-\infty}^{\infty} \frac{dp}{(1+\beta p^2)^{2}} \frac{p^{2}}{2m}= \frac{1}{2 m \beta}.
\eea

\subsubsection{Transformation to quasi-position wavefunctions}

\begin{figure}[htb]
\centering{
\includegraphics[width=10cm]{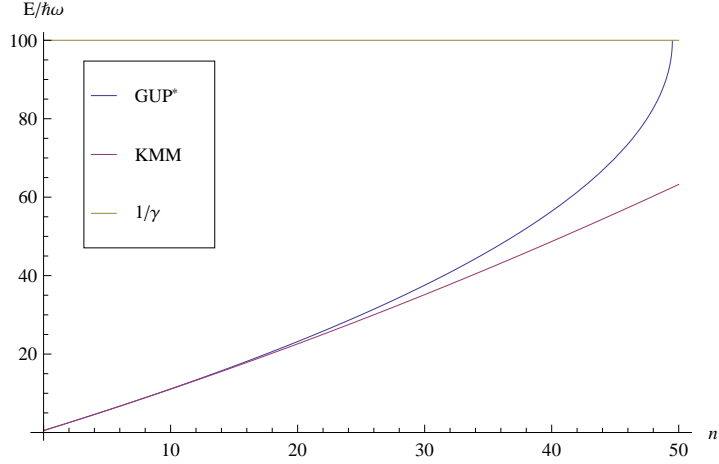}
\caption{The wavelength of quasiposition wave is given as function of the momentum eigenstate, Eq. (\ref{quiB}), in ordinary QM and GUP approach at $\beta=0.2$. The graph taken from Ref. \cite{pedram}.  \label{figwave3}
}
}
\end{figure} 

Through projecting arbitrary states on maximally localized states,  the probability amplitude for the particle being maximally localized around a position can be obtained. For quasiposition wavefunction $\phi(\zeta)=\langle \phi^{ml} _{\zeta} | \phi \rangle$ \cite{16},  where in the limit $\beta \rightarrow 0$, the ordinary position wave function $\phi(\zeta)=\langle\zeta|\phi\rangle$. The quasiposition wavefunction of a momentum eigenstate $\phi_{\widetilde p} (P) = \delta (p-\widetilde p)$ with energy $E = {\widetilde p}^{2}/2\, m$ is characterized as a plane wave. The transformation of the wavefunction in momentum representation into its counterpart quasiposition wavefunction is given as \cite{16}
\bea 
\phi (\zeta) &=& \sqrt{\frac{2\, \sqrt{\beta}}{\pi}} \int_{-\infty}^{\infty} \frac{d p}{(1+\beta\, p^2)^{\frac{3}{2}}}\; \exp\left[i\, \frac{\zeta\, \tan^{-1}(\sqrt{\beta}\, p)}{ \hbar\, \sqrt{\beta}}\right]\, \phi(p). \label{ddd}
\eea
In terms of modified dispersion relation, the wavelength is given as  \cite{16}
\bea \label{quiB}
\lambda (E) =\frac{2\, \pi\, \hbar\, \sqrt{\beta}}{\tan^{-1}\sqrt{2\, m\, \beta\, E}}. 
\eea  
In absence of GUP, we get $\lambda_{0} = 4\, \hbar\, \sqrt{\beta}$, no wavelength components is allowed which is smaller than $\lambda_{0}$.
Furthermore, no arbitrarily fine ripples are possible, because the energy of short wavelength diverges when the wavelength approaches $\lambda_{0}$  
\bea \label{qui}
 E(\lambda) &=& \frac{1}{2 m \beta}\, \left(\tan \frac{2 \pi \hbar \sqrt{\beta}}{\lambda} \right)^{2}. 
\eea

Fig.~\ref{figwave3} illustrates $\lambda(E)$ in dependence on $m\, E$ for ordinary QM and GUP approach at $\beta=0.2$. It is obvious that Eq. (\ref{quiB}) is bounded from below. Therefore, a nonzero minimal wavelength is likely \cite{16}. The transformation, Eq. (\ref{ddd}), is Fourier type. The transformation of a quasiposition wavefunction into a momentum-space wavefunction  \cite{16} $\phi(p) = (8\, \pi\,  \sqrt{\beta}\, \hbar)^{-1} \int_{-\infty}^{\infty} d \zeta\, (1+\beta\, p^2)^{1/2}\; \exp\left[-i\, \frac{\zeta\, \tan^{-1}(\sqrt{\beta}\, p)}{\hbar\, \sqrt{\beta}}\right]\;  \phi(\zeta)$.

\section{Minimal length uncertainty: maximal momentum}
 \label{sec:max}

\subsection{Momentum modification}

Based on DSR, the GUP approach suggests modifications in the commutators \cite{12}
\bea 
\left[x_{i}\,, p_{j}\right] &=& i\, \hbar\, \left(\delta_{ij}\, (1+\beta p^{2})+2 \beta p_{i}p_{j} \right)  =  i\, \hbar\, \left[(1- l_{pl} |\textbf{p}|) \delta_{i j}+ l_{pl}^2  p_{i} p_{j} \right]. \label{dsr2} \\
\left[x_{i} ,p_{j}\right] &=& i\, \hbar\, \left[\delta_{i j} +\alpha_{1}\, p\, \delta_{i j}+ \alpha_{2} \frac{p_{i}\, p_{j}}{p} + \beta_{1}\, p^{2}\, \delta_{i j} + \beta_{2}\, p_{i}\, p_{j} \right].
\eea 
Then from Jacobi identity, it follows that
\bea
- \left[[x_{i}, x_{j}], p_{k} \right] = \left[[x_{j}, p_{k}], x_{i}\right] + \left[[p_{k}, x_{i}], x_{j}\right] & = & 0, \label{uuu2}\\
\left[\left(\frac{\alpha_{1} - \alpha_{2}}{p}\right) + \left(\alpha_{1}^{2} + 2\beta_{1} - \beta_{2}\right) \right] \Delta_{j k i } &=& 0,
\eea
where $\Delta _{jki} = p_{i} \delta_{jk} - p_{j} \delta_{ik}$. It was assumed that $\alpha_{1}=\alpha_{2}=-\alpha$, where the negative sign appearing in Eq. (\ref{uuu2}) or Eq. (\ref{dsr2}). At $\alpha>0$, then $\alpha_{1}^{2}+2\,\beta_{1}-\beta_{2}=0$ has the roots $\beta_{1}=\alpha^{2}$ and $\beta_{2}=3 \alpha ^{2}$ with $\alpha^{2}=\beta$. The resulting commutators are consistent with the string theory, black holes physics and DSR
\bea
\left[x_{i}, p_{j} \right]=i \hbar \left[ \delta _{i j} -\alpha \left( p \delta _{i j} +\frac{p_{i} p_{j}}{p} \right)+\alpha ^{2} \left( p^{2} \delta _{ij} +3 p_{i} p_{j} \right)\right]. \label{ali1}
\eea
By using the Jacobi identity, 
\bea
\left[x_{i} ,x_{j}\right]=\left[p_{i} ,p_{j}\right] &=& 0, \label{ali2}
\eea
where $\alpha = \alpha_{0} \, \ell_{pl}/\hbar= \alpha_{0}/ (M_{pl}\, c)$ and the Planck length  $\ell _{pl} \approx 10^{-35}~$m and energy $\epsilon_{pl} = M_{pl} c^{2} \approx 10^{19}~$GeV.

In $1$D, this GUP approach was formulated as \cite{advplb,Das:2010zf}
\bea 
\Delta x\, \Delta p &\geq & \frac{\hbar}{2} \left(1-2\, \alpha\, \langle p \rangle +4\, \alpha^{2} \langle p^{2} \rangle \right). \label{ali3}
\eea
It is obvious that $(\Delta p)^{2}=\langle p^{2}\rangle -\langle p\rangle^{2}$ and therefore 
\bea 
\Delta x\, \Delta p &\geq &\frac{\hbar}{2} \left[1+\left(\frac{\alpha}{\sqrt{\langle p^{2} \rangle}}+4\, \alpha^{2} \right)(\Delta p)^{2} + 4\, \alpha^{2}\, \langle p \rangle^{2} - 2\, \alpha\, \sqrt{\langle p^{2} \rangle} \right].
\eea
The commutators and inequalities similar to the ones given in Eqs. (\ref{ali1})  and (\ref{ali3}) have been proposed and derived in Ref. \cite{advplb,Das:2010zf}. This implies a minimum measurable length and a maximum measurable momentum, simultaneously
\bea
\Delta x &\geq & (\Delta x )_{min} \approx \alpha\, \hbar \approx \alpha_{0}\, \ell_{pl}, \label{app1} \\
\Delta p &\le &  (\Delta p )_{max} \approx \frac{1}{\alpha} \approx \frac{M_{pl}\, c}{\alpha _{0}}, \label{app2}
\eea
and defines 
\bea 
X_{i} &=& x_{0 i}, \label{ali555}\\ 
P_{j} &=& p_{0 j} (1 - \alpha\, p_{0} +2\, \alpha^{2}\, p_{0}^{2}). \label{ali222}
\eea 
We note that $p_{0}^{2}=\sum_{j}^3\,p_{0 j}\,p_{0 j}$ satisfies the canonical commutation relations $\left[x_{0 i}, p_{0 j}\right] = i \hbar \delta_{i j}$ and $p_{0j}$ is defined as the momentum at low-energy scale, which is represented by  $p_{0 j}=-\,i\,\hbar\;\de/\de x_{0 j}$, while $p_{j}$ is considered as the momentum at high-energy scale. It is assumed that the dimensionless parameter $\alpha_{0}$ has value very close to unity. In this case, the $\alpha$-dependent terms are important only when the energies (momenta) are comparable to the Planck energy (momentum), and the lengths are comparable to the Planck length. This implies the

\subsubsection{Main difficulties with the GUP approach \cite{advplb}}
Regardless the wide range of applications in different physical systems, crucial difficulties are listed out in Ref. \cite{pedram}:
\begin{itemize}
\item It contains linear and quadratic terms of momenta with a minimum measurable length and a maximum measurable momentum. 
\item It was claimed that when the energy becomes close the Planck limit, there should be a modification in Eq. (\ref{ali222}) and this should ensure commutators of space, Eq. (\ref{ali2}), as the canonical system, which can predict the measurable length and a maximum measurable momentum, simultaneously. 
\item it is a perturbative approach. Therefore, it is only valid for small values of the GUP parameter $\alpha$,
\item it can not approach the non-commutative geometry, see Eq. (\ref{ali2}),
\item it suggests a minimal length uncertainty which can be interpreted as the minimal length. The maximal momentum uncertainty differs from the idea of the maximal momentum which is required in DSR theories, where the maximal momentum given in uncertainty not on the value of the observed momentum, see Eq. (\ref{app2}), 
\item it suggests momentum modification given in Eq. (\ref{ali222}), but does not achieve the commutator relation of the momentum space $[p_{i},p_{j}] \ne 0$, 
\item its minimal length uncertainty with maximal momentum results in uncertainty instead of  maximum observed momentum, see Eqs. (\ref{app1}) and (\ref{app2}), and
\item the introduction of the minimal length (non-varnishing value) allows the study for  the Hilbert space representation corresponding to the momentum wavefunction $\psi(p)$. 
\end{itemize}

\subsection{Hilbert space representation}

The first term in Eq. (\ref{ali3}) is related to the momentum and refers to maximal momentum. In this term, various differences between the Hilbert space representation and the work of KMM \cite{16} can be originated. Assuming that the minimal observable length has a non-vanishing uncertainty, one should construct a new Hilbert space representation, which is compatible with the commutation relation accompanied with the GUP approach $[x_{i}\,, p_{j}] = i\, \hbar\,  \delta_{i j}\, \left(1-\alpha\,  p  +2\, \alpha^{2}\,  \vec{p}^{2} \right)$.
But, when neglecting the minimal momentum uncertainty, there would still exist a continuous momentum space representation. This means that various physical applications of the minimal length by implementing convenient representation of the commutation relations on momentum-space wavefunctions can be explored \cite{amir}
\bea 
\textbf{X}_{i}\, \phi (p) &=& x_{0 i}(1 -\alpha p_{0} +2\, \alpha^{2}\, \vec{p_{0}}^{2})\, \phi (p), \label{amm1}\\
\textbf{P}_{j}\, \phi (p) &=&  p_{0 j}\, \phi (p), \label{amm2}
\eea
where $p_{0}^{2} = \sum_{j}^3\, p_{0 j}\, p_{0 j}$ satisfying the canonical commutation relations $\left[x_{0i },\, p_{0 j}\right] = i\, \hbar\, \delta_{i j}$ and $p_{0 j}$ is defined as the momentum at low-energy scale which is represented by  $x_{0 i}=i\, \hbar\, \de_{p_{i}}$. 

These commutation relations imply a nonzero minimal uncertainty in each position coordinate (in ordinary QM,  $\left[p_{i},\, p_{j}\right]=0$). Then, it is straightforward to show that 
\bea
\left[x_{i},\, x_{j}\right] &=& i\, \hbar\, \alpha\, \left(4\, \alpha - \frac{1}{P}\right)\; \left(\textbf{P}_{i} \textbf{X}_{j} - \textbf{P}_{j}\, \textbf{X}_{i} \right).
\eea
In light of this, one should be worry about the divergence in the KMM formalism \cite{16}, at vanishing momentum. Therefore, ''Singularity'' is likely, because the derivative diverges at $p=0$. The commutation relations do not violate the rotational symmetry. In fact, the rotation generators can still be expressed in terms of position and momentum operators $L_{i j} = (\textbf{X}_{i}\,  \textbf{P}_{j} -\textbf{X}_{j}\, \textbf{P}_{i})/(1 -\alpha\, p_{0} +2\, \alpha^{2}\, \vec{p_{0}}^{2})$.
The action on a momentum-space wave function $L_{i j} \phi(p) = -\, i\, \hbar\,  \left( p_{i}\, \partial_{p_{j}} - p_{j}\, \partial_{p_{i}}\right)\, \phi(p)$.
 However, the main difference with the ordinary QM appears in the relation
\bea
\left[x_{i},\, x_{j} \right] &=& i\, \hbar\, \alpha\, \left(4\, \alpha - \frac{1}{P}\right)\, \left(1 - \alpha\, p_{0} +2 \alpha^{2}\, \vec{p_{0}}^{2}\right)\; L_{i j}.  \label{eq:xixj}
\eea

In the original KMM formalism \cite{16}, the $1/P$-term, which represents trace of effect of the maximal momentum, does not exist. The previous equation (\ref{eq:xixj}) express the noncommutative nature of the spacetime manifold at the Planck scale. 
\begin{itemize}
\item The existence of an upper bound of momentum fits well with DSR. In this representation, the scalar product should be modified due to the presence of the additional factor $(1-\alpha\, p_{0} +2 \, \alpha^{2}\, \vec{p_{0}}^{2})$ and the maximal momentum. 
\item The integrals are calculated between $-p_{pl}$ and  $+p_{pl}$, Planck momenta. This differs from the integration region in the KMM formalism \cite{16,amir} and thus implies the existence of a maximal Planck momentum, $p_{pl} \equiv M_{pl}\, c$.
\bea
\langle \phi | \psi \rangle &=& \int_{- p_{pl}}^{+p_{pl}}\, \frac{\phi^{*} (p)\; \psi (p)}{(1 -\alpha\, p_{0} +2\, \alpha^{2} p_{0}^{2} )}\, d p.
\eea
\item  Accordingly, the identity operator  is given as \cite{amir}
\be 
\int_{- p_{pl}}^{+p_{pl}} \frac{|p \rangle \langle p|}{\left(1 -\alpha\, p_{0} +2\, \alpha ^{2}\, p_{0}^{2}\right)}\, d p = 1,
\ee
and the scalar product of the momentum eigenstates should be changed to 
\be 
\langle p | p^{'} \rangle = \left(1 -\alpha\, p_{0} +2\, \alpha^{2}\, p_{0}^{2} \right)\, \delta \left(p - p^{'} \right).
\ee
\end{itemize}

\subsubsection{Eigenstates of position operator in momentum space}

It was proposed  \cite{16,amir} that the position operator acting on the momentum-space eigenstates 
$\textbf{X}\,.\,\phi_{\xi} (p) = \xi\, \phi_{\xi} (p)$, where $\phi_{\xi}(p)=\langle \xi | p \rangle$ is the position eigenstate with $|\xi \rangle$ being an arbitrary state
$i\, \hbar\, \left(1 -\alpha p_{0} +2 \alpha ^{2} p_{0} ^{2} \right)\, \de_{p}\,  \phi_{\xi}(p) = \xi\, \phi_{\xi}(p)$.
By solving this differential equation, the formal position eigenvectors can be derived \cite{amir}
\bea 
\phi_{\xi} (p) &=& C\, \exp \left[-i\, \frac{2\, \xi}{\alpha\, \hbar\, \sqrt{7}} \left(\tan^{-1}\, \frac{1}{\sqrt{7}}+\tan^{-1}\, \frac{4\, \alpha\, p -1}{\sqrt{7}} \right) \right].
\eea
The formal position eigenvectors in the momentum-space can be deduced when the factor  $C$ is extracted and  normalized condition is applied \cite{amir}
\bea
\phi_{\xi} (p) &=&\sqrt{\frac{\alpha\, \sqrt{7}}{2}} \left[\tan^{-1}\, \left(\frac{4\, \alpha\, p_{pl} -1}{\sqrt{7}} \right) +\tan^{-1} \left(\frac{4\, \alpha\, p_{pl} +1}{\sqrt{7}} \right)\right]^{-\frac{1}{2}}\nonumber \\ 
&& \hspace*{7mm}\exp \left[-i\, \frac{2\, \xi}{\alpha\, \hbar\, \sqrt{7}} \left(\tan^{-1} \left(\frac{1}{\sqrt{7}}\right)+\tan^{-1}\left(\frac{4\, \alpha\, p -1}{\sqrt{7}}\right) \right) \right]. \label{eq:gmse}
\eea

The previous expression (\ref{eq:gmse}) represents generalized momentum-space eigenstate of the position operator in presence of both minimal length and maximal momentum. The scalar product of the formal position eigenstates \cite{amir}
\bea
\langle \phi_{\xi^{'}} |\phi_{\xi} \rangle &=& \int_{- p_{pl}}^{+p_{pl}} \frac{dp}{(1 -\alpha\, p_{0} +2\, \alpha^{2}\, p_{0}^{2})} \phi_{\xi^{'}}^{*}(p)\, \phi_{\xi}(p), \nn\\
 &=&  {\frac{\alpha\, \sqrt{7}}{2}}\, \rho_{0}\, \exp\left(-i\, \frac{2\, \left(\xi - \xi^{'}\right)}{\alpha\, \hbar\, \sqrt{7}}\, \tan^{-1}\left(\frac{1}{\sqrt{7}}\right)\right) 
 \int_{- p_{pl}}^{+p_{pl}}\, \frac{\exp\left[{ -i\, \frac{2\, \left(\xi - \xi^{'}\right)}{\alpha\, \hbar\, \sqrt{7}}}\,  \tan^{-1}\left(\frac{4\, \alpha\, p -1}{\sqrt{7}}\right)\right]} {(1 -\alpha\, p_{0} +2\, \alpha^{2}\, p_{0}^{2} )}\, dp, \hspace*{10mm}
\eea
where $\rho_{0}=\left[\tan^{-1} \left(\frac{4\, \alpha\, p_{pl} -1}{\sqrt{7}} \right) +\tan^{-1}\left(\frac{4\, \alpha\, p_{pl} +1}{\sqrt{7}} \right)\right]^{-1}$ and therefore,
\bea
\langle \phi_{\xi^{'}} |\phi_{\xi} \rangle &=&  \Omega\, \left( \exp\left\{-i\, \left[\frac{2 (\xi - \xi ^{'})}{\alpha \hbar \sqrt{7}}\, \tan^{-1}\left(\frac{4 \alpha p_{pl} -1}{\sqrt{7}}\right) -\frac{\pi}{2} \right]\right\}  -   \exp\left\{ i\, \left[ \frac{2 (\xi - \xi ^{'})}{\alpha \hbar \sqrt{7}}\, \tan^{-1}\left(\frac{4 \alpha p_{pl} +1}{\sqrt{7}}\right)+\frac{\pi}{2} \right]\right\}\right), \hspace*{10mm}
\eea
with $\Omega = \frac{\rho_{0}\, \hbar\, \alpha\, \sqrt{7}}{2\, \left(\xi - \xi ^{'}\right)}\; \exp\left[-i\, \frac{2\, \left(\xi - \xi ^{'}\right)}{\alpha\, \hbar\, \sqrt{7}}\; \tan^{-1}\, \left(\frac{1}{\sqrt{7}}\right)\right]$.

\begin{figure}[htb]
\centering{
\includegraphics[scale={.35}]{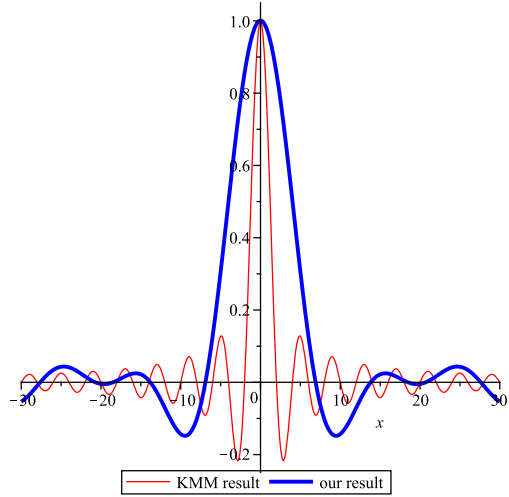}
\caption{The dependence of $\langle \phi_{\xi ^{'}} |\phi_{\xi} \rangle $ on $\xi - \xi ^{'}$ as calculated in two GUP approaches \cite{amir,16}. The graph taken from Ref. \cite{amir}, to which {''our result''} refers. }
\label{fig:4}
}
\end{figure}

Fig. \ref{fig:4} compares between the behavior of $\langle \phi_{\xi^{'}} |\phi_{\xi} \rangle$ and that of $\xi - \xi^{'}$, which are calculated using two GUP approaches \cite{amir,16}. The red (thin) curve shows an oscillating behavior. It is obvious that the scalar product used in the GUP approach \cite{amir} causes   flattening (less oscillation) relative to the one which was introduced in Ref. \cite{16}.

For the formal position eigenvectors, the expectation value of the energy reads $\left\langle \phi_{\xi} \left| \textbf{p}^{2}/(2\, m) \right| \phi_{\xi} \right\rangle = \int_{- p_{pl}}^{+p_{pl}} \phi_{\xi^{'}}^{*}\, p\, \frac{p^{2}}{2\, m} \phi_{\xi}(p)/(1 -\alpha\, p +2\, \alpha^{2}\, p ^{2} )\, d p$.
This leads to $\left\langle \phi_{\xi} \left|\textbf{p}^{2}/(2\, m) \right| \phi_{\xi} \right\rangle = \frac{\alpha\, \sqrt{7}\, \rho_{0}}{4\, m} \int_{- p_{pl}}^{+p_{pl}} \phi_{\xi ^{'}}^{*}\, p\, \frac{p^{2}}{2 m} \phi_{\xi} (p)/(1 -\alpha p +2 \alpha ^{2} p ^{2}) \; p^{2}\, dp,$ 
and therefore  \cite{amir}
\bea
\left\langle \frac{\textbf{p}^{2}}{2\, m} \right\rangle &=& \left[\frac{\sqrt{7} \rho P_{pl}}{4\, m} + \frac{\sqrt{7}\, \rho}{32\, m\, \alpha}\; \ln \left(\frac{1 -\alpha\, p_{pl} +2\, \alpha^{2}\, p_{pl}^{2}}{1 +\alpha\, p_{pl} +2\, \alpha^{2}\, p_{pl}^{2}} \right) -\frac{3}{16\, m\, \alpha} \right].
\eea

About this GUP approach, few remarks are on order now
\begin{itemize}
\item As shown in previous sections, the energy spectrum is not divergent as the one related to the framework of KMM GUP-approach \cite{16}, especially in the presence of both minimal length and maximal momentum,
\item but, it turns out also that the expectation values of the energy as calculated by the GUP approach \cite{advplb,Das:2010zf,afa2} are no longer divergent \cite{amir}. 
\item It should be highlighted that the expectation values of energy are not lying within the domain of $P$, which physically means that they have infinite momentum uncertainty.  
\end{itemize}

\subsubsection{Maximal localization states}

In order to calculate the states $|\phi_{\zeta}^{ml} \rangle$ of the maximum localization around the position $\zeta$, it should be assumed that $\left\langle \phi_{\zeta}^{ml} \left| \hat{X} \right| \phi_{\zeta}^{ml} \right\rangle = \zeta$   \cite{16}.  As in section \ref{Localization} and by using Eqs. (\ref{amm1}) and (\ref{amm2}) and the differential equation in momentum space, Eq. (\ref{equaa}), then  
\bea 
\left[\left(i\hbar(1-\alpha p+2\alpha ^{2} p^{2} \right) \partial_{p}  - \langle \textbf{X} \rangle)+ i \hbar \frac{1+2\alpha ^{2} (\Delta p)^{2} +2\alpha ^{2} \langle p \rangle ^{2} -\alpha \langle p \rangle}{2 (\Delta p)^{2} } (p-\langle p \rangle)\right] \phi (p) &\approx& 0. \hspace*{1cm}
\eea
When taking into account that $\langle \textbf{X} \rangle=\zeta$, $\langle p \rangle=0$ and $\Delta p=\alpha/2$, the minimal position uncertainty can be deduced from the solution of this differential equation, which are correspondent to the states of absolutely maximal localization and critical momentum uncertainty.  By normalization where the Planck momentum is of the order of magnitude as that of $P_{pl}=\alpha/2$, then $\eta=(4 \alpha p_{pl} - 1)/\sqrt{7}=3/\sqrt{7}$. Therefore, the momentum-space wavefunctions $\phi_{\zeta}^{ml} (p)$ of states, which are maximally localized around $\langle \textbf{X} \rangle = \zeta$ \cite{amir}
\bea 
\phi_{\zeta}^{ml} (p) &=& \frac{\sqrt{6 \alpha}\left[\sqrt{8}\, e^{\eta \tan^{-1}(\eta)} - e^{-\eta \tan^{-1}\left(\frac{\eta}{3}\right)}\right]^{-\frac{1}{2}}}{(1 +\alpha p +2 \alpha^{2} p^{2} )^{\frac{3}{4}}} \; e^{\frac{-\eta}{2} \tan^{-1}\left(\frac{4 \alpha p - 1 }{\sqrt{7}}\right)}\; e^{{-i \frac{2 \zeta}{\alpha \hbar \sqrt{7}} } \left(\tan^{-1}\left(\frac{\eta}{3}\right) + \tan^{-1}\left(\frac{4 \alpha p - 1 }{\sqrt{7}}\right) \right)}. \hspace*{1cm}
\eea

It is apparent that the difference between this result and the one which was obtained in framework of  KMM GUP  \cite{16} is due the presence of first-order  momentum, Eq. (\ref{amm1}), which implies the existence of a maximal momentum. The  maximal localization states are now the proper physical states of the finite energy \cite{amir}
\bea 
\left\langle \phi_{\zeta}^{ml}\left| \frac{\hat{\textbf{P}}^{2}}{2\, m} \right|\phi_{\zeta}^{ml} \right\rangle &=& \frac{2\, \sqrt{\beta}}{\pi} \int_{- p_{pl}}^{+p_{pl}} \frac{\phi_{\zeta}^{ml*}(p)\, \frac{p^{2}}{2\, m}\, \phi_{\zeta}^{ml}(p)}{(1 -\alpha p +2 \alpha^{2} p^{2} )}\, d p. 
\eea
This can be approximated as $\approx (32\, m\, \alpha^{2})^{-1}$.  

\subsubsection{Quasiposition wavefunction transformation}

When projecting arbitrary states to maximally localized states, the probability amplitude for the particle can be deduced. This is maximally localized around a concrete position \cite{16,amir}. The transformation of a state in momentum wavefunction representation into its quasiposition wavefunction looks as \cite{amir} 
\bea
\phi (\zeta) &=& A \int_{- p_{pl}}^{+p_{pl}} \frac{\exp\left[\frac{-\eta}{2} \tan^{-1}\left(\frac{4\, \alpha\, p - 1 }{\sqrt{7}}\right)\right]}{\left[1 +\alpha\, p +2 \alpha^{2}\left(p ^{2}\right)\right]^{\frac{7}{4}}}\; \exp(i\, H\, \zeta),
\eea
where $A = \sqrt{6 \alpha}\; \left[\sqrt{8}\, \exp(\eta \tan^{-1}(\eta))  - \exp\left(-\eta \tan^{-1}(\eta/3)\right)\right]^{-\frac{1}{2}}$ and 
$H = 2/(\alpha \hbar \sqrt{7})\, \left[\tan^{-1}(\eta/3) + \tan^{-1}((4\, \alpha\, p - 1)/\sqrt{7}) \right]$ are modified wavenumbers.
Then, the modified wavelength in quasiposition wavefunction representation for the physical states reads $\lambda (p) = \pi\, \alpha\, \hbar\, \sqrt{7}/[\tan^{-1}\left(\frac{\eta}{3}\right) + \tan^{-1}\left((4\, \alpha\, p - 1)/\sqrt{7}\right) ]$. 
Because $\alpha$ is non-vanishing and $p$ is limited to the Planck momentum, there should be no wavelength smaller than $\lambda_{0} = \lambda (p_{pl}) = (\pi\, \alpha\, \hbar\, \sqrt{7})/(\tan^{-1}(\eta/3) + \tan^{-1}[(4\, \alpha\, p_{pl} - 1)/\sqrt{7}])$.
By implementing the relation between energy and momentum, for instance through $E=p^{2}/2 m$, we get the energy
\bea
E(\lambda) &=& \frac{2}{m \alpha^{2}} \left( \frac{\tan\left(\frac{\hbar \pi \alpha \sqrt{7}}{\lambda }\right)}{ \tan\left(\frac{\hbar  \pi \alpha \sqrt{7} }{\lambda }\right) + \sqrt{7}} \right)^{2},
\eea
and $E(\lambda_{0})=(P_{pl}^{2})/(2\, m)$ which apparently agrees well with ordinary QM. 

In this approach, 
\begin{itemize}
\item all these expressions do not diverge, 
\item they are important that they are distinguishable from the KMM \cite{16}, where the quasiposition wavefunctions in contrast to the ordinary QM {\it ripples}, because the energy of the short wavelength modes is divergent and
\item similar to the ordinary QM, those wavefunctions have ordinary {\it fine ripples}, because no longer divergence in the energy at $\lambda_{0}$ takes place. 
\end{itemize}
These are important results from this new GUP approach, especially the one, which guarantees both minimal length and maximal momentum.

\newpage

\section{Applications of quadratic GUP approach}
\label{implquadratic}

\subsection{Early Universe: Friedmann equations}
\label{sec:2GUPFRIED}

At very short distances, some aspects are likely,
\begin{itemize}
\item the holographic principle for gravity is assumed to relate the gravitational quantum theory to QFT,
\item the entropy of black hole is related to the area of the apparent horizon \cite{entr1,entr3},
\item the covariant entropy bound in the Friedmann-Lemaitre-Robertson-Walker  (FLRW) matric is found indicating to a holographic nature in terms of temperature and entropy \cite{entr4} and 
\item the cosmological boundary can be chosen as the cosmological apparent horizon instead of the event horizon of a black hole. 
\end{itemize}
In light of this, we recall that the statistical (informational) entropy of a black hole can be calculated using the brick wall method \cite{entr5}. In order to avoid the divergence near the event horizon, a cutoff parameter would be utilized. Since the degrees of freedom would be dominant near the horizon, the brick wall method is usually replaced by a thin-layer model making the calculation of entropy possible \cite{entr6a,entr6b,entr6c,entr6d,entr6e,entr6f,entr6g,entr6h}. The entropy of FLRW Universe is given by time-dependent metric. 

We propose that  GUP approach can be utilized in calculating the entropy of various black holes \cite{entr7a,entr7b,entr7c,entr7d,entr7e,entr7f,entr7g,entr7h,entr7i,app1,entr7k,entr7l}. We highlight that the effect of GUP on the reheating phase (after the inflation) of the Universe has been studied \cite{reheat}.  Therefore, the influence of GUP on the thermodynamics of the FLRW Universe likely provides a deep  understanding of the QG corrections to the dynamics of the FLRW Universe \cite{app1}. For instance, the entropy of the apparent horizon of the FLRW Universe gets corrections if the GUP effects are taken into consideration\cite{app11q}. 

\subsubsection{Some basic features of FLRW Universe}
\label{sec:flrw}

In FLRW Universe, the standard $(n+1)$-dimensional metric reads
\bea
d s^2 &=& h_{ab}\, d x^a\, dx^b + r^2\, d\Omega^2_{n-1}, \label{eq:mtrc}
\eea
where $x^a=(t,r)$ and $h_{ab}=\mathtt{diag}(-1,a^2/(1-k r^2))$ with $\vec{r}=a(t) r$ and $ x^{0}=t $. $d\Omega^2_{n-1}$ is the line element of an $(n+1)$-dimensional unit sphere. $a(t)$ and $k$ are scale factor and curvature parameter $k=-1, 0, +1$, respectively. Then, the radius of the apparent horizon is given by
\bea
R_A &=& \left(H^2+\frac{k}{a^2}\right)^{-1/2}. \label{eq:eins0}
\eea
It is obvious that the time evolution of the scale factor strongly depends on the background equation of state. Seeking for simplicity, we utilize \cite{zhu},  $a(t) = t^{2/3 \bar{k}}$,
where $t$ is the cosmic time, $\bar{k}=1-(b\, c)^2/(1 - c^2)$ and $b$ and $c$ are free and dimensionless parameters. The Hubble parameter $H=\dot{a}/a$ and radius of the apparent horizon, respectively, where the dot represents derivative with respect to the cosmic time $t$
\bea
H(t) &=& \frac{2}{3} \frac{1}{\bar{k}\, a^{3 \bar{k}/2}},\\
R_A &=& \left(H \sqrt{1 + \left(\frac{3}{2}  \bar{k} \right)^{4/3  \bar{k}}\, H^{4/3  \bar{k}-2} k }\right)^{-1/2}.
\eea
From the metric given in Eq. (\ref{eq:mtrc}) and the Einstein in non-viscous background equations, we get
\bea
H^2 +\frac{k}{a^2} &=& \frac{8\, \pi\, G}{3} \rho +\frac{\Lambda}{3}, \label{eq:eins1}\\
\dot H - \frac{k}{a^2} &=& - 4\, \pi\, G (\rho + p), \label{eq:eins2}
\eea
where $\Lambda$ is the cosmological parameter.  Then, the total energy density $\rho$ and temperature $T$ inside the sphere of radius $R_A$ can be evaluated 
\bea
\rho &=& \frac{\pi^{n/2}}{\Gamma\, \left(\frac{n}{2}\right)+1} \, \frac{n(n-1)}{16\, \pi\, G} \, R_A^{n-1},\\
T &=& \frac{R_A}{2 \pi} H^2 \left|1 + \frac{1}{2\, H^2} \left(\dot H + \frac{k}{a^2}\right)\right|,
\eea
where $n$ gives the dimension of the Universe and $p$ stands for the pressure. From Eq. (\ref{eq:eins0}) and (\ref{eq:eins1}), it is obvious that the inverse radius (of the apparent horizon) is to be determined by the energy-momentum tensor, i.e. matter and cosmological constant $\Lambda$. When taking into consideration the viscous nature of the background geometry makes the treatment of thermodynamics of FLRW considerably becomes complicated \cite{Tawfik:2011gh,Tawfik:2011sh,Tawfik:2011mw,Tawfik:2010pm,Tawfik:2010mb,Tawfik:2010bm,Tawfik:2010ht,Tawfik:2009nh,Tawfik:2009mk}.
For completeness, we give the cross section of particle production $\sigma = M_{p}^{-2} \left[\rho M_{p}^{-1} (8 \Gamma(n/2)/(n-2))\right]^{2/(n-2)}$.
The continuity equation, time evolution of energy density, will be given in Eq. (\ref{continuity}).

\subsubsection{GUP and Friedmann equation}
\label{qGUPfriedm}

When ($n+1$)-dimensional FLRW Universe is considered, the metric field equation, Eq. (\ref{eq:mtrc}), is rewritten as  \cite{app11q}
\begin{eqnarray}
d s^2 = - d t^2 + a^2 \left(\frac{d r^2}{1-k\, r^2}+r^2\, d \Omega_{n-1}^2\right),
\end{eqnarray}
where $d \Omega_{n-1}^2$ represents the line element of an ($n-1$)-dimensional unit sphere. In the FLRW spacetime, the dynamical apparent horizon is marginally trapped surface with a vanishing expansion \cite{app11q}. By using $\tilde{r}=a\, r$, then the radius of the apparent horizon $\tilde{r}_A = 1/\sqrt{H^2+k/a^2}$. The apparent horizon has associated entropy $S = A/(4\, G)$  and temperature $T = 1/(2\, \pi\, \tilde{r}_A)$, where $A=n\Omega_n\tilde{r}_A^{n-1}$ is the apparent horizon area and $\Omega_n=\pi^{n/2}/\Gamma(n/2+1)$ gives the volume of an $n$-dimensional unit sphere \cite{app11q}. 
The Friedmann equations, Eqs. (\ref{eq:eins1}) and (\ref{eq:eins2}) reads \cite{app1,app10q} 
\begin{eqnarray}
\dot{H}-\frac{k}{a^2} &=& - \frac{8 \pi G}{n-1}\, (\rho + p),\\
H^2+\frac{k}{a^2} &=& \frac{16 \pi G}{n (n-1)}\, \rho. \label{fr}
\end{eqnarray}
The first law of thermodynamics reads $dE=TdS$ \cite{app10q,app1}. To derive Eq.(\ref{fr}), the continuity equation of the perfect fluid should be used \cite{app1,app10q}
\begin{eqnarray} \label{continuity}
\dot{\rho}+n\, H(\rho + p)=0.
\end{eqnarray}
The energy density $\rho$ is related to the pressure of the cosmic fluid, $p=\omega\, \rho$, i.e. through the equation of state. With this regard, we highlight that he implementation of viscous equations of state in early Universe was analysed, systematically \cite{Tawfik:2011sh,Tawfik:2011mw,Tawfik:2010pm,Tawfik:2010mb,Tawfik:2010ht,Tawfik:2010bm,Tawfik:2009mk}. 

When we start up with the GUP approach which was introduced in Ref. \cite{16}
\begin{eqnarray}
\Delta x\, \Delta p \geq 1+\alpha^2\, l_p^2\,  \Delta\, p^2, \label{GUP}
\end{eqnarray}
after some simple manipulations,  the momentum uncertainty reads 
\begin{eqnarray}
\Delta p \geq \frac{1}{\Delta x}\left[\frac{\Delta x^2}{2\, \alpha^2\, l_p^2}-\frac{\Delta x^2}{2\, \alpha^2\, l_p^2} \, \sqrt{1-\frac{4\, \alpha^2\, l_p^2}{(\Delta x)^2}}\right]=\frac{1}{\delta x}\, f_G(\Delta x^2), \label{gup3}
\end{eqnarray}
where $f_G(\Delta x^2)=\frac{\Delta x^2}{2\, \alpha^2\, l_p^2} - \frac{\Delta
x^2}{2\, \alpha^2\, l_p^2} \sqrt{1-\frac{4\, \alpha^2\, l_p^2}{(\Delta
x)^2}}$.

The energy of absorbed or emitted  particle as uncertainty of the momentum can be  identified as $dE \simeq \Delta p$ \cite{gup-hawking1q}. From quantum properties of absorbed or emitted particle, the Heisenberg uncertainty principle $\delta p \geq \hbar/\Delta x$ can be implemented. In natural units $c=\hbar=k_B=1$, we find that the increase or decrease in the area of the apparent horizon can be expressed as
\begin{eqnarray}
dA=\frac{4\, G}{T}\, dE \simeq \frac{4\, G}{T}\frac{1}{\Delta x}. \label{area1}
\end{eqnarray}
When the GUP effect, Eq. (\ref{gup3}), is taken into consideration, the change in the apparent horizon area can be modified as
\bea
d A_G &=& \frac{4\, G}{T} d E \simeq \frac{4\, G}{T} \frac{1}{\delta x} f_G(\delta x^2). \label{area2}
\eea 
Using Eq. (\ref{area1}), we get
\begin{eqnarray}
dA_G=f_G (\delta x^2) dA. \label{area3}
\end{eqnarray}

The position uncertainty $\delta x$ of absorbed or emitted particle can be chosen as the particle's Compton length, which is equivalent the inverse of Hawking temperature, $\delta x\simeq 2\, \tilde{r}_A=2 (A/(n\, \Omega_n))^{1/(n-1)}$ \cite{gup-entropy1q}. Thus, the departure function $f_G(\delta x^2)$ can be re-expressed in terms of $A$, $f_G(A) = 2/(\alpha^2\, l_p^2) (A/(n\, \Omega_n))^{2/(n-1)}\; [1-\sqrt{1-\alpha^2\, l_p^2((n\, \Omega_n)/A)^{2/(n-1)}}]$ \cite{app1}.

Hereafter, we use $f_G(A)$ to represent the departure function $f_G(\delta x^2)$. At $\alpha=0$, the Taylor series of $f_G(A)$ gives
\begin{eqnarray}
f_G(A)&=&1+\frac{\alpha^2\, l_p^2}{4} \left(\frac{n\, \Omega_n}{A}\right)^{\frac{2}{n-1}} + 
\frac{(\alpha^2\, l_p^2)^2}{8} \left(\frac{n\, \Omega_n}{A}\right)^{\frac{4}{n-1}}
+ \sum_{d=3} c_d (\alpha l_p)^{2d} \left(\frac{n\Omega_n}{A}\right)^{\frac{2\, d}{n-1}}, \hspace*{10mm} \label{f} 
\end{eqnarray}
where $c_d$ is a constant. If Eq. (\ref{f}) is substituted in Eq. (\ref{area3}) and then integrated, we get a  modified area $A_G$. Also, we get the correction to the entropy-area relation by using $S_G=A_G/4G$ \cite{app11q}. But integrating Eq. (\ref{area3}) might be complicated and dimension dependent. Therefore, as anticipated in Ref. \cite{app1}, the discussions should be divided into three cases. From $2/(n-1)$ in Eq. (\ref{f}), we find that $3 \leq n \, ;\, n\,=\, 3,\,4,\,  \dots $ \cite{app1}. In order to obtain the modified Friedmann equations from the modified entropy-area relation \cite{app11q} throughout $(n+1)$-dimensional FLRW space-time, firstly we are needing to estimate the modified apparent horizon area appears in Eq. (\ref{area3}) and use of  the Bekenstein-Hawking area law \cite{app11q}, $S=A/4G$. 
\begin{eqnarray}
\left(\dot{H}-\frac{k}{a}\right) S'_G(A) &=& - \pi\, (\rho+p), \label{fridemann1}\\
\frac{8\, \pi\, G}{3} \rho &=& - \frac{\pi}{G} \int S'_G(A) \left(\frac{4G}{A}\right)^2\, dA. \label{fridemann2}
\end{eqnarray}
This implies that some correction terms in the entropy of the apparent horizon is dimension-dependent \cite{app1}. Since the modified Friedmann equations in an ($n+1$)-dimensional FLRW Universe is not relevant to that whether the value of $n$ is an even or odd number \cite{app1}.

\subsubsection{Entropic corrections and modified Friedmann equations}

The entropic corrections in modified Friedmann equations appear in two types:
\begin{itemize}
\item{\bf Logarithmic-type corrections:} 
We start with the corrected entropy-area relation \cite{Banerjee:2008cfq,Modesto:2010rmq,Fang:2013rmq}
\begin{equation}
S= \frac{A}{4 G} + \alpha\, \ln \frac{A}{4 G}+\beta\, \frac{4 G}{A},
\label{correctionLogarithmic}
\end{equation}
where the Newton's constant $G=L^{2}_{p}$ and the area $A=4\, \pi {\widetilde r_A}^{2}$. The relevant effective area of the holographic surface is defined as \cite{Banerjee:2008cfq,Modesto:2010rmq,Fang:2013rmq}
\bea
{\widetilde A}= A + 4\, \alpha\, L^{2}_{p}\, \ln\left(\frac{A}{4\, L^2_p}\right) + \frac{16\, \beta\, L^4_p}{A}.
\eea
We propose that the effective degrees of freedom \cite{Fang:2013rmq} (at apparent horizon) are
\be  \label{la ns}   
\widetilde N_{\mathrm{sur}} = \frac{4\, \pi\, {\widetilde r_A}^2}{L^2_p} \left(1 + \frac{\alpha L^2_p}{2 \pi {\widetilde r_A}^2} - \frac{\beta L^4_p}{3 \pi^2 {\widetilde r_A}^4} \right).
\ee
The modified Friedmann equation due to the Logarithmic-type corrections Eq. (\ref{correctionLogarithmic}) \cite{Fang:2013rmq}
\bea
\left(H^2 +\frac{k}{a^2}\right) + \frac{\alpha L^2_p}{2\pi} \left( H^2+\frac{k}{a^2}\right)^2
 - \frac{\beta L^4_p}{3\pi^2} \left(H^2+\frac{k}{a^2}\right)^3
 = \frac{8\,\pi\, L^2_p}{3}\,\rho. 
\eea

\item{\bf Power-law corrections:}
The entropy with power-law corrections \cite{Das:2007mjq} can be deduced as follows.
\be
S = \frac{A}{4\, L^2 _p} \left( 1 - K_{\alpha}\, A^{1-\frac{\alpha}{2}} \right).
\ee
From the definition $K_\alpha =\alpha (4 \pi)^{\frac{\alpha}{2}-1}/(4-\alpha ) r^{2-\alpha}_c$,  where $r_c$ is the crossover scale, the effective degrees of freedom at the apparent horizon read \cite{Fang:2013rmq} 
\be     \label{pl ns}
\widetilde N_{{sur}} = \frac{4\, \pi\, {\widetilde r_A}^2}{L^2_p} \left[ 1 - \left(\frac{r_c}{\widetilde r_A} \right) ^{\alpha -2} + C\, {\widetilde r_A}^2 \right].
\ee
In the limit $\alpha \longrightarrow 0$ and with the constant constant $C= 1/r^2_c$, no entropic correction should appear. The exactly-modified Friedmann equation has been derived  \cite{Sheykhi:2010yqq,Karami:2010bgq,Fang:2013rmq}
\be
H^2 +\frac{k}{a^2} - \frac{1}{r^2_c} \left[r^\alpha_c \left(H^2 +\frac{k}{a^2} \right)^{\frac{\alpha}{2}} - 1 \right] = \frac{8\, \pi\, L^2_p}{3}\, \rho.
\ee
\end{itemize}

\subsubsection{conclusion}

The influence of GUP on the thermodynamics of the FLRW Universe shows that the GUP contributes with some corrections to the entropy-area relation at the apparent horizon of the FLRW Universe as well as to the Friedmann equations. This latter implies that the GUP affects the dynamics of the FLRW Universe. The leading logarithmic correction term exists only for odd number in one-dimensional FLRW spacetime. This term gives a positive contribution to the entropy of the apparent horizon. For even number in one-dimensional FLRW spacetime, the  logarithmic correction term disappears from the entropy. The expansion of the Universe is attributed to the difference between the degrees of freedom on a holographic surface and the one in the bulk \cite{Fang:2013rmq} . The idea taken from the modification of holographic screen in both ways ''power-law corrections'' or ''logarithmic corrections'' implies an additional term due the introduction of the minimal length to the entropy-area relation which will be modify the Friedmann equations \cite{Fang:2013rmq}.

\subsection{Inflationary parameters}
\label{sec:inflt1}

\subsubsection{Hybrid inflation and black hole production}

In a scenario of semi-classical black hole combining hybrid inflation \cite{app2q} and characterized by the hybrid inflation model, the inflation fields $(\phi, \psi)$ are governed by the inflation potential, 
\be
V(\phi, \psi ) = \frac{1}{2}\, m^{2}\, \phi^{2}+\frac{1}{2}\, \gamma\, \phi^{2}\, \psi^{2} + \left(M^{2} -\frac{\sqrt{\lambda}}{2}\, \psi^{2} \right)^{2},
\ee
where $M$ being the mass of the black hole. There are two conditions on $\phi$:
\begin{itemize}
\item When $\phi$ executes a \textit{"slow-roll"} down \cite{Inflation1q}, then the potential is responsible for more than 60 $e$-folds expansion while $\psi$ remains zero.
\item But if $\phi$ is reduced to a critical value, $\phi_{c}= \sqrt{2 M^{2} \sqrt{\lambda}/\gamma}$, the phase transition which results in a \textit{''rapid-fall''} \cite{Inflation1q} of the energy density of the $\psi$ field,  ends the inflation. The latter lasts only for a few $e$-folds.
\end{itemize}

The equations of motion (EoM) for these fields read \cite{Inflation1q}.
\bea
\ddot{\phi}+3\, H\, \dot{\phi} + \left(m^{2} + \gamma\, \psi^{2}\right) &=& 0,\\
\ddot{\psi}+3\, H\, \dot{\psi}+ \left(\lambda\, \psi^{2} + \gamma\, \phi^{2} - 2\, \sqrt{\lambda} M^{2}\right) &=& 0.
\eea
The Hubble parameter can be taken into consideration from the Friedmann equation \cite{Inflation1q}
\be 
H^{2}=\frac{8\, \pi}{3\, m_{p}^{2}} \left(\frac{1}{2}\, \dot{\phi}^{2}+\frac{1}{2}\, \dot{\psi}^{2}+V(\phi, \psi)\right).
\ee
The solution for the $\psi$ field in the small $\phi$ regime \cite{Inflation1q} measured backwards from the end of the inflation is $\psi\, (N(t)) = \psi_{e}\, e^{-\kappa\, N(t)}$, where $\kappa=-3/2+\sqrt{9/2+2 \sqrt{\lambda}\, M^{2}/H_{*}^{2}}$ is the angular momentum and $N(t)=H_{*} (t_{e} - t)$ is the number of $e$-folds existing from $t_{e}$ to $t$ with $H_{*}=\sqrt{8 \pi /3}\, M^{2}/m_{p}$.

It was assumed that the Universe was inflated during the second stage of the inflation era  as $\exp(N_c) \sim [(2\, m_p)/(\kappa\, H_{*})]^{1/\kappa}$  \cite{bellidoq}. If the second stage of the inflation is short, i.e. $N_c\sim {\mathcal O}(1)$, then direct after the inflation the energy may still be dominated by the oscillations of $\psi$ with $p=0$. After the inflation, the scale factor of the Universe  would grow as $(t\, H_*)^{2/3}$ \cite{Inflation1q}. When the scale $(t\, H_*)^{2/3}\, H_*^{-1}\, e^{N_c}$ became comparable to the particle horizon $t$ or $t \sim (t\, H_*)^{2/3}\, H_*^{-1}\, \exp(N_c)$ \cite{Inflation1q}, then
\begin{equation}
t \sim t_h = H_*^{-1}\, \exp\left(3~N_{c} \right).
\end{equation}
The initial mass of black hole with size $r_s\sim H_*^{-1}\, \exp(3\, N_c)$ would be formed as  \cite{Inflation1q}
\begin{equation}
\mu_i \sim \frac{m_p}{H_*}\, \exp\left(3~N_{c}\right) \equiv g\,  \frac{m_p}{H_*} \left(\frac{2\, m_p}{\kappa\,  H_*}\right)^{3/\kappa}.
\end{equation}
A dimensionless parameter $g$ is introduced to account for the dynamic range of the gravitational collapse \cite{Inflation1q}. Since $H_*$ depends on $M$, while $s$ depends on both $M$ and $\lambda$, the initial mass of the black hole depends only on $M$ and the coupling in $\psi$-sector of the hybrid inflation  \cite{Inflation1q}.

\subsubsection{Randall-Sundrum II model on inflationary dynamics}

The scalar field $\phi$ which drives the inflation has energy density and pressure, respectively \cite{Inflation2q}
\bea
\rho &=& \frac{1}{2} \dot{\phi}^2 + V(\phi), \label{eq:rhoEU}\\
P &=& \frac{1}{2} \dot{\phi}^2 - V(\phi), \label{eq:pEU}
\eea
where $V(\phi)$ is the inflation potential. The inflationary scalar density perturbations are calculated in the presence of the minimal length \cite{Inflation2q}. The slow-roll regime is characterized by \cite{Inflation20q}
\bea
\frac{1}{2} \dot{\phi}^{2} &\ll & V(\phi),\\
3\, H\, \dot{\phi} &\approx & - V^{'}(\phi).
\eea
A fundamental energy scale $\epsilon$ in the order of Planck energy, which adds to the correction \cite{Inflation21q} seems to define the conformal time, $\tau = - 1/(a\, H)$, and the comoving momentum, $\kappa$. The latter is related to the physical momentum, $\kappa = a\, p = - p/(\tau\, H)$. 
The conformal time can be given as function of the energy density $\tau _0=-\epsilon/ H\kappa$. 

\begin{figure}[htb]
\includegraphics[scale={0.5}]{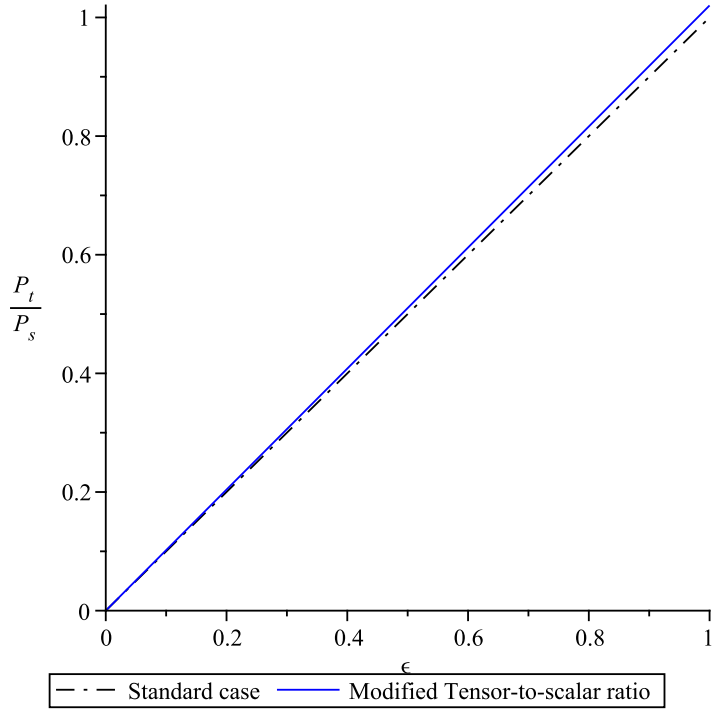}
\caption{The difference between tensor-to-scalar ratio in standard and GUP-modified inflation is given in dependence on the energy $\epsilon$ at fixed $\kappa$ and $\beta=10^{-2}$. The graph taken from \cite{Inflation200q}. }
\label{figure}
\end{figure}

By using the quadratic GUP \cite{16}, we express the change in the comoving momentum from $\kappa$ to $\kappa (1+\beta \kappa ^{2})$ \cite{Inflation200q}. This modifies the dispersion relation, which was approved in loop quantum gravity and noncommutative geometry \cite{Amelino2q}. The evolution of the perturbations in the inflation reads \cite{Inflation200q}, $\mu^{''}_{\kappa}+\mu_{\kappa}\,  (\kappa^{2} -a^{''}/a) = 0$, 
where $\mu$ is related to the scalar field $\mu = a\, \delta \phi$. The scalar spectral index in the presence of the minimal length cutoff is given as \cite{Inflation200q},
\bea
n_{s} &=& \frac{d~ \ln(\mathcal{P}_{s})}{d~ \ln(\kappa) (1+\beta\, \kappa^{2})}+1 = \frac{(1+\beta\, \kappa^{2})}{(1+ 3\, \beta\, \kappa^{2})}\, \frac{d~ \ln(\mathcal{P}_{s})}{d~ \ln(\kappa)} +1   \approx  (1-2\,\beta\, \kappa^{2})\, \frac{d~ \ln(\mathcal{P}_{s}) }{d~ \ln(\kappa)}+1,
\eea
where $\mathcal{P}_{s}$ is the amplitude of the scalar density perturbation. The change in the Hubble parameter due to the GUP will be realized using the slow-roll parameters \cite{Inflation20q,Inflation21q}. At the horizon crossing epoch, we have $d H/d \kappa =-\epsilon H/\kappa$ \cite{Inflation20q}.
When $\kappa$ is replaced by $\kappa(1+ \beta\, \kappa^{2})$, then we get 
\be \label{kabba}
H \approx \kappa^{-\epsilon}\, \exp\left(- \beta\, \epsilon\, \kappa^{2}\right).
\ee
By using $\mu^{''}_{\kappa}+\mu_{\kappa}\,   (\kappa^{2} -a^{''}/a) = 0$, then the tensorial density fluctuations are given as \cite{Inflation21q}
\be
\mathcal{P}_{t} (\kappa) = \left(\frac{H}{2\, \pi} \right)^{2}\, \left[1-\frac{H}{\epsilon_{p}}\, \sin\left(\frac{2\,  \epsilon_{p}}{H}\right)\right],
\ee
where the second term on the right hand side expressed the direct contribution from the quantum gravity effect and $\epsilon_{p}$ being the Planck energy. But for scalar density fluctuations, one should multiply the tensorial density fluctuations by an extra term $(H/\dot{\phi})^{2}$  \cite{Inflation2q}
\be
\mathcal{P}_{s} = \left(\frac{H}{\dot{\phi}}\right)^{2} \left(\frac{H}{2\, \pi}\right)^{2} \left[1-\frac{H}{\epsilon_{p}}\, \sin\left(\frac{2\, \epsilon_{p}}{H} \right)\right],
\ee
where $H$ was given in Eq. (\ref{kabba}). The variation of $\beta$, which is essentially a fixed quantity related to the minimal length, means a control on the strength of the quantum gravity effect. Then, the ratio tensor-to-scalar  reads \cite{Inflation200q}  
\bea
r_s\,=\,\frac{\mathcal{P}_{t}}{\mathcal{P}_{s}}=\left(\frac{\dot{\phi}}{H}\right)^{2} = \left(\frac{16\, \pi\, \sqrt{\epsilon}  V(\phi)} {{M_{4}\, H} }\right)^{2},
\eea
where $M_{4}$ is  $4$-dimensional (fundamental) Planck scale. The difference between tensor-to-scalar ratio in standard and  modified case is shown in Fig. (\ref{figure})  \cite{Inflation200q}. It is obvious that the ratio increases linearly with the incorporation of the quantum gravity effects.

\subsubsection{conclusion}

By studying the effect of GUP  on the inflationary dynamics of both the standard 4D theory and the Randall-Sundrum II braneworld setup, it was shown that in the presence of the strong quantum gravity effects the spectral index is not scale invariant \cite{Inflation2q}. In this sense, any deviation from the scale invariance of the spectral index essentially contains a footprint of these high-energy effects \cite{Inflation2q}. There is an oscillatory behavior in the $\kappa$-dependence of the density fluctuations which  essentially can be detected in the CMB spectrum \cite{Inflation200q}. Another possible signature may be some imprints on the cosmic microwave background (CMB) fluctuations due to the thermodynamics of primordial black hole (PBH)-CMB interactions \cite{Inflation2q}.

\subsection{Black hole thermodynamics}

\subsubsection{Black-hole entropy and GUP approaches}

The deformation of generalized commutation relations reads
\begin{eqnarray}\label{4,1,2}
\left[X_i,\, P_j\right] = i\, \hbar ~f_{i j}(X,P),\nonumber \\
\left[P_i,\, P_j\right] = i\, \hbar ~h_{i j}(X,P),\nonumber \\
\left[X_i,\, X_j\right] = i\, \hbar ~g_{ij}(X,P),
\end{eqnarray}
where operators $X_{i}$ and $P_{j}$ are coordinates and momentum variables, respectively. The deformation functions $f_{i j}$, $g_{i j}$ and $h_{i j}$ possessing properties like bilineary, Libniz rules and Jacobi identity \cite{Bina1q}. The given relations can be reduced to the deformed Poisson brackets and Dirac notation $\left[X_i,\, P_j\right] = i\, \hbar \{X_i,\, P_j\}$.
Accordingly,  the quadratic GUP approach with $\zeta$ is the GUP parameter  \cite{16},
\begin{eqnarray}\label{4,1,2}
\{X_i,\, P_j\} &=&  (1+\zeta \,P^{2}),\nonumber \\
\{P_i,\, P_j\} &=& \{X_i,\, X_j\} \,=\,0 
\end{eqnarray}

The thermodynamics of  the system of interest  is easily accessible through the partition function 
\begin{eqnarray}\label{4,1,1}
Z=\frac{1}{2\, \pi\, \hbar}\, \int \exp\left[-\beta\, H(x,p)\right]\; dx\, dp.
\end{eqnarray}
The corrected partition-function of quantum black-hole  leads to \cite{Bina1q}
\begin{eqnarray}\label{4,1,8}
Z^{GUP}_{Q}=\sqrt{\frac{2\, \pi}{3}}~\frac{1}{\beta\, {E_{p}}}\, \exp\left(-{\frac{\beta^{2}{E_{p}}^{2}}{16\, \pi}}-\frac{\hbar\,\zeta}{c\,\ell_{p}\,\beta}\right).
\end{eqnarray}
Similar to the non-deformed case,  the free energy is defined as \cite{Bina1q}
\bea
\epsilon &=& -\frac{\partial \ln (Z^{\it{GUP}}_{\it {Q}})}{\partial \beta}=\frac{{E_{p}}^2}{8\pi}\beta+\frac{1}{\beta}-\frac{\hbar}{c\,\ell_{p}\,\beta^{2}}\,\zeta=Mc^{2}.
\eea  
In framework of GUP, the temperature of  quantum black-hole $\beta$ can be given in term of  Hawking temperature $\beta_H$ \cite{Bina1q}
\begin{eqnarray}\label{4,1,9}
\beta = \beta _{H}\left[ 1-\frac{1}{\beta_{H}\, M\, c^{2}}+\frac{M\, E_{p}}
{\left(\beta_{H}\, M\, c^{2}-1\right)\, \left(\beta_{H}\, M\, c^{2}-2\right)}\, \zeta \right].
\end{eqnarray}
Then, the entropy reads \cite{Bina1q}
\begin{eqnarray} \label{4,1,12}
\frac{S^{GUP}}{k} = \frac{S^{GUP}_{BH}}{k} - \frac{1}{2}\, \ln \left(\frac{S^{GUP}_{BH}}{k}\right) - 2\, M\, c^{2}\, \left(\frac{S^{GUP}_{BH}}{S_{BH}}-1\right)+{\cal O}\left(S^{GUP\,-1}_{BH}\right),
\end{eqnarray}
where the Hawking-Bekenstein entropy  \cite{Bina1q},
\begin{eqnarray}\label{4,1,gjug}
\frac{S^{GUP}_{BH}}{k} = \frac{S_{BH}}{k}\, \left(1+\frac{E_{p}^{3}}{8\, \pi\, M^{2}{c}^{6}}\,\zeta \right).
\end{eqnarray}

This correction \cite{Bina1q} is similar to ones derived from other methods \cite{gup-entropy1q,G2q,Zhaoq}. Furthermore, it was shown that this result has the same form as that of the non-deformed case, the logarithmic correction to the entropy appears with a $-1/2$ factor \cite{Bina1q}.

\subsubsection{Black hole remnant}

Due to GUP there should exist a Planck size  at the end of the black-hole evaporation \cite{Inflation1q}. In other words, the GUP may prevent the black hole from complete evaporating, i.e. there should exist a black-hole remnant with the Planck mass. The stability of such a Planck size containing the remnant mass may be further protected by supersymmetry \cite{Inflation1q}. Thus, the uncertainty relation in position is given as
\be
\Delta x \geq \frac{\hbar}{\Delta p}+\zeta^{2}  \ell_{p}^{2} \frac{\Delta p}{\hbar},
\ee
where $\ell _{p}$ is the Planck length and $\zeta$ is a factor originated in the String theory \cite{16}. In the vicinity of the black-hole surface, there should be an intrinsic uncertainty in the position, which is nearly equal to the Schwarzschild radius \cite{Inflation10q}, $\Delta x \approx r_{s}= 2 G M_{BH}/c^{2}$. Under the GUP effect, the photon emitted by black hole is characterized by temperature, which is related to the Hawking one \cite{Inflation10q}, $T_H= \hbar c^{3}/(8 \pi G M_{BH})$. The modified black-hole temperature is given by $T_{H}=\mu\, m_{p}\, c^{2}/(4\, \pi\, \zeta ^{2}) [1 \mp \sqrt{1- \zeta^{2}/\mu^{2}}]$, where $\mu=M_{BH}/m_{p}$ is the mass of the black hole $M_{BH}$ normalized to the Planck mass $m_{p}$. We note that the temperature is complex and unphysical when the mass  $<\zeta\, m_{p}$ and the Schwarzschild radius  $<2\, \zeta \ell _{p}$. The minimum length allowed by the GUP approach is given at $\mu\, \zeta$ \cite{Inflation1q}. The Hawking temperature $T_{H}$ is finite. That its slope is infinite due to vanishing heat capacity. The black-hole evaporation should be stopped. In Stefan-Boltzmann law, the rate of evaporation is given as
\begin{equation}
\dot{\mu} = -\frac{16\, g}{t_{ch}}\, \frac{\mu^6}{\xi^8} \left[1-\sqrt{1-\xi^2/\mu^2}\right]^4, \label{eq:I}
\end{equation}
where $t_{ch}=60 (16)^2\, \pi\, t_p \approx 4.8\, \times 10^4\, t_p$ is a characteristic time for the black-hole evaporation, and $t_p=(\hbar\, G/c^5)^{1/2} \approx 0.54 \times 10^{-43}~$s is the Planck time. We find that the energy output, which is given by Eq. (\ref{eq:I}) becomes finite, where $\mu=\xi$, i.e. $d \mu/dt\, \vert_{\mu=\xi} = -16\, g/(\xi^2\, t_{ch})$. Thus, the black hole with an initial mass $\mu_i$ can evaporate during 
\begin{eqnarray}
\tau &=& \frac{t_{ch}}{16\, g} \, \left[\frac{8}{3}\, \mu_i^3 + 
\frac{8}{3} (\mu_i^2 - \xi^2)^{3/2} - 4\, \xi^2 (\mu_i^2 - \xi^2)^{1/2}  - 8\, \xi^2\, \mu_i+4\, \xi^3\, \cos^{-1}\frac{\xi}{\mu_i} +\frac{19}{3}\, \xi^3-\frac{\xi^4}{\mu_i}
\right] \approx \frac{\mu_i^3}{3\, g}\, t_{ch}, \label{eq:J} \hspace*{10mm}
\end{eqnarray}
where $\mu_i \gg 1$. This continues till a concrete remnant is left.  

\subsubsection{Conclusion}

The thermodynamic properties of black hole is determined. In doing this the relevant Bekenstein-Hawking entropy is introduced in GUP framework and concluded that again the logarithmic correction of the entropy appears with a prefactor $1/2$. Furthermore, the value of the entropy diminishes. This can be comprehended from the fact that the GUP reduces the available physical states in the black-hole remnant \cite{Inflation1q}. Since $H_*$ depends on $M$ while $s$ depends on $M$ and $\lambda$, the initial black hole mass depends  on the mass and the coupling in the sector of the hybrid inflation \cite{Inflation1q}.

\subsection{Compact stellar objects}

\subsubsection{Compact stars and Tolman-Oppenheimer-Volkoff equation}

The configuration of a spherically symmetric static star composed of perfect fluids is determined by the Tolman-Oppenheimer-Volkoff (TOV) equation \cite{Tolman1939PRq,Oppenheimer1939PRq}
\bea
\frac{dP}{dr} &=& -\left(\rho+P/c^2\right)\, \frac{G\, m(r)+4\, \pi\, G\, r^3\, P/c^2}{r \left[r - 2\, G\, m(r)/c^2\right]},\label{TOV1-PR} \\
\frac{d m(r)}{d r} &=& 4\, \pi\, r^2\, \rho(r), \label{TOV1-mR}
\eea
where $P$ and $\rho$ are respectively the pressure and the macroscopic energy density measured in proper coordinates. The equation of state (EoS) and appropriate boundary conditions, Eqs. (\ref{TOV1-PR}) and (\ref{TOV1-mR}) can be supplied to determine $P(r)$, $m(r)$ and $\rho(r)$. If the pressure and the gravitational potential remain small forever, i.e. $P(r)\ll \rho\, c^2,\;2Gm(r)/c^2r\ll 1$, then the TOV equation reduces to the fundamental equation of Newtonian gravity 
\bea
\frac{dP}{dr} = - \rho(r)\, \frac{G\, m(r)}{r^2}, \label{TOV1-PR-New}
\eea
which is suitable in describing the low-density compact-stars. For compact stars (such as neutron stars), GR plays an important role \cite{Wiley}. An ideal neutron star is the simplest model, in which the nuclear interactions are ignored and the pressure of degenerate neutrons apparently acts against the gravitational collapse \cite{Oppenheimer1939PRq}. Various EoS are introduced to represent strong and nuclear interactions. It is worthwhile to mention that the QG effects have been studied by many authors \cite{
Santos2012q,Deliduman2012q,ArapogluJCAP2011q,Pani2011PRDq,Wiseman2002PRDq,GERMANNI2001PRDq}.

In Fermi stars, based on GUP approaches the ideal gas statistics has been discussed in various models \cite{RAMA2001PLB103q,LNC2002PRD125028q,NozariMehdipour2007Chaos1637q,
Fityo2008PLA5872q,WYZ2010JHEP043q}. The ultra-relativistic Fermi gas was analysed under the effects of GUP \cite{WYZ2010JHEP043q} and the proper particle number, energy density and pressure  are determined \cite{WYZ2010JHEP043q}
\begin{eqnarray}
\frac{N}{V} &=& \frac{8 \pi}{(h c)^3} E_H^3 f(\kappa), \label{particlenumb1density}\\
\rho &=& \frac{8 \pi}{c^2 (h c)^3} E_H^4 h(\kappa), \label{properENERGYDENSITY}\\
P &=& \frac{8 \pi}{(h c)^3} E_H^4 g(\kappa), \label{properpressure1}
\end{eqnarray}
where $E_H=c/{\sqrt \beta}=M_p c^2/\sqrt{\beta_0}$ stands for the Hagedorn energy \cite{WYZ2010JHEP043q} and $\kappa=\varepsilon_F\sqrt{\beta/c^2}=\varepsilon_F/E_H$. Moreover
\begin{eqnarray}
h(\kappa) &\equiv & \frac 14\frac{\kappa^4}{(1+\kappa^2)^2},\label{hkapp1}\\
f(\kappa) &\equiv & \frac{1}{8}\left[\frac{\kappa(\kappa^2-1)}{(1+\kappa^2)^2}
+\textrm{tan}^{-1}(\kappa)\right],\\
g(\kappa) & \equiv & \kappa f(\kappa)-h(\kappa), \label{gkapadingy}
\end{eqnarray}
which are derived, when the GUP effects from quadratic of momenta are taken into consideration. It is apparent that when $\kappa$ increases, the proper energy density and the proper number density are bounded \cite{Zhangq}, while the proper pressure blows up. Based on precise measurement of Lamb shift, an upper bound on $\beta_0$ is deduced as $\beta_0< 10^{36}$ \cite{Das1}, on which a restriction was claimed \cite{BrauB2006PRDq}. However, another better bound ($\beta_0<10^{34}$) can be calculated from electroweak consideration. 

At $\beta_0=10^{34}$, Eqs. (\ref{properENERGYDENSITY}) and (\ref{properpressure1}) can be rewritten as
\bea
\rho &=& 5.24 \times 10^{95}\frac{h(\kappa)}{\beta_0^2} \sim 10^{27} h(\kappa)\;  \textrm{kg}\cdot \textrm{m}^{-3},\label{densitykapa1} \\
P &=& 4.73\times 10^{112}\frac{1}{\beta_0^2}\; g(\kappa) \sim 10^{44} g(\kappa)\; \textrm{Pascals}.
\eea
For density higher than the nuclear one, the QG plays a main role due to degeneracy pressure even when the interaction correction is disregarded. 
Therefore, it is of great interest to investigate the cores of the compact stars. For nuclear matter at equilibrium, the QG effects are negligible.
From radio binary pulsars, several accurate masses determinations for neutron stars are reported \cite{Zhangq}. Accordingly, the related difference densities should be taken into consideration.

By applying the Newtonian limit, Eq. (\ref{TOV1-PR-New}), with normal nuclear density, two configurations of the compact stars can be addressed: \cite{WYZ2010JHEP043q}.
\begin{itemize}
\item The neutron star is almost composed of ultra-relativistic particles and
\item The major contribution to its mass is coming from non-relativistic cold nuclei. 
\end{itemize}
TOV Eqs. (\ref{TOV1-PR}) and (\ref{TOV1-mR}) are applicable for ultra-compact stars. By setting $r=r_0\tilde{r},\,m=m_0\tilde{m}$, $P=P_0 \tilde{P}$ and 
\bea
\rho &=& \frac{m_0}{4\, \pi\, r_0^3} \tilde{\rho} \equiv \rho_0\, \tilde{\rho}, \nn \\
P_0 &=&{\rho_0}{c^2}, \label{TOVscale-1}\\
\frac{G\, m_0}{c^2\, r_0} & \equiv & 1, \nn
\eea
the TOV Eqs (\ref{TOV1-PR}) and (\ref{TOV1-mR}) are reduced to 
\bea
\frac{d\tilde{P}}{d \tilde{r}} &=& -(\tilde{\rho}+\tilde{P})\frac{\tilde{m}+\tilde{r}^3\tilde{P}}
{\tilde{r}(\tilde{r}-2\tilde{m})},\label{tov-fermnoqg1dem} \\
\frac{d\tilde{m}}{d\tilde{r}} &=& \tilde{r}^2\tilde{\rho}. \label{tov-fermnoqg2dem}
\eea
At vanishing $r$, EoS are given by Eqs. (\ref{particlenumb1density}), (\ref{properENERGYDENSITY}) and (\ref{properpressure1}). At $\kappa\to 0$, $P = \rho/3c^2$. When defining $r=r_0\tilde{r}$ and $m=m_0\tilde{m}$, where 
\bea
r_0^{-2} &\equiv & \frac{4\, \pi\, G}{c^4}\, \frac{8\, \pi}{(h\, c)^3}{E_H^4},\label{r01minimall3} \\
m_0 &\equiv & 4\, \pi\, r_0^3\, \frac{8\, \pi}{c^2\, (h\, c)^3}\, E_H^4=1.93 \times 10^{-8}\, \beta_0\;\; \text{kg}, ,\label{M01minimall3} \\
P_0 &=& \rho_0\, c^2 =\frac{8\, \pi}{(h\, c)^3}{E_H^4},\label{TOVScale-2f}
\eea
then, $r_0$ is the minimum radius \cite{WYZ2010JHEP043q} and $r_0=\sqrt{\frac\pi 4}{\beta_0}\, l_p=\sqrt{\frac\pi 4} \sqrt{\beta_0}\, \Delta_{\textrm{min}}=1.43\times 10^{-35}\, \beta_0\;\; \text{m}$, which appears also Eqs. (\ref{tov-fermnoqg1dem})--(\ref{TOVScale-2f}). It is obvious that the system can not have an arbitrary scale. The scale is entirely determined by $\beta_0$.

\subsubsection{Conclusion}

Based on GUP approaches, and  by using TOV equations and suitable EoS for zero-temperature ultra-relativistic Fermi gas, the QG effects on compact stars can be studied. It was shown that ${2m(r)}/{r}$ varies with $r$ \cite{Zhangq}. The QG plays an important role in the region $r\sim 10^3 r_0$, where $r_0\sim \beta_0 l_p $ is close to the center of compact stars. The metric components $g_{tt}\sim r^4$, then $g_{rr}=[1-{r}^2/(6r_0^2)]^{-1}$ \cite{Zhangq}. All these effects are different from those obtained from the classical gravitational aspects.

\subsection{Saleker-Wigner inequalities}
\label{sec:SW1}

\subsubsection{Saleker-Wigner inequalities and Heisenberg uncertainty principle }

At the event horizon and based on HUP, the scale $R_{\rm g}$ implies conventional derivation of Hawking lifetime. When assuming that the black hole is a black body, the Stefan-Boltzmann law can be implemented \cite{Townsendq,Adlerq}.
If a clock of mass $M$ has the uncertainty in its quantum position $\Delta x$, then the momentum uncertainty $\hbar\, \Delta x^{-1}$. Such a clock shall have an accuracy $\tau$ and be able to measure time intervals up to $T$. After $t$ time, the position uncertainty becomes $\Delta x'=\Delta x+\hbar tM^{-1}\Delta x^{-1}$ \cite{swLitr}.

If effects of the clock mass are neglected, the minimum position uncertainty $\Delta x=\sqrt{\hbar t/M}$. In order to keep the clock accurate over the total running time, $T$, the linear spread of the clock $\lambda$ must be limited \cite{swLitr}
\begin{eqnarray}
\lambda \geq 2 \sqrt{\hbar\, T/M}. \label{1}
\end{eqnarray}
When the position uncertainty gets of the same order of magnitude, the clock's size becomes larger than the position uncertainty \cite{swLitr}. This is nothing but the Salecker-Wigner first clock-inequality \cite{Wignerq}. The quantum position uncertainty must not be larger than the minimum wavelength of the quanta striking it in order to read out the time. This constrain is expressed as $\Delta x'\leq c\tau$. 

When a signal with nonzero rest mass is utilized, a bound on the minimum mass of the clock is defined throughout \cite{swLitr}
\begin{eqnarray} 
M \geq \frac{4\hbar}{c^2\tau}\left(\frac{T}{\tau} \right). \label{2}
\end{eqnarray}
The expression (\ref{2}) defines the Salecker-Wigner second clock inequality \cite{Wignerq}. It is apparent that this inequality is more restricted than the one imposed by HUP for energy and time. Furthermore, the uncertainty in quantum position should not result in significant inaccuracies in measuring the time, even over the whole running time of the clock \cite{swLitr}.  In deriving Salecker-Wigner clock inequalities, Eqs. (\ref{1}) and (\ref{2}),  unsqueezed, unentangled, and Gaussian wave packets should conjectured. With this regard, the black holes represent as analogue (not digital) quantum-clocks \cite{Ng01q}. 

Assuming that the Schwarzschild radius $R_{\rm g}=2GM/c^2$ is the minimum size, then the maximum time $T$ reads \cite{barrow96}
\begin{eqnarray}
\label{4}
T \sim \frac{G^2 M^3}{\hbar c^4}=\frac{M^3}{m^3_{\rm p}} t_{\rm p},
\end{eqnarray}
where $t_{\rm p}$ and $m_{\rm p}$ are the Planck time and mass, respectively. For black hole, $T$ is given by the Hawking lifetime \cite{Hawkingq}. Compared to  conventional methods, the application of Salecker-Wigner inequality, Eq. (\ref{1}), to the event horizon scale helps in estimating Eq. (\ref{4}). This prediction seems to be valid even without the assumption of black-body radiator. 

Based on GUP, a modified clock inequalities shall be deduced in Sec. \ref{sec:mswi}. Also, modified black-hole lifetime can be found \cite{swLitr}.

\subsubsection{Modified Salecker-Wigner  inequalities}
\label{sec:mswi}

Using HUP and some properties of black holes, Scardigli illustrated how GUP can be explained in a {\it gedanken} (thought) experiment \cite{7,Scardigli}
\begin{eqnarray}
\label{11}
\Delta x \geq \frac{\hbar}{\Delta p}+l^2_{\rm p} \frac{\Delta p}{\hbar},
\end{eqnarray}
where $l^2_{\rm p}=\sqrt{G\hbar/c^3}$ is the Planck length. The GUP approach, Eq. (\ref{11}), can be rewritten as $\Delta x\geq \hbar(1/\Delta p+\beta \Delta p)$, where $\beta$ is a constant \cite{Changq}. Accordingly,  a modified black hole lifetime was estimated \cite{swLitr} in a conventional way \cite{Adlerq}.
\begin{eqnarray}
\label{ad} 
T_{\rm ACS} &=& \frac{1}{16} \Big\{\frac{8}{3}\Big(\frac{M}{m_{\rm
p}}\Big)^3-8\frac{M}{m_{\rm p}}-\frac{m_{\rm
p}}{M}+\frac{8}{3}\Big[\Big(\frac{M}{m_{\rm p}}\Big)^2-1\Big]^{3/2}
\nonumber\\&&-4\sqrt{\Big(\frac{M}{m_{\rm
p}}\Big)^2-1}+4\arccos\Big(\frac{m_{\rm
p}}{M}\Big)+\frac{19}{3}\Big\}\; t_{\rm ch}.
\end{eqnarray}
The subscript $ABC$ stands for Adler-Chen-Santiago \cite{Adlerq} and $t_{\rm ch}=16^2\times60\, \pi\, t_{\rm p}$. In deriving this expression, the authors of Ref. \cite{Adlerq} assumed a black-body radiator and relativistic dispersion-relation $E=pc$. As noted in Ref. \cite{Nozari06q}, modified uncertainty principle is connected with modified dispersion-relation.

Because the space-time fluctuation will be significant when the measured length scale approaching Planck length, it is reasonable to expect that the linear spread of a clock must not be less than the Planck distance. According to GUP approaches, the Salecker-Wigner  inequalities gain modification. 

The GUP approach, Eq. (\ref{11}), implies a minimum length, $2\, l_{\rm p}$. This is a rigorous  limit on the linear spread of the clock \cite{Wignerq}. From Eq. (\ref{11}), it is obvious that a clock with mass $M$ has position $\Delta x$ and momentum uncertainty $\Delta p\sim\frac{\Delta x\hbar}{2l^2_{\rm p}}\left[1-\sqrt{1-4l^2_{\rm p}/\Delta x^2}\right]$ \cite{Adlerq}. According to Ref. \cite{swLitr}, the Salecker-Wigner inequality can be derived
\begin{eqnarray}
\lambda \geq 2\, l_{\rm p} \sqrt{1+\frac{\hbar T}{M l^2_{\rm p}}}. \label{13}
\end{eqnarray}
This limit is stronger than the one given in Eq. (\ref{1}) and can reproduce Eq. (\ref{1}) for $\hbar T\gg Ml^2_{\rm p}$. The condition that the position uncertainty created by the time measurement must not be larger than the minimum wavelength is still valid. 

Also, the Salecker-Wigner second  inequality, Eq. (\ref{2}), gets modification \cite{swLitr}
\begin{eqnarray} 
M \geq \frac{4\hbar T}{c^2\tau^2}\frac{1}{1-4 t^2_{\rm p}/\tau^2}. \label{20}
\end{eqnarray}
Accordingly, the mass, the running time, the accuracy of the clock, and the Planck time are related with each others. The concepts of gravity and the quantum uncertainty might be connected with each other. Obviously, expression (\ref{20}) defines a lower limit on the accuracy of $\tau > 2 t_{\rm p}$. Similar to the Salecker-Wigner inequalities, Eqs. (\ref{1}) and (\ref{2}), Eqs. (\ref{13}) and (\ref{20}) are valid for single analogue clocks, merely. 

The maximum running/life time of the black hole is also modified \cite{swLitr}
\begin{eqnarray}
\label{22}
T_{\rm MB} \sim \frac{M R^2_{\rm g}}{4 \hbar} \left(1-4 l^2_{\rm p}/R^2_{\rm g}\right) = \frac{M^3}{m^3_{\rm p}} \left(1-m^2_{\rm p}/M^2\right)\; t_{\rm p}.
\end{eqnarray}
Expression (\ref{22}) includes term $M\, t_{\rm p}/m_{\rm p}$, which distinguishes this it from the Hawking lifetime, Eq. (\ref{4}). This expression is valid for $M\geq m_{\rm p}$. Using the GUP approach, Eq. (\ref{11}), Adler, Chen and Santiago \cite{Adlerq} concluded that the thermal radiation of the black hole should stops at the Planck length. This is consistent with the results obtained by the modified clock inequalities \cite{swLitr}. 

\begin{figure}
\centering{
\includegraphics[width=10cm]{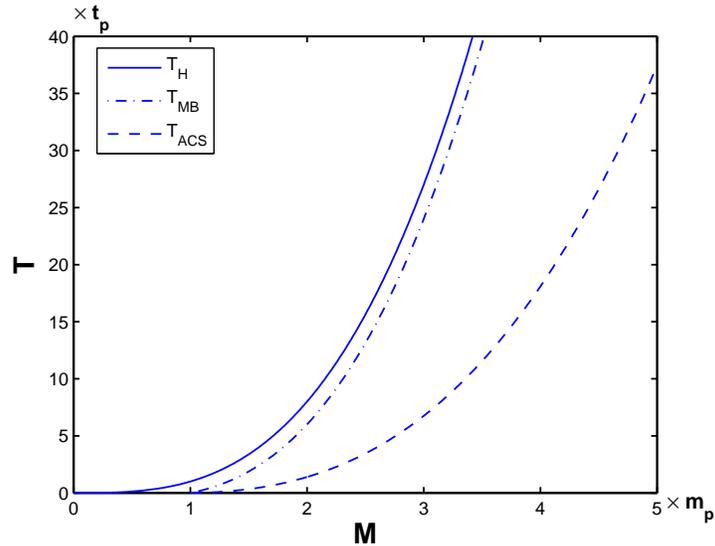}
\caption{A comparison between the Hawking lifetime $T_{\rm H}$, the modified clock inequality lifetime $T_{\rm MB}$, and the Adler-Chen-Santiago lifetime $T_{\rm ACS}$, where the numerical factor $16^2 \times 60 \pi$ is omitted. The graph taken from \cite{swLitr}. \label{Fig:SW1}}
}
\end{figure}

We find that the first two terms in $T_{\rm ACS}$ are consistent with $T_{\rm MB}$, aside from the factor $16^2\times 60 \pi$. A comparison between $T_{\rm H}$ and $T_{\rm MB}$ and $T_{\rm ACS}$ is presented in Fig. \ref{Fig:SW1}.

Tor the travel time of photons, the minimum time interval measured by the black hole  is given as  \cite{barrow96,Ng01q} 
\bea
\label{06}
\tau &\sim & 2\, G\, \frac{M}{c^3}=\frac{R_{\rm g}}{c}.
\eea 
Thus, the black hole can be utilized as an information-processing system. The number of computational steps is given by $T_{\rm MB}/\tau$ \cite{swLitr} 
\begin{eqnarray}
\label{223}
N  \sim \frac{M^2}{m^2_{\rm p}}\, \left(1 - \frac{m^2_{\rm p}}{M^2}\right).
\end{eqnarray}
In Planck units, the number of bits required in specifying the information content of the black hole at the event horizon area  can be estimated from the black hole entropy \cite{entr1,entr3} or the holographic principle \cite{Hooftq,Susskindq}.

\subsubsection{Conclusion}

The modified clock inequalities give bounds on the size and the accuracy of the quantum analogue clock. These must be lager than $2 l_{\rm p}$ and $t_{\rm p}$, 
respectively. A modified black-hole lifetime is obtained $T_{\rm MB} \sim \frac{M^3}{m^3_{\rm p}}t_{\rm p}(1-m^2_{\rm p}/M^2)$. Apparently, this differs from the Hawking lifetime \cite{Hawkingq}. By viewing a black hole as an information-processing system \cite{swLitr}, the number of bits required to specify the information content of the black hole is given as $N\sim M^2/m^2_{\rm p} (1-m^2_{\rm p}/M^2)$.

\subsection{Entropic Nature of the gravitational force}

\subsubsection{Newton's law of entropic nature }
\label{sec:verld1}

Based on the holographic principle, Verlinde proposed a radical revision to the nature of gravitational force \cite{everlind1}. It is assumed that: \cite{everlind1}
\begin{itemize}
\item in the vicinity of surface $\Omega$, the change of the surface entropy $\Delta S_\Omega$ should be related to $\Delta x$ and to the change of the radial distance from the surface $\lambda$
\be
\Delta S_\Omega = 2\pi k_B \frac{\Delta x}{\lambda_m}.,
\ee
\item a force $F$ is resulted in from such a generic expression for the thermodynamics,
\be
\label{nature}
F \Delta x = T \Delta S_\Omega,
\ee
\item the number of information-bits $N$ stored on the surface $\Omega$ reads $N = A_\Omega/\ell_P^2$,
where $A_\Omega$ is the area of surface $\Omega$ and $\ell_P$ is the Planck length,
\item at the temperature $T$, the surface $\Omega$ is in thermal equilibrium. Therefore, all bytes are equally likely. The available energy is equipartitioned among all them, i.e. 
\be 
U_\Omega = \frac{1}{2}\, N\, k_B\, T = M\, c^2,
\ee
where $M$ is the rest mass. 
\item then, the Newton's law for gravity can be derived \cite{Nicolini:2010nb}.
\be
 F= G\, \frac{M\, m}{r^2}.
\ee 
\end{itemize}

\subsubsection{Non-commutative geometry implying a modification in Newton's law}
\label{sec:newtnVerl2}

From Verlinde's procedure \cite{everlind1}, section \ref{sec:verld1}, modifications on the Newtonian law can be deduced. The basic assumption is that the entropy offer a description for the gravity in the underlying microstructure of a quantum spacetime \cite{everlind1,Nicolini:2010nb},
\be
\Delta S_{\Omega} = k_B\, \Delta A\, \left(\frac{c^3}{4\,\hbar\, G}+\frac{\partial s}{\partial A}\right). \label{genentropy}
\ee
Here $s$ is a function of the area $A$. The description of the microscopic structure of quantum system can be given by non-commutative geometry  \cite{Nicolini:2010nb}, which encodes spacetime microscopic dof by means of modified uncertainty relation between the coordinates, $\Delta x^\mu \Delta x^\nu\ge \theta$ \cite{Nicolini:2010nb}, where the parameter $\theta$ has the dimension of length squared. When the coordinate operators fail to commute,  $\theta$ serves as a natural ultraviolet cutoff from the geometry \cite{Nicolini:2010nb}.
\begin{equation}
[x^\mu, x^\nu]=i\Theta^{\mu\nu}. \label{nc}
\end{equation}
In order to introduce the non-commutative scale, $\theta$ should be equivalent to $|\Theta^{\mu\nu}|$.
Because of uncertainty on $\Omega$, there should exist a fundamental unit $\Delta S_{\theta}$, which is perceived at $\Delta x_{min}\propto \lambda_m$. Therefore, the change in the entropy reads  $\Delta S_\Omega=\Delta S_\theta \left(\Delta x/\Delta x_{min}\right)$ \cite{Nicolini:2010nb} or $\Delta x_{min}=\alpha^2\,\lambda_m /(8\, \pi)$. The number of information-bits $N=A_\Omega/\theta$. In non-commutative geometry, the Planck scale and the GUP parameter $\alpha$ come up with corrections to the entropy  $\Delta S_{\theta}=k_B \theta \left(\frac{c^3}{4 \hbar G}+\frac{\partial s}{\partial A}\right)$ \cite{Nicolini:2010nb}.
The temperature is given as 
\begin{equation}
T=\frac{M}{r^2}\frac{\theta\, c^2}{2\, \pi\, k_B}. \label{temp2}
\end{equation}
By combining all these equations in Eq. (\ref{nature}), a correction to the Newton's law will be implied  \cite{Nicolini:2010nb}, 
\begin{equation}
F=\frac{Mm}{r^2}\left(\frac{4 c^3 \theta^2}{\hbar\alpha^2}\right)\left[\frac{c^3}{4\hbar G}+\frac{\partial s}{\partial A}\right]. \label{ncforce}
\end{equation}
Equation (\ref{ncforce}) coincides with the Newton's law for gravity to first term \cite{Nicolini:2010nb}, if $\theta=\alpha\ell_P^2$. The modified Newton's law reads \cite{Nicolini:2010nb} 
\begin{equation}
\Delta F = \frac{GMm}{r^2}\left[1+4\ell_P^2\frac{\partial s}{\partial A}\right]. \label{Newtoncorr}
\end{equation}

\subsubsection{Conclusion}

Verlinde considered that the gravitation force has an entropic nature. This would allow to deal with the gravity by means of the thermodynamic mechanics \cite{everlind1}. On the surface $\Omega$, the introduction of the non-commutative geometry implies change in the entropy as function of area $A$. Also the  introduction of Planck scale to the GUP parameter, $\alpha$, leads to corrections in the Newtonian law of gravity. The appearance of a linear term reflecting the effect of linear GUP approach results in another modification in the number of bits and in the temperature of the black hole. All these modify the gravitational force with a negative correction, which is inversely proportional to the cubic of the apparent black hole radius.

\subsection{Measurement of minimal time intervals}

Because of absence of a QG theory, there is no theory of the gravity at very short distances. At larger distances comparable with the Planck length, GR would be a good approximation \cite{itzhaki}. In order to measure the time intervals, a clock should be located at a distance $x$ away from the observer. The clock should emit - at least - one photon, from which the observer gains his information about the clock. Accordingly, three sources for uncertainty are expected \cite{itzhaki}:
\begin{itemize}
\item The accuracy of the time measurement, $\Delta t$.
\item The time taken by the photon to reach the observer, due to the uncertainty of the metric caused by the energy uncertainty $\Delta E$.
\item The uncertainty in the distance that the photon should cover to reach the observer is $2 R$, where $R$ is the clock's radius. This last source contributes with $2 R/c$ to the total uncertainty.
\end{itemize}

\subsubsection{Time uncertainty at shortest distance $x_c$}

At position $R\le x_{c}$, the shortest length is $x_{c}=\alpha \sqrt{G\hbar/c^{3}}$.  Accordingly, the uncertainty inequality \cite{ahq}, $\Delta t \Delta E \geq \hbar$  \cite{itzhaki} implies that there exists time uncertainty 
\begin{equation} \label{delta t}
\Delta T_{tot}(\Delta E)>\frac{\hbar}{\Delta E}+\frac{2\Delta EG}{c^{5}}\, \log\left(\frac{x}{x_{c}}\right), 
\end{equation}
where $\Delta T_{tot}$ is the error in the whole process.
As mentioned above $\Delta E<c^{4} x_{c} / (2 G)$  is satisfied only for $x>e^{2/\alpha^{2}} x_{c}$.

If $x_{c}<x<e^{2/\alpha^{2}} x_{c}$, then the minimum time uncertainty \cite{itzhaki} 
\begin{equation}
\Delta T_{min}=\frac{x_{c}}{c}\left[\frac{2}{\alpha^{2}}+\log\left(\frac{x}{x_{c}}\right)\right]. 
\end{equation}
And the  related energy uncertainty $\Delta E=x_{c}c^{4}/(2G)$.

\subsubsection{Time uncertainty at largest distance $x_c$}

At distances $R$, which exceed $x_{c}$, the GR can be utilized inside the proposed clock.  
In order to measure time as accurate as possible,  a clock with $R=x_{c}$ \cite{itzhaki} should be utilized. The total time uncertainty is given as \cite{itzhaki}  
\begin{equation}
\Delta T_{tot}(\Delta E)>\frac{\hbar}{\Delta E}+\frac{2\Delta
EG\log\left[\frac{x}{x_{c}}\right]}{c^{5}}+\frac{2}{c}\, x_{c}>\frac{\hbar}{\Delta
E}+\frac{2\Delta E\, G\, \log\left[\frac{x}{x_{c}}\right]}{c^{5}}.
\end{equation}

\subsubsection{Consequences on physics of early Universe}

The effects of GUP on the inflationary dynamics and the thermodynamics of the early Universe have been studied \cite{Tawfik:2012he}, section \ref{sec:inflt1}. The tensorial and scalar density fluctuations are evaluated and compared with the standard case. A convincing agreement with Wilkinson Microwave Anisotropy Probe data has been reported \cite{Tawfik:2012he}. When assuming that a quantum gas of scalar particles should be confined in a thin layer near the apparent horizon of the Friedmann-Lemaitre-Robertson-Walker (FLRW) Universe which satisfies the boundary condition, we find that the number density and entropy densities and the free energy can be reproduced by using the GUP approach. A qualitative estimation for the QG effects on all these thermodynamic quantities was introduced \cite{Tawfik:2012he}.

\subsubsection{Conclusion}

The possibility of defining a measurable maximal energy and a minimal time interval is estimated in different quantum aspects. First, we find that the quadratic GUP approach gives non-physical results \cite{DahabTaw}, as the resulting maximal energy $\Delta\, E$ violates the conservation of energy. The minimal time interval $\Delta\, t$ shows that the direction of time arrow is backwards.  Calculations at the shortest distance, at which GR is assumed to be a good approximation for QG and at larger distances are performed. It is found that both maximal energy and minimal time have the order of the Planck time. The uncertainties in both quantities are bounded, accordingly.  It is found that the quadratic GUP approach results in finite $\Delta\, E$ and positive $\Delta\, t$. Based on the Schwarzshild solution, this result is utilized in calculating the maximal energy and minimal  time.

\newpage

\section{Applications of linear GUP approach}
\label{implLinear}

\subsection{Inflationary parameters}
\label{sec:1GUPFRIED}

The study of GUP effects on the inflationary era represents an essential ingredient to many investigations \cite{Tawfik:2012he}. Some of observations have been elaborated in sections \ref{sec:flrw} and \ref{qGUPfriedm}. We first start from the number density arising from the quantum states in the early Universe. Then, we calculate the free energy and entropy density. The recipe of calculating thermodynamic quantities from the quantum nature of physical systems dates back to a about one decade \cite{qentra,qentrb,qentrc,qentrd,qentre,qentrf}, where the entropy arising from mixing quantum states of degenerate quarks in a very simple hadronic model has been estimated and applied to different physical systems.

\subsubsection{Inflation parameters and linear GUP approach}
\label{sec:gup}

As discussed in previous sections, the linear GUP approach \cite{advplb,Das:2010zf} predicts a maximum observable momentum and a minimal measurable length. Furthermore, the standard commutation relations are conjectured to be changed. In order to relate this with the inflation era, we define $\phi$ as the scaler field deriving the inflation in the early Universe \cite{Tawfik:2012he}. The pressure and energy density, are given in Eq. (\ref{eq:rhoEU}) and (\ref{eq:pEU}), respectively.

The slow-roll parameters \cite{slowroll} 
\begin{eqnarray}
    \epsilon &=& \frac{{M_{p}}^2}{2}\,\left(\frac{\acute{V}(\phi)}{V(\phi)}\right)^2,\label{7}\\
    \eta &=& {M_{p}}^2\,\frac{\acute{\acute{V}}(\phi)}{V(\phi)}, \label{8}
\end{eqnarray}
where $M_{p}=m_{p}/\sqrt{8 \pi}$ is the four dimensional reduced Planck mass can be approximated to guarantee that the quantities in Eq. (\ref{7}) and (\ref{8}) are much smaller than unity. These conditions are supposed to ensure an inflationary phase, in which the expansion of the Universe is accelerating, where the conformal time is $\tau = - 1/(a\,H)$.

To distinguish it from $k$ \cite{Tawfik:2012he},  the curvature parameter, the wave number is labelled as $j$. Thus, $j$ represents the {\it comoving} momentum and thus $j \longrightarrow j(1-\alpha\,j)$ and $a= j(1-\alpha\,j)/P$. The scalar spectral index is 
\begin{eqnarray}
n_s  &=& \frac{d\,\ln p_s}{d\,\ln j(1-\alpha\,j)} +1
  = \frac{(1-2\alpha\,j)}{(1-\alpha\,j)}\,\frac{d\,\ln p_s}{d\,\ln j}+1 \simeq (1-\alpha\,j)\,\frac{d\,\ln p_s}{d\,\ln j}+1. \label{12}
\end{eqnarray}
where $p_s$ is the amplitude of the scalar density fluctuations. The derivative of  $H$ with respect to $j$ \cite{slowroll,dHdK}
\begin{equation}\label{15}
d H/d j = - \epsilon\, H/j.
\end{equation}
This is valid at the horizon crossing epoch.
From momentum modification, an approximative expression for $H$ as a function of  the modified momentum is obtained
\begin{equation}\label{17}
H \simeq j^{-\epsilon}\, \exp(\epsilon\, \alpha\, j).
\end{equation}
Therefore, it can be concluded that the linear GUP approach enhances the Hubble parameter so that $H(\alpha=0)/H(\alpha\, \neq 0)<1$.

\subsubsection{Tensorial and scalar density fluctuations}

One of main consequences of the inflation is the generation of primordial cosmological perturbations \cite{tensorialscalarD} and long wavelength gravitational waves (tensor perturbations) \cite{Tawfik:2012he}. Therefore, the tensorial density perturbations (gravitational waves) can serve essential tools in distinguishing between different inflationary models \cite{tensorialscalarB}. For completeness, we mention that the perturbations typically give a much smaller contribution to the cosmic microwave background (CMB) radiation anisotropy than the inflationary adiabatic scalar perturbations \cite{tensorialscalarC}.

The tensorial and scalar density fluctuations, respectively, read \cite{Tawfik:2012he}
\begin{eqnarray}
  p_t &=& \left(\frac{H}{2\pi}\right)^{2} \left[1-\frac{H}{\Lambda}\,\sin\left(\frac{2\Lambda}{H}\right)\right] \nn\\
 & =& \left(\frac{k^{-\epsilon} e^{\epsilon\,\alpha\,k}}{2\pi}\right)^{2} \left[1-\frac {k^{\epsilon-1} e^{-\epsilon\,\alpha\,k}}{a}\,\sin\left(\frac{2}{a k^{1-\epsilon} e^{\epsilon\,\alpha\,k}}\right)\right], \\
  p_s &=& \left(\frac{H}{\dot{\phi}}\right)^2\left(\frac{H}{2\pi}\right)^{2} \left[1-\frac{H}{\Lambda}\,\sin\left(\frac{2\Lambda}{H}\right)\right] \nn\\
  & =& \left(\frac{H}{\dot{\phi}}\right)^2\left(\frac{k^{-\epsilon} e^{\epsilon\,\alpha\,k}}{2\pi}\right)^{2} \left[1-\frac {k^{\epsilon-1} e^{-\epsilon\,\alpha\,k}}{a}\,\sin\left(\frac{2}{a k^{1-\epsilon} e^{\epsilon\,\alpha\,k}}\right)\right].
\end{eqnarray}
Then, the ratio tensor-to-scalar fluctuations, \cite{tensorRa1,tensorialscalar,tensorialscalarB} 
\begin{equation} \label{18}
    \frac{p_t}{p_s} = \left(\frac{\dot{\phi}}{H}\right)^{2}= \left(\frac{16\pi\,\sqrt{\epsilon}\,V}{M_4\,k^{-\epsilon} e^{\epsilon\,\alpha\,k}}\right)^2.
    \end{equation}
In the standard case, i.e. without GUP, this ratio is assumed to linearly depend on the inflation slow-roll parameters \cite{tensorRa1} $p_t/p_s = {\cal O}(\epsilon)$. The exact dependence shall be given in Eq. (\ref{eq:cal2}). The potential itself is model-dependent, for instance, $V(\phi) =
M_{p} \exp[- \sqrt{2/H_0 p}\, \phi]$ \cite{effectss}. For Klein-Gordon equation and according to the model presented given in Ref. \cite{Tawfik:2012he}, 
$\dot{\phi} = \left(\frac{\sqrt{2\epsilon}\,V}{M_{p}\,H}\right)^2$.
Then, the tensor-to-scalar fluctuations read \cite{Tawfik:2012he}
\bea
\frac{p_t}{p_s} &=& \left[\frac{\sqrt{2}\,V}{M_{p}}\, \frac{\sqrt{\epsilon}}{j^{-2\epsilon}\, \exp\left(2\, \epsilon\, \alpha\, j\right)}\right]^2.
\eea

\begin{figure}[htb]
\centering{
\includegraphics[width=5.5cm,angle=-90]{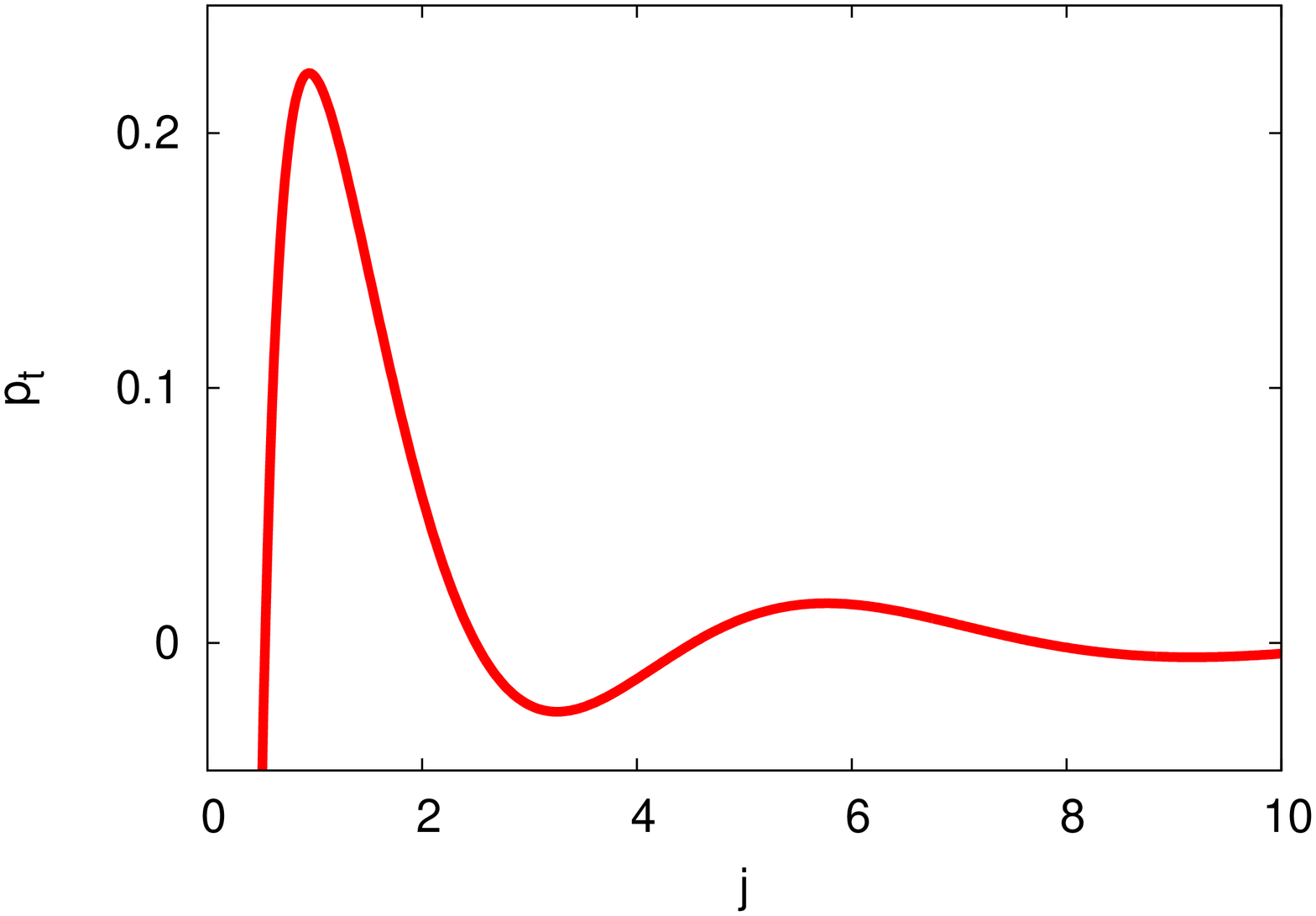}
\includegraphics[width=5.5cm,angle=-90]{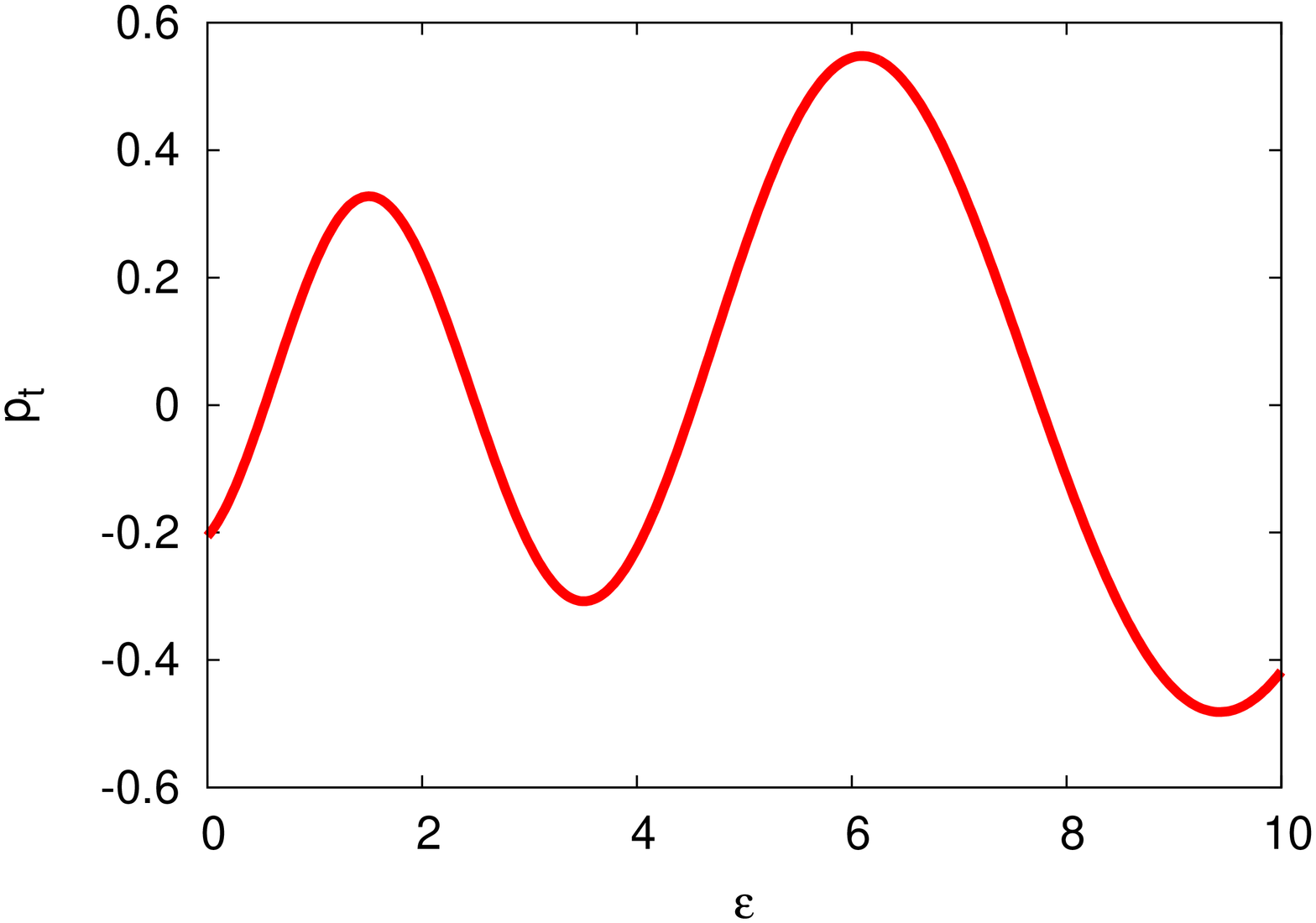}
\caption{The tensorial density fluctuations $p_t$ is given in dependence on the wave number $j$ (left panel) and on the slow-roll parameter $\epsilon$ (right panel). The GUP parameter $\alpha$ is kept constant, $\alpha=10^{-2}~$GeV$^{-1}$ (lower bound). It is assumed the $\sqrt{2}\,  V/M_{p}$ remains constant, (nearly unity). These two assumptions set the physical scale. The graphs taken from Ref. \cite{Tawfik:2012he}.
\label{fig:Pt1}
}}
\end{figure}

Fig. \ref{fig:Pt1} gives the tensorial density fluctuations $p_t$ in dependence on the wave number $j$ (left panel) and on the slow-roll parameter $\epsilon$ (right panel). In both graphs, the GUP parameter $\alpha$ is kept constant, $\alpha=10^{-2}~$GeV$^{-1}$, i.e. an upper bound is selected. Also, it is assumed that the potential is nearly of order of the reduced mass $M_{p}$, i.e. $\sqrt{2} V/M_{p}\sim 1$. It is obvious that $p_t$ diverges to negative values at low $j$. Increasing $j$ flips $p_t$ to positive values. When reaching a maximum value, it decreases almost exponentially and simultaneously oscillates around the abscissa. The amplitude of oscillation drastically decreases with increasing $j$. The right-hand panel shows that $p_t(\epsilon)$ oscillates around the abscissa. Here, the amplitude of the oscillation raises with increasing $\epsilon$. The oscillation can be detected in the CMB spectrum quantizing the primordial residuals of the quantum gravity effects.

\begin{figure}[htb]
\centering{
\includegraphics[width=5.5cm,angle=-90]{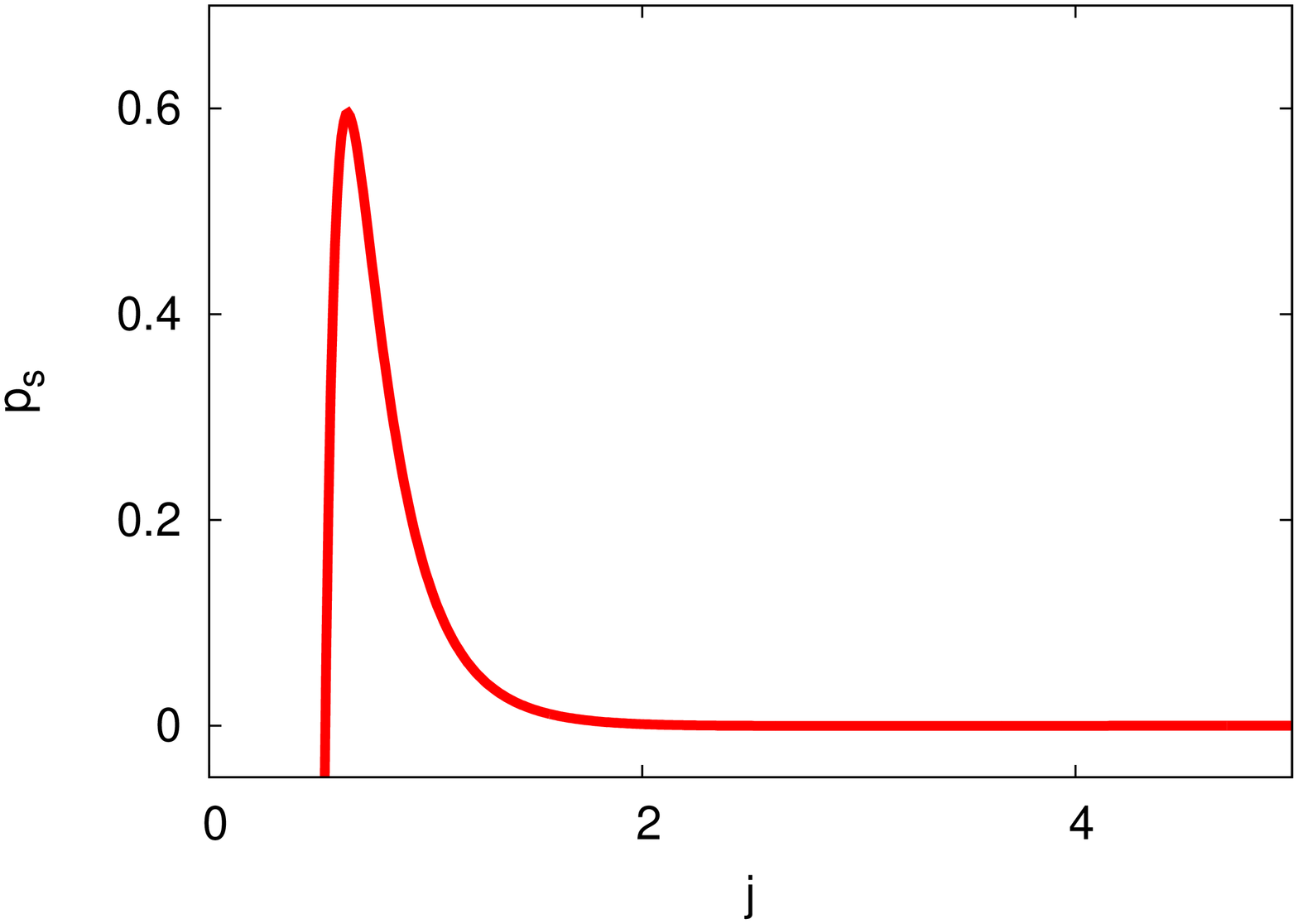}
\includegraphics[width=5.5cm,angle=-90]{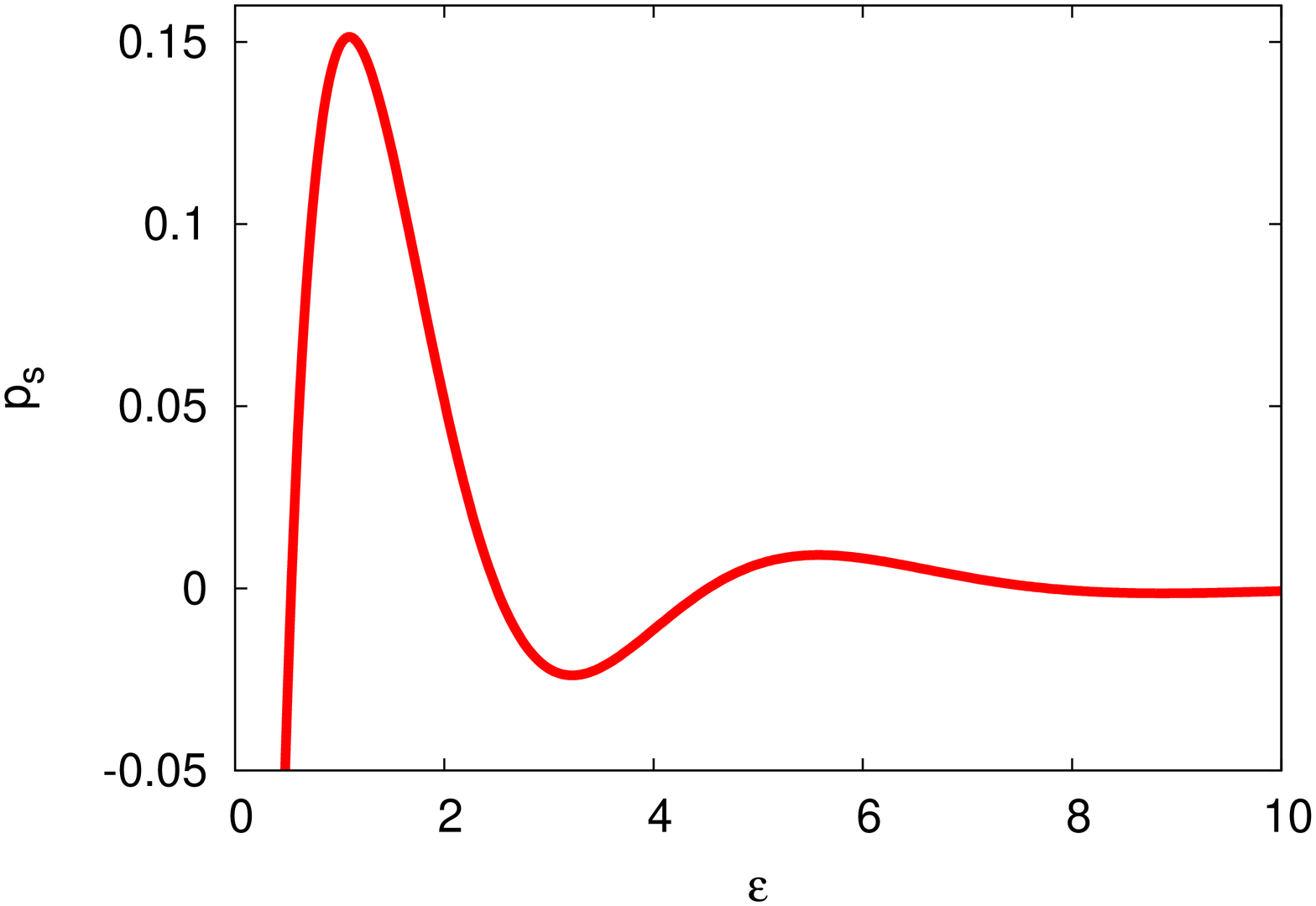}
\caption{The scalar density fluctuations $p_s$ is given in dependence on $j$ (left panel) and on slow-roll parameter $\epsilon$ (right panel). $\alpha$ and $\sqrt{2} V/M_{p}$ have the same values as in Fig. \ref{fig:Pt1}. They set the physical scale. The graphs taken from Ref. \cite{Tawfik:2012he}. 
\label{fig:Ps1}
}}
\end{figure}

Fig. \ref{fig:Ps1} refers to nearly the same behavior as that of the dependence of scalar density fluctuations $p_s$ on the wave number $j$ and $\epsilon$. It is apparent that $p_s$ diverges to negative value at low $j$. Increasing $j$ brings $p_s$ to positive values. But after reaching a maximum value, it decreases almost exponentially. Nevertheless its values remain positive. Oscillation of $p_s(\epsilon)$ is also observed. Here, $p_s(\epsilon)$ behaves almost similar to $p_t(k)$. After reaching a maximum value, it almost exponentially decreases and simultaneously oscillates around the abscissa. The amplitude of oscillation drastically decreases with increasing $\epsilon$.

\begin{figure}[htb]
\includegraphics[width=7.cm,angle=-90]{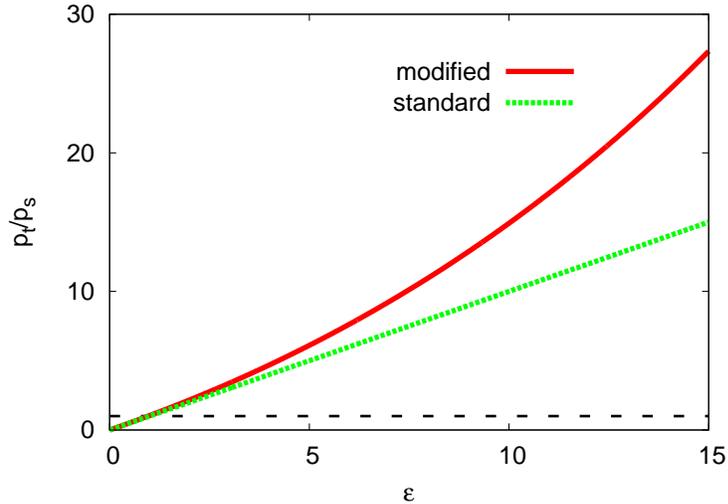}
\caption{The dependence of the ratio $p_t/p_s$ on the slow-roll parameter $\epsilon$ is given in ''standard'' and ''modified'' cases. The GUP parameter $\alpha$ (in ''modified'' case) and $\sqrt{2} V/M_{p}$ have the same values as in Fig. \ref{fig:Pt1} and therefore the physical scale is defined. The horizontal dashed line represents constant ratio $p_t/p_s$. The graph taken from Ref. \cite{Tawfik:2012he}. }
\label{fig:PtPs1}
\end{figure}

Fig. \ref{fig:PtPs1} gives the ratio $p_t/p_s$ in dependence on $\epsilon$ in two cases. The first case, the ''standard'' one, is represented by the solid curve. The second case, the ''modified'' case, is given by the dashed curve. The latter is characterized by finite $\alpha$, while in the earlier case, $\alpha$ vanishes. Compared to the ''standard'' case, there is a considerable increase in the values of  $p_t/p_s$ with raising $\epsilon$. For the ''modified'' case, i.e. upper bound of $\alpha=10^{-2}$ GeV$^{-1}$, the best fit results in an exponential function \cite{Tawfik:2012he}
\bea \label{eq:cal2}
\frac{p_t}{p_s} &=& \mu \, \epsilon^{\nu},
\eea
where $\mu=0.875\pm0.023$ and $\nu=1.217\pm0.014$. All these quantities are given in natural units. For the ''standard'' case, the results can be fitted by
\bea \label{eq:cal3}
\frac{p_t}{p_s} &=&  \epsilon.
\eea
The difference between Eqs. (\ref{eq:cal2}) and (\ref{eq:cal3}) is stemming from the factor in the denominator, which reflects the correction due to the GUP approach.

\subsubsection{Scalar spectral index and linear GUP approach}

The CMB results and many other astrophysical observations give constrains on the standard cosmological parameters such as  $H$,
baryon density $n_b$ and even the age of the Universe \cite{refff0,refff1}. Therefore, it important to suggest constrains on the power spectrum of the primordial fluctuations \cite{refff2}. This is doable via the spectral index. From Eq. (\ref{12}),  at $\sqrt{2} V/M=1$, the scalar spectral index reads \cite{Tawfik:2012he}
\bea
n_s &=& 1+ \left\{4 e^{-6 j \alpha  \epsilon} j^{6 \epsilon} \pi ^2 (1-j \alpha) \right. \nonumber \\
&& \epsilon  \left[-\frac{3}{2 \pi ^2} e^{6 j \alpha  \epsilon} j^{-6 \epsilon}\, \left(1-\frac{e^{-j \alpha  \epsilon} j^{-1+\epsilon
}}{a} \sin \left(\frac{2 e^{-j \alpha  \epsilon} j^{-1+\epsilon}}{a}\right)\right)+\right. \nonumber \\
&& \frac{3}{2 \pi ^2} e^{6 j \alpha  \epsilon} j^{1-6 \epsilon}\, \alpha  \left(1-\frac{e^{-j \alpha  \epsilon} j^{-1+\epsilon}}{a} \sin
\left(\frac{2 e^{-j \alpha  \epsilon} j^{-1+\epsilon}}{a}\right)\right)+  \nonumber \\
&& \frac{1}{4 \pi ^2\, \epsilon}e^{6 j \alpha  \epsilon} j^{-6 \epsilon}\left(-\frac{1}{a}e^{-j \alpha  \epsilon} j^{-1+\epsilon} \left(\frac{2
e^{-j \alpha  \epsilon} j^{-1+\epsilon} (-1+\epsilon)}{a}-\frac{2 e^{-j \alpha  \epsilon} j^{\epsilon} \alpha  \epsilon }{a}\right) \cos \left(\frac{2
e^{-j \alpha  \epsilon} j^{-1+\epsilon}}{a}\right)-\right. \nonumber \\
&& \left.\left.\left.\frac{e^{-j \alpha  \epsilon} j^{-1+\epsilon}}{a} (-1+\epsilon ) \sin \left(\frac{2 e^{-j \alpha  \epsilon} j^{-1+\epsilon
}}{a}\right)+\frac{e^{-j \alpha  \epsilon} j^{\epsilon} \alpha  \epsilon}{a} \sin \left(\frac{2 e^{-j \alpha  \epsilon} j^{-1+\epsilon}}{a}\right)\right)\right]\right\}/ \nonumber \\
&& \left[1-\frac{e^{-j \alpha  \epsilon} j^{-1+\epsilon}}{a} \sin \left(\frac{2 e^{-j \alpha  \epsilon } j^{-1+\epsilon }}{a}\right)\right].
\eea

The results of $n_r=d\, n_s/d\, \ln\, j$ are illustrated in the right-hand panel of Fig. \ref{fig:NsNr}. Early analysis of the Wilkinson Microwave Anisotropy Probe (WMAP) data \cite{wmap1,wmap2} indicates that $n_r = -0.03\pm 0.018$. As noticed in Ref. \cite{wmap2}, such an analysis may require modification, as the statistical significance seems to be questionable. On the other hand, it is indicated that the spectral index quantity $n_s-1$ seems to run from positive values on the long length scales to negative values on the short length ones. This is also noticed in left-hand panel of Fig. \ref{fig:NsNr}, where $n_s$ vs. $\omega$ is graphically illustrated. This agreement can be interpreted as an evidence that model agrees well with WMAP. Recent WMAP analysis shows that $n_s=0.97\pm0.017$ \cite{wmap3}. The importance of this agreement is a firm prediction of the inflationary cosmology through the  relation between scalar and tensor spectra. For modes which are larger than the current horizon, the tensor spectral index is positive \cite{runnn}.

\subsubsection{Consequences on later eras of the cosmological history}
\label{eras}

\begin{figure}[htb]
\centering{
\includegraphics[width=7.cm]{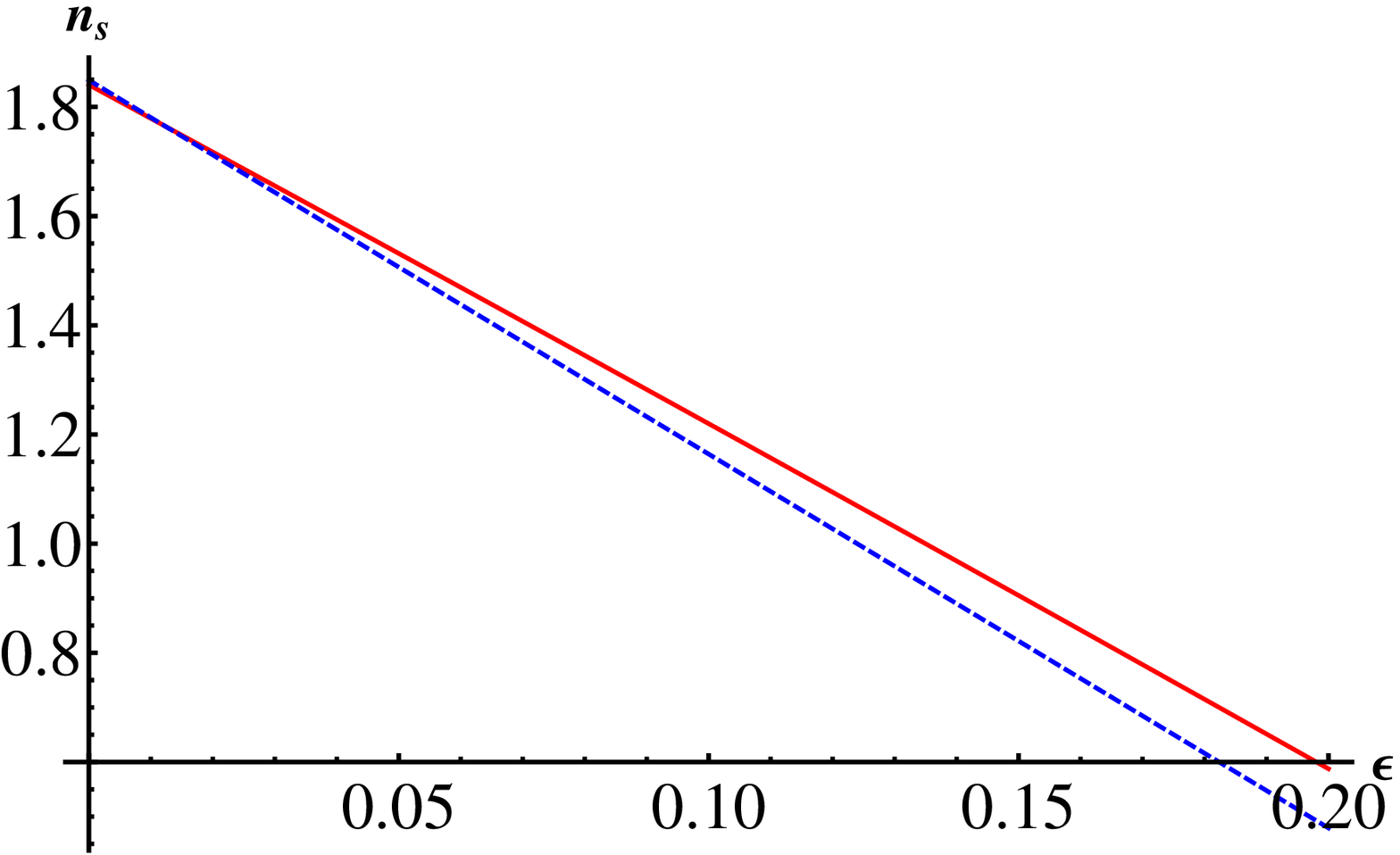}
\includegraphics[width=7.cm]{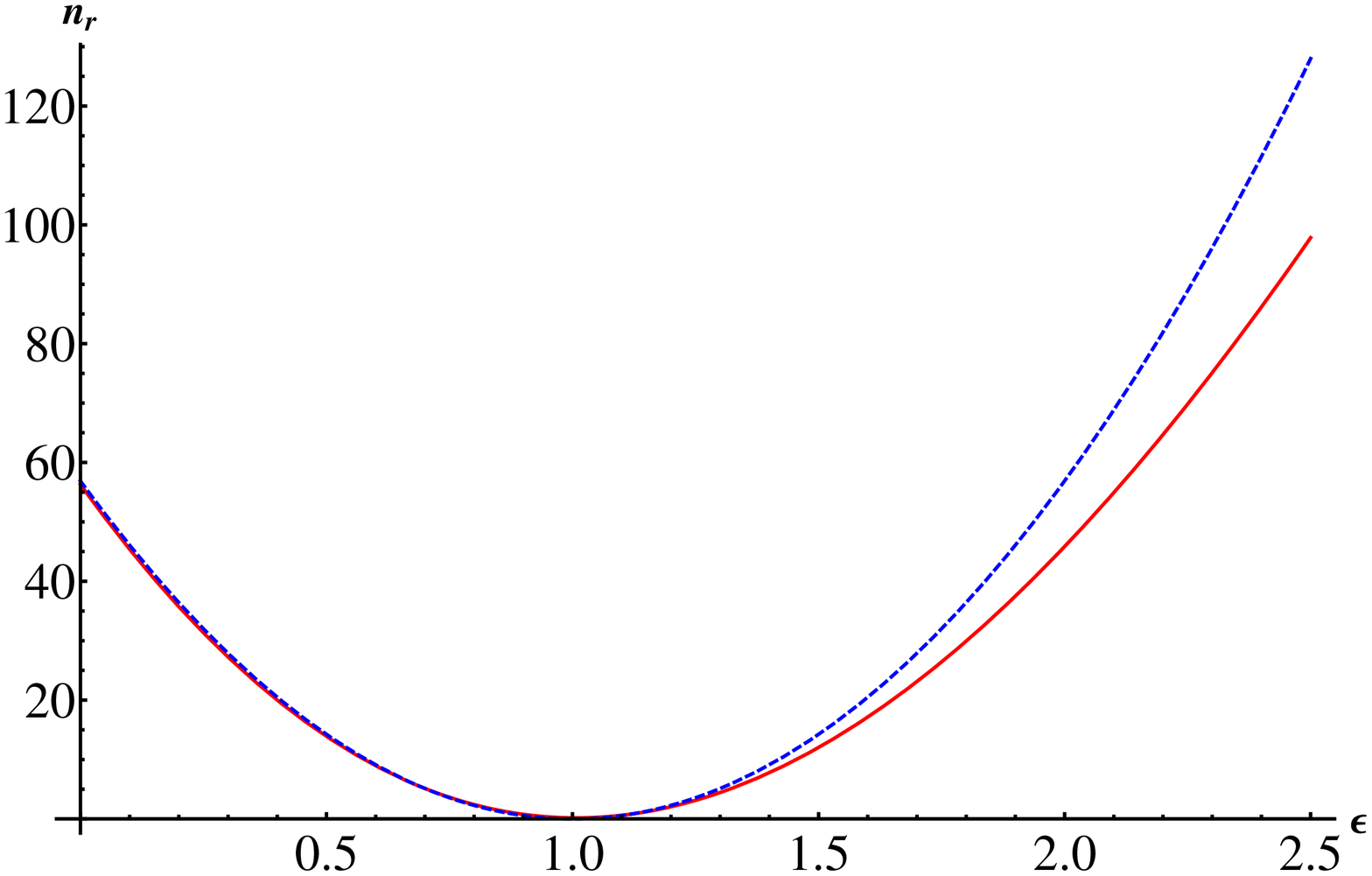}
\caption{Left-hand panel: the spectral index $n_s$ is given in dependence on $\epsilon$, where $j$ and $a$ are kept constant (equal $1$). The {\it ''running''} of $n_s$ is shown in the right-hand panel. The solid curves represent the results from the modified momentum $j\rightarrow j(1-\alpha j)$, i.e. applying the GUP approach. The dashed curves represent the standard case (unchanged momentum), i.e. $\alpha=0$.  All these quantities are given in natural units. The graphs taken from Ref. \cite{Tawfik:2012he}.
\label{fig:NsNr}
}}
\end{figure}

In describing the primordial power spectrum, many if not all inflation models implement three independent parameters: \begin{itemize}
\item the amplitude of the scalar fluctuations, 
\item the tensor-to-scalar ratio $n_r$ and 
\item the scalar spectral index $n_s$. 
\end{itemize}
They are observationally measurable and allow connection between the high-energy physics and the observational cosmology, in particular CMB.

The dependence of tensor-to-scalar, $p_t/p_s$, on $\epsilon$ is drawn in Fig. \ref{fig:PtPs1}. The ''modified'' momentum characterized by finite $\alpha$ and reflecting the quantum gravity effects, shows a considerable increase with raising $\epsilon$. Accordingly, the best fit was given in Eq. (\ref{eq:cal3}). The ''standard'' case can be fitted  by
\bea \label{eq:cal3b}
\left.\frac{p_t}{p_s}\right|_{s} &=& \epsilon.
\eea
The relation between Eqs. (\ref{eq:cal3}) and (\ref{eq:cal3b}) can be given as
\bea
\left.\frac{p_t}{p_s}\right|_{qc} &=& \left(\frac{\mu}{\left.\frac{p_t}{p_s}\right|_{s}}\right)^{\nu},
\eea
where the fitting parameters $\mu$ and $\nu$ are given in Eq. (\ref{eq:cal2}).

The dependence of $n_s$ on $\epsilon$ is presented in the left-hand panel of Fig. \ref{fig:NsNr}, while the dependence of its ''running'' in $n_s$ is illustrated in the right-hand panel. Including quantum gravity effects keeps the linear dependence of  $n_s(\epsilon)$ unchanged, but makes it slower than in the standard case of unchanged momentum. Increasing $\epsilon$ leads to an increase in the difference between modified and unmodified momentum. The running $n_s$ is not affected by quantum gravity at $\epsilon<1$. At higher $\epsilon$ values, $n_r$ in modified momentum gets slower than the one in standard case.

The spectral index $n_s$ describes the initial density ripples in the Universe. If $n_s$ is small, the ripples with longer wavelengths are strong, and vice versa. This has the effect of raising the CMB power spectrum on one side and lowering it on the other side. $n_s$  is like a fingerprint of the very beginning of the universe in that first trillionth of a second after the Big Bang called inflation. The way of distributing matter during the initial expansion reflects the nature of the energy field controlling the inflation. The current observations of $n_s$ are in agreement with the inflation prediction of a nearly scale-invariant power spectrum, corresponding to a slowly rolling inflation field and a slowly varying Hubble parameter during inflation. Based on Eq. (\ref{17}), the GUP approach apparently enhances the Hubble parameter so that $H(\alpha=0)<H(\alpha\neq 0)$.

\subsubsection{Conclusions}

The tensorial  $p_t$ and scalar density fluctuations $p_s$ are given in dependence on the wave number $j$ and on the slow-roll parameter $\epsilon$. For a systematic comparison, the GUP parameter $\alpha$ is kept constant, $\alpha=10^{-2}~$GeV$^{-1}$. Also, it is assumed the $\sqrt{2} V/M_{p}\sim 1$. We conclude that $p_t$ diverges to negative value at low values of $j$. When increasing $j$, $p_t$ gets positive values. After reaching a maximum value, it almost exponentially decreases but simultaneously oscillates around the abscissa. The amplitude of oscillation drastically decreases with increasing $j$. Also, $p_t(\epsilon)$ is found to oscillate around the abscissa. Here, the amplitude of the oscillation raises with increasing $\epsilon$. The oscillation can be detected essentially in the CMB spectrum quantizing the primordial residuals of the quantum gravity effects.

The running  $n_s$ is utilized to shed light on the scaling of spectral index. The WMAP data indicates that the spectral index quantity $n_s-1$ seems to run from positive values in long length scales to negative values in short length scales \cite{wmapns1}. This behavior was already confirmed in Ref.  \cite{Tawfik:2012he}. The importance of such agreement would be the firm prediction of inflationary cosmology through consistent relation between scalar and tensor spectra. The Planck scale physics is conjectured to modify this relation, considerably, and leads to running in the spectral index. For modes which are larger than the current horizon, the tensor spectral index is likely positive. 

\subsection{Lorentz invariance violation}
\label{sec:liv}

The combination of HUP and finiteness of the speed of light $c$ would lead to creation and annihilation processes, especially when studying the Compton wavelength of the particle of interest \cite{garay1,Scardigli}. Another consequence of the space-time foamy structure at small scales is the Lorentz invariance violation (LIV), which is originated in the proposal that the Lorentz invariance (LI) may represent an approximate symmetry of the Nature (dates back to about four decades) \cite{LI1}. A self-consistent framework for analyzing possible violation of LI was suggested by Coleman and Glashow \cite{cg1,stecker1}. In gamma-ray bursts (GRB), the energy-dependent time-offsets are investigated in different energy bands assuming standard cosmological model \cite{jellis1}. A kind of weak indication for the redshift dependence of the time delays suggestive of LIV has been found. A comprehensive review on the main theoretical motivations and observational constraints on the Planck scale suppressed Lorentz invariance violation is given in Ref. \cite{revw} and the references therein. Recently, the Planck scale itself turns to be accessible in quantum optics \cite{nature2012}.

The modified dispersion relationship likely leads to further predictions which obviously have feasibilities in experiments, such as an energy dependent speed of light. The gamma-ray observations \cite{dr7} might  imply that the speed of light was faster in the very early Universe, when the average energy was comparable to the Planck scale \cite{dr10}. As pointed out by Moffat \cite{dr8}, and Albrecht and Magueijo \cite{dr9}, such an effect could provide an alternative solution to the horizon problem and other problems addressed by the inflation. Such modified dispersion relations may also lead to corrections in the predictions of inflationary cosmology, observable in future high precision measurements of the CMB spectrum. Finally, a modified dispersion relation may lead to an explanation of the dark energy in terms of  very high momentum and low-energy quanta, as pointed out by Mersini {\it et al.} \cite{dr12}.

The linear GUP approach assumes that the momentum of a particle with mass $M$ having distant origin and an energy scale which is comparable to the Planck scale would be a subject of a tiny modification \cite{advplb,Das:2010zf,afa2} so that the comoving momenta are $p_{\nu} = p_{\nu} \left(1-\alpha\, p_0 + 2\, \alpha^2\, p_0^2\right)$ and $p_{\nu}^2 = p_{\nu}^2 \left(1-2\, \alpha p_0 + 10\, \alpha^2\, p_0^2\right)$,
where $p_{0}$ is the momentum at low energy \cite{Tawfik:2012hz}. The parameter $\alpha=\alpha_0/(c\, M_{pl}) =\alpha_0 l_{pl}/\hbar$ \cite{advplb,Das:2010zf,afa2}, where $\alpha_0$ is dimensionless parameter of order one. Then in comoving frame, the dispersion relation  reads
\bea \label{eq:disp}
E_{\nu}^2 = p_{\nu}^2\, c^2 \left(1-2\,\alpha\,p_0\right) + M_{\nu}^2\, c^4.
\eea
When  a linear dependence of $p$ on $\alpha$ is taken into consideration and the higher orders of $\alpha$ are ignored, then the Hamiltonian is
\bea \label{eq:Hh1}
\textbf{H} &=& \left(p_{\nu}^2\, c^2 - 2\, \alpha\, p_{\nu}^3 \, c^2 + M_{\nu}^2\, c^4 \right)^{1/2}.
\eea
There are several experimental and theoretical developments \cite{dr1,dr6} showing threshold anomalies in ultra high-energy cosmic ray protons \cite{dr3,dr4} and possible TeV photons \cite{dr45A,dr45B}.

\subsubsection{Comoving velocity and time of arrival}

The derivative of Eq. (\ref{eq:Hh1}) with respect to the momentum results in a comoving time-dependent velocity, i.e.  Hamilton equation
\bea
v(t) &=& \frac{1}{a(t)} \left(P_{\nu{_0}}^2\, c^2 -3 \alpha\, P_{\nu{_0}}^2 \, c^2\right)\; \left(P_{\nu{_0}}^2\, c^2 -2\,\alpha\,P_0^3 + M_{\nu}^2\, c^4\right)^{-1/2}, \\
 &=& \frac{c}{a(t)} \left(1-2\alpha p_0 -\frac{M_{\nu}^2 c^2}{2 p_{\nu}^2} + 
\alpha p_0 \left[\frac{ M_{\nu}^2 c^2}{p_{\nu}^2} -
 \frac{ M_{\nu}^2 c^4}{p_{\nu}^2 c^2 + M_{\nu}^2 c^4} +
 \frac{ M_{\nu}^2 c^4}{p_{\nu}^2 c^2 + M_{\nu}^2 c^4} \frac{M_{\nu}^2 c^2}{2 p_{\nu}^2}\right]
\right). \hspace*{1cm} \label{eq:vt}
\eea 
The comoving momentum is related to the physical one through the relation \hbox{$p_{\nu}=p_{\nu_{0}}(t_0)/a(t)$} and the scale factor $a$ is related to the redshift $z$,
\bea \label{eq:at1}
a(z) &=& \frac{1}{1+z}.
\eea  
In the relativistic limit, $p\gg M$, the fourth and fifth terms in Eq. (\ref{eq:vt}) simply cancel each other. Then  \cite{Tawfik:2012hz}
\bea
v(z) &=& c\,(1+z)\left[1-2\, \alpha\, (1+z)\, p_{\nu_0}  - \frac{M_{\nu}^2\, c^2}{2 (1+z)^2 p_{\nu_0}^2} +
\alpha\, \frac{ M_{\nu}^4 c^4}{2\, (1+z)^3\, p_{\nu_0}^3} 
\right], \label{eq:vz}
\eea
in which $p_0$ is treated as a comoving momentum.

The comoving redshift-dependent distance travelled by the particle of interest is defined as
\bea \label{eq:rz}
r(z) &=& \int_0^z \frac{v(z)}{(1+z)\, H(z)} dz,
\eea 
where $H(z)$ is the Hubble parameter depending on $z$. From Eqs. (\ref{eq:vz}) and (\ref{eq:rz}), the time of flight is given as \cite{Tawfik:2012hz}
\bea \label{eq:tnu}
t_{\nu} &=& \int_0^z \left[1-2\,\alpha\, (1+z)\, p_{\nu_0}  - \frac{M_{\nu}^2\, c^2}{2 (1+z)^2 p_{\nu_0}^2} +
\alpha\, \frac{ M_{\nu}^4 c^4}{2\, (1+z)^3\, p_{\nu_0}^3} \right] \frac{d z}{H(z)},
\eea
which counts for the well-known time of flight of a prompt low-energetic photon (first term), i.e. the time of flight is invariant in Lorentz symmetry. Furthermore, it is apparent that Eq. (\ref{eq:tnu}) contains a time of flight delay 
\bea \label{eq:deltat1}
\Delta t_{\nu} &=&  \int_0^z \left[2\alpha \left( (1+z)\, p_{\nu_0} - \frac{ M_{\nu}^4 c^4}{4\, (1+z)^3\, p_{\nu_0}^3} \right) + \frac{M_{\nu}^2\, c^2}{2\, (1+z)^2 p_{\nu_0}^2}  \right] \frac{d z}{H(z)}.
\eea
\begin{itemize}
\item The first and second terms are due to LIV effects stemming from GUP (both have $\alpha$ parameter). 
\item The third term gives the effects of the particle mass on the time of flight delay. Furthermore, the second term alone seems to contain a mixed effects from LIV (GUP) and rest mass.
\end{itemize}

Having observational results and/or reliable theoretical models for the redshift-dependence of the Hubble parameter $H$ is necessary to determine $\Delta t_{\nu}$, Eq. (\ref{eq:deltat1}),  $H(z) = 1/a(z)\, \left((d a(z)/d z) \; (d z/d t)\right) = - 1/(1+z)\, d z/d t$, which can be deduced from Eq. (\ref{eq:at1}). In general, the expansion rate of the Universe varies with the cosmological time \cite{Tawfik:2009mk,Tawfik:2009nh,Tawfik:2010bm,Tawfik:2010ht,Tawfik:2010pm,Tawfik:2011mw,Tawfik:2011sh,Tawfik:2011gh}. It depends on the background matter/radiation and its dynamics \cite{Tawfik:2011sh}. The cosmological constant can among others stand for the dark matter content and likely affects the temporal evolution of $H$ \cite{Tawfik:2011mw}. Fortunately, the redshift $z$ itself can be measured in the spectroscopic redshifts of galaxies with certain uncertainties ($\sigma_z\leq 0.001$). Based on this, a differential measurement of time at a given redshift interval automatically provides a direct measurement for $H(z)$ \cite{hz1,hz2,hz3}, which can be used to deduce constraints on the essential cosmological parameters \cite{const}, for instance, the measurements of the expansion rate and their constrains in evaluating the integrals given in Eq. (\ref{eq:deltat1}) are implemented \cite{Tawfik:2012hz}.

\begin{figure}[htb]
\centering{
\includegraphics[angle=-90,width=7.cm]{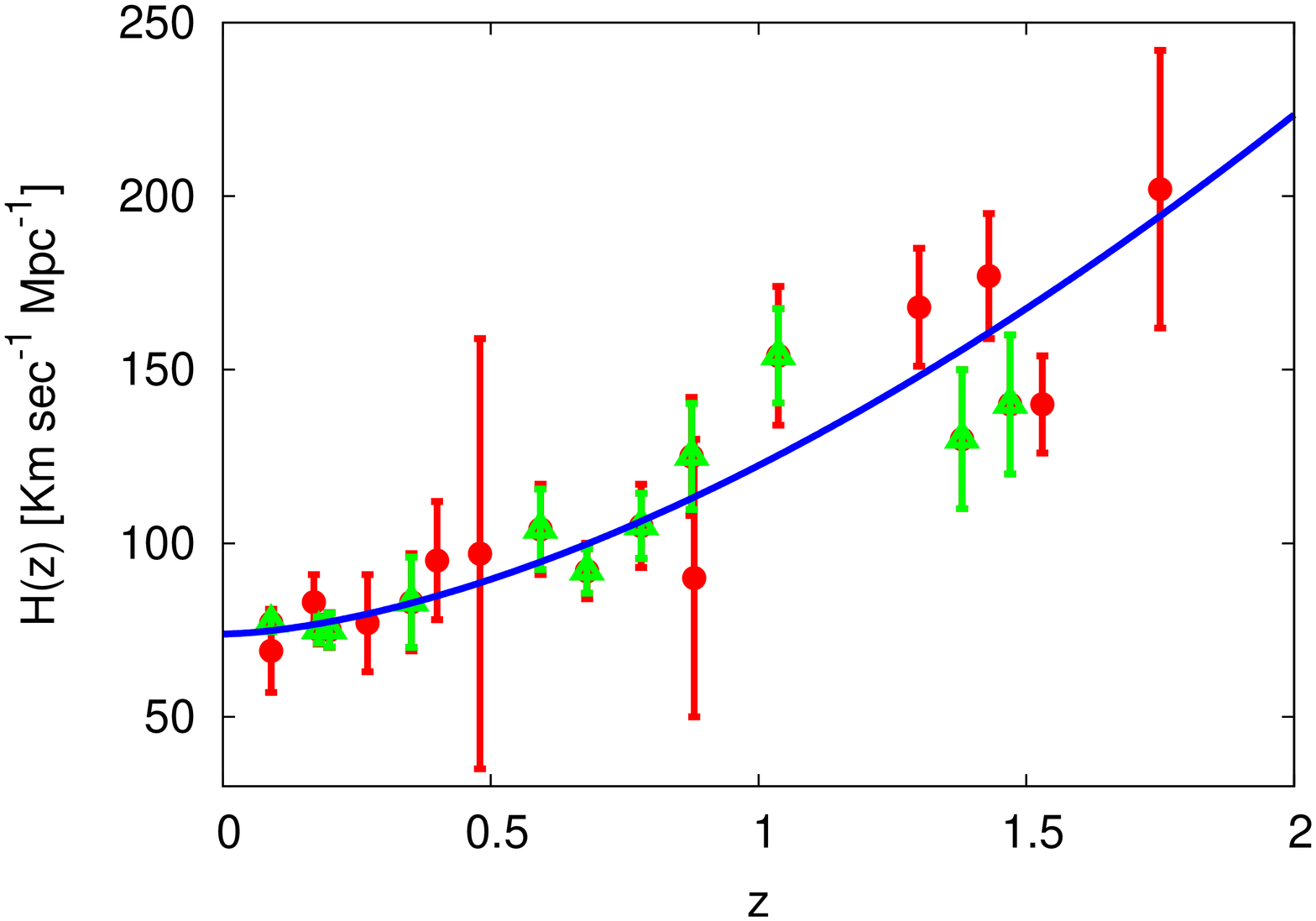}
\includegraphics[angle=-90,width=7.cm]{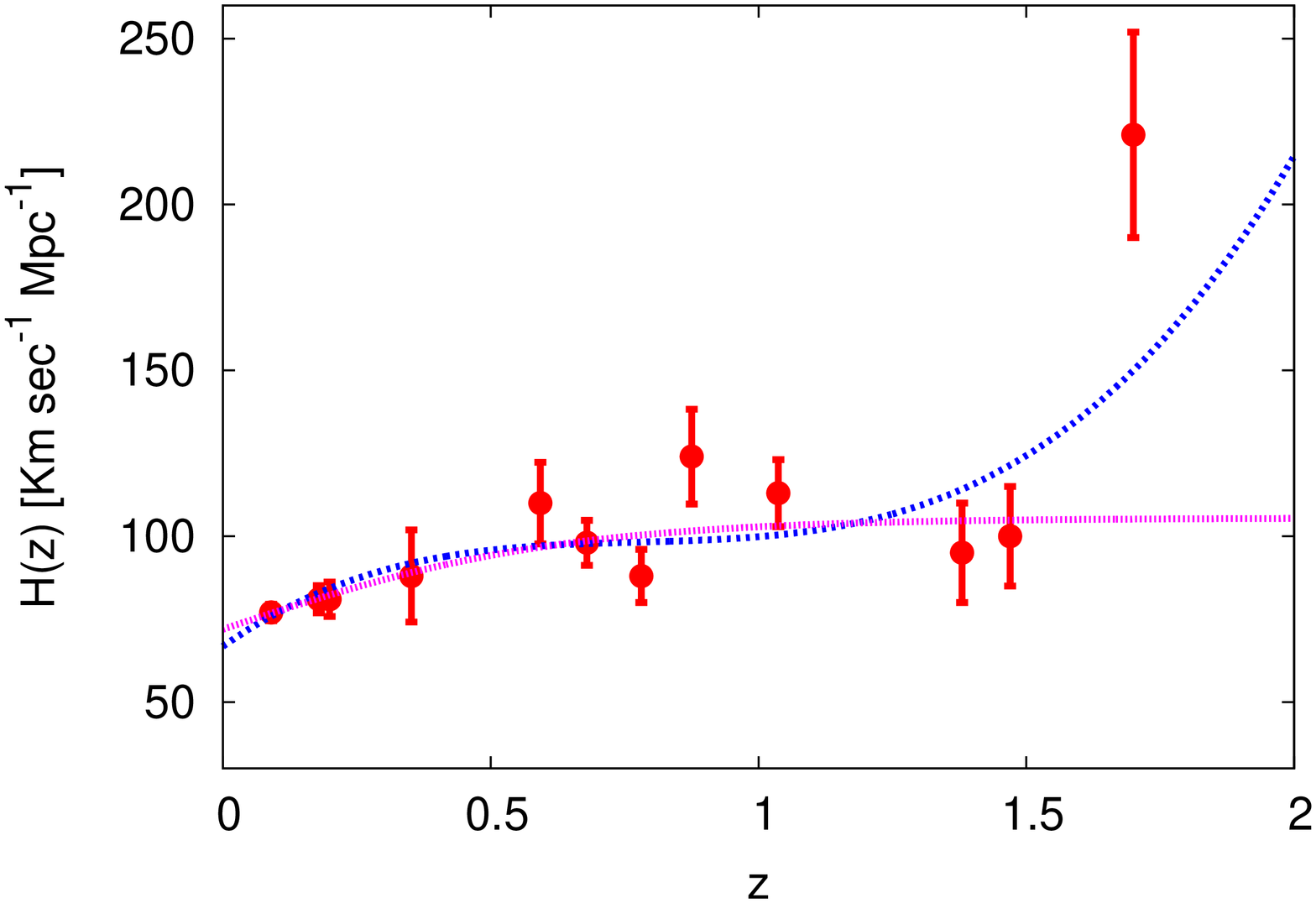}
\caption{Left-hand panel: the Hubble parameter $H$  calculated in BC03 model (open triangle) and in combination with CMB data, that gives constrains of the possible deviations from the standard (minimal) flat $\Lambda$CDM model (solid circles), is given in dependence on the redshift $z$. The results from MS model are drawn in the right-hand panel. The curves represent the fitting parameters (see text for details). The graphs taken from Ref. \cite{Tawfik:2012hz}. 
\label{fig:Hofz1} 
}}
\end{figure}

 For the observational results which are obtained from BC03 model \cite{bc03} and by using a combination with CMB data and setting constrains on possible deviations from the standard (minimal) flat $\Lambda$CDM model \cite{hz3}, the expression \cite{Tawfik:2012hz}
\bea \label{eq:measr_hz1}
H(z)=\beta_1 + \gamma_1\, z + \delta_1\, z^2,
\eea
where $\beta_1=72.68\pm3.03$, $\gamma_1=19.14\pm 5.4$ and $\delta_1=29.71\pm 6.44$, fits well with the observational measurements. The solid curve in left-hand panel of Fig. \ref{fig:Hofz1} represents the results from this expression. For MS model \cite{mastro}, two expressions are suggested \cite{Tawfik:2012hz}:
\bea 
H(z)&=&\beta_2 + \gamma_2\, z + \delta_2\, z^2 + \epsilon_2\, z^3, \label{eq:measr_hz2}\\
H(z)&=& \beta_3+\gamma_3\, \text{tanh}(\delta_3\, z), \label{eq:measr_hz3}
\eea
where $\beta_2=66.78\pm8.19$, $\gamma_2=113.27\pm7.5$, $\delta_2=-140.72\pm 12.6$, $\epsilon_2=60.61\pm5.48$, $\beta_3=71.94\pm4.35$, $\gamma_3=33.51\pm 7.94$ and $\delta_3=1.6\pm0.1$. The results of Eq. (\ref{eq:measr_hz2}) are given by the dashed curve in the right-hand panel of Fig. \ref{fig:Hofz1}. Eq. (\ref{eq:measr_hz3}) is drawn as the dotted curve, where the largest point is excluded while remaining points build up the ensemble used in the fitting \cite{Tawfik:2012hz}.
It is obvious that the implementation of Eq. (\ref{eq:measr_hz2}), which is a rational function, in Eq. (\ref{eq:deltat1}) results in a non-analytic integral. On the other hand, implementing Eq. (\ref{eq:measr_hz3}) in Eq. (\ref{eq:deltat1}) makes the second and third integrals non-solvable, while the first term is.

\subsubsection{Conclusions}

With varying the redshift, the relative change in the speed of massive muon neutrino and its time of flight delays is calculated. The redshift depends on the temporal evolution of $H$, which can be estimated from a large sample of early-type galaxies extracted from several spectroscopic surveys spanning over $\sim 8\times 10^9$ years of cosmic lookback, most massive, red elliptical galaxies, passively evolving and without signature of ongoing star formation are picked up and used as standard cosmic chronometers giving a cosmic time directly probe for $H(z)$. The measurements according to BC03 model and in combination with CMB data constraining the possible deviations from the standard (minimal) flat $\Lambda$CDM model are used to estimate the $z$-dependence of the Hubble parameter. The measurements based on MS model are used to show that the results are model-dependent.

\subsection{Black hole thermodynamics}
\label{sec:bhs}

The finding that black holes should have well-defined entropy and temperature represented one of the greatest achievements in recent astrophysics \cite{entr3,entr1}. In statistical physics and thermodynamics, the thermal evolution of entropy relates the number of thermal macrostates to that of microstates of the system of interest in thermal medium. In GR, the BH entropy is a pure geometric quantity so that when comparing BH with a thermodynamic system, we find an important difference. Whether BH has interior  degrees of freedom corresponding to its entropy, the Bekenstein-Hawking entropy delivered an answer to this and characterized the statistical meaning \cite{entr3,entr1}.  Counting the microstates was proposed by Medved and Vagenas \cite{refff2new}, that this presumably lies within the framework of QG. For example, the String theory \cite{hist4} and the loop quantum gravity \cite{hist5} succeeded in presenting an statistical explanation formulated in an entropy-area law. The proportionality relating BH entropy with area was derived from classical thermodynamics, as well \cite{sAclass}.

\subsubsection{Number of quantum states, entropy and free energy}

In brick wall model, the entropy can be calculated as follows.
\bea
S_0 &=& \beta^2 \left.\frac{\partial F_0}{\partial \beta}\right|_{\beta=\beta_H}
= \frac{\beta^2}{\pi} \int_{r_++\epsilon}^{L} dr\; \frac{1}{\sqrt{f}} \int_{m \sqrt{f}}^{\infty} d \omega \left.\frac{\omega e^{\beta \omega} \left(\frac{\omega^2}{f} - m^2\right)^{1/2}}{\left(e^{\beta \omega}-1\right)^{2}} \right|_{\beta=\beta_H},
\eea
where $\beta$ is the inverse temperature, $F_0$ is the free energy and $L$ and $\epsilon$ are infrared and ultraviolet regulators, respectively. $\beta_H$ is the inverse Hawking temperature. In a zero-temperature quantum mechanical system around the black hole, the entropy 
$S_0^{ext} \approx \ln (1/(2\, \Lambda\, \epsilon))$, which can be interpreted as the physical limit that $\Lambda$ should be less than $1/(2\epsilon)$.

In natural units, the modified uncertainty relation
\bea
\Delta x\, \Delta p \geq \frac{\hbar}{2} \left[1-2\, \alpha\, \langle p\rangle + 4\, \alpha^2\, \langle p^2\rangle \right],
\eea
leads to a modification in the volume of phase cell in $(1+1)$-dimensions from $2\, \pi$ to 
$ 2\pi\, \left(1 - 2\, \alpha\; p + 4\, \alpha^2\; p^2\right)$.
The number of quantum states with energy less than $\epsilon$ is given as \cite{reffff1}
\bea
n_0(\omega) &=& \frac{1}{2 \pi} \int d r \; d p_r
= \frac{1}{\pi} \int^L_{r_+\, + \epsilon} dr \, \frac{1}{\sqrt{f}}\; \left(\frac{\omega^2}{f}-m^2\right)^{1/2}, \label{eq:n11}
\eea
where $m$ in the mass of the scalar field and $\omega$ is a parameter of the substitution of Klein-Gordon equation. The expression equation (\ref{eq:n11}) will be changed to
\bea
n_I(\omega) &=& \frac{1}{2 \pi} \int dr \; d p_r\; \frac{1}{1-2\, \alpha\, p + 4\, \alpha^2\,  p^2} \label{eq:nz1} \nonumber \\
&=& \frac{1}{2 \pi} \int dr \; \frac{1}{\sqrt{f}} \frac{\left(\frac{\omega^2}{f}-m^2\right)^{1/2}}{1-2\, \alpha\, \left(\frac{\omega^2}{f}-m^2\right)^{1/2} + 4\, \alpha^2\,  \left(\frac{\omega^2}{f}-m^2\right)}, \label{eq:nz2}
\eea
where $r$ and $f$ are estimated as follows. In Schwarzschild  gauge, the metric and field tensors, respectively,  read
\bea
d s^2 &=& - f(r) d t^2 + \frac{1}{f(r)} d r^2, \label{eq:ds2}\\
F_{rt} &=& F_{rt} (r). \label{eq:frt}
\eea
The function $f(r)$ in the static solution is defined as $f(r) = 1- (M/\Lambda) \exp(-2 \Lambda r) + (Q^2/4 \Lambda^2) \exp(-4 \Lambda r)$, 
where $M$ is the mass of black hole and $Q$ gives its charge. The outer event horizon has the radius $r_{+} = [1/(2 \Lambda)]\, \ln\left[M/(2 \Lambda) + \sqrt{\left(M/(2 \Lambda)\right)^2 - \left(Q/(2 \Lambda)\right)^2} \right]$. 
In light of this, its derivative vanishes and the Klein-Gordon equation is reduced to
\bea
\frac{d^2 R}{d r^2} +\frac{1}{f} \frac{d f}{d r} \frac{d R}{d r} + \frac{1}{f} \left(\frac{\omega^2}{f} - m^2\right) R &=& 0,
\eea
where $\phi(r) = \exp\left(-i \omega t\right) R(r)$. Using WKB approximation, then $R \sim \exp\left(i S(r)\right)$, $p_r^2 = \frac{1}{f} \left(\frac{\omega^2}{f} - m^2\right)$, and $p_r=d S/d r$ and $p^2 = \frac{\omega^2}{f} - m^2$.

The free energy at Hawking temperature can be derived from  Eq. (\ref{eq:nz2})  \cite{reffff1} $F_0 = - (1/\pi) \int^L_{r_++\epsilon}\; dr \,(1/\sqrt{f}) \, \int_{m \sqrt{f}}^{\infty} \, \left(\omega^2/f-m^2\right)^{1/2}/(\exp(\beta \omega)-1)\,  d\omega$,
which turns is a subject of change $F_I = -\int_{m \sqrt{f}}^{\infty} d \omega\, n_I(\omega)/(e^{\beta \omega}-1)$,
\bea \label{eq:fe2}
F_I &=&     -\frac{1}{\pi} \int dr \frac{1}{\sqrt{f}} \int_{m \sqrt{f}}^{\infty}  \frac{\left(\frac{\omega^2}{f}-m^2\right)^{1/2}}{\left(e^{\beta \omega}-1\right) \left[1-2 \alpha \left(\frac{\omega^2}{f}-m^2\right)^{1/2} + 4 \alpha^2 \left(\frac{\omega^2}{f}-m^2\right)\right]}\, d \omega.
\hspace*{10mm} 
\eea

\subsubsection{Black hole entropy and linear GUP approach}

The entropy can be deduced near the event horizon, i.e. within the range $(r_+, r_++\epsilon)$, $f\rightarrow 0$, from Eq. (\ref{eq:fe2})  
\bea
S_0 &=& \frac{\beta^2}{\pi} \int^{L}_{r_++\epsilon} d r \frac{1}{\sqrt{f}} \int^{\infty}_{m \sqrt{f}} d\omega \left.\frac{\omega e^{\beta \omega} \left(\frac{\omega^2}{f}-m^2\right)^{1/2}}{\left(e^{\beta \omega}-1\right)^2}\right|_{\beta=\beta_H}. \label{eq:S01}
\eea
Again, due to GUP, it shall be subject of change
\bea
S_I &=& \frac{\beta^2}{\pi} \int dr \frac{1}{\sqrt{f}} \int_{m \sqrt{f}}^{\infty} \frac{\omega \left(\frac{\omega^2}{f}-m^2\right)^{1/2} e^{\beta \omega}}{e^{2\beta \omega - 2} \left[1-2 \alpha \left(\frac{\omega^2}{f}-m^2\right)^{1/2} + 4 \alpha^2 \left(\frac{\omega^2}{f}-m^2\right)\right]} d\omega \nonumber \\
&=& \frac{1}{\pi} \int_{r_+}^{r_++\epsilon} dr \frac{1}{\sqrt{f}} \int_{0}^{\infty} \frac{f^{-1/2}\, \beta^{-1}\, x^2}{(1-e^{-x})(e^x-1) \left[1-2 \alpha \frac{x}{\beta \sqrt{f}} + 4 \alpha^2 \frac{x^2}{\beta^2 f}\right]} d x,
\eea
where $x=\beta\, \omega$. We note that as $f\rightarrow 0$, then $\omega^2/f$ is the dominant term in the bracket containing $\omega^2/f - m^2$. We are interested in the thermodynamic contributions of  just vicinity near horizon $r_+,r_++\epsilon$, which corresponds to a proper distance of the order of the minimal length. The latter can be related to the GUP parameter $\alpha$. So we have from Eq. (\ref{eq:ds2}) $\alpha = \int_{r_+}^{r_++\epsilon} \frac{d r}{\sqrt{f(r)}}$,
which apparently sets the lower bound to the GUP parameter $\alpha$. Then, the entropy reads
\bea
S_I &=& \frac{1}{\pi\; \alpha}\, \int_{r_+}^{r_++\epsilon} \frac{d r}{\sqrt{f(r)}} \int_0^{\infty} \, \frac{a^2\, X^2}{\left(e^{\frac{a\, X}{2}} - e^{-\frac{a\, X}{2}}\right)^2\; \left(1-2\, X + 4\, X^2\right)}\,d X,
\eea
where $x = (\beta/\alpha)\, \sqrt{f}\, X = a\, X$.
Then
\bea
S_I &=& \frac{1}{\pi}\; \Sigma_I =
 \frac{1}{\pi}\; \int_0^{\infty}\, \frac{a^2\, X^2}{\left(e^{\frac{a\, X}{2}} - e^{-\frac{a\, X}{2}}\right)^2\; \left(1-2\, X + 4\, X^2\right)}\, d\, X.
\eea
We note  that as $r\rightarrow r_+$, $f\rightarrow 0$, then $a\rightarrow 0$ and $\lim_{a\rightarrow 0}  a^2\, X^2/\left(e^{a \, X/2}-e^{-a \, X/2}\right)^2=1$.
Therefore, 
\bea
\Sigma_I &=& \int_0^{\infty} \frac{d\, X}{1-2 \, X + 4 \, X^2} 
= \frac{2\, \pi}{3 \sqrt{3}},
\eea
and
\bea
S_I &=& \frac{1}{\pi} \; \Sigma_I = \frac{2}{3 \, \sqrt{3}}.
\eea
It is obvious that $S_I$ is finite and does not depend on any other parameter. We notice that in contrast to the case of brick wall method, there is no divergence within the just vicinity near the horizon due to the effect of the generalized uncertainty relation on the quantum states.

\subsubsection{Linear GUP approach and entropy of Schwarzshild black hole}

In natural units,  the line element in Schwarzschild black hole is given as
\bea
d s^2 &=& -\left(1-2\, \frac{M}{r}\right) d\, t^2 + \left(1-2\, \frac{M}{r}\right)^{-1}\, d\, r^2 + r^2\, d\,\Omega_2^2.
\eea
Then, the Hawking radiation temperature $T$, the horizon area $A$ and the entropy $S$, respectively, read 
\bea
T &=& \frac{1}{4 \pi r_H} = \frac{1}{8 \pi M},\\
A &=& 4 \pi r_H^2 = 16 \pi M^2,\\ 
S &=& \pi r_H^2 = 4 \pi M^2,
\eea
where $r_H = 2 M$ is the location of the black-hole horizon. The increase (decrease) in the horizon area due to absorbing (radiating) a particle of energy $d\, M$ can be expressed as $d\, A = 8\, \pi\, r_H\, d r_H = 32\, \pi\, M\, d M$. 

This particle is conjectured to satisfy Heisenberg's uncertainty relation $\Delta\, x_i \, \Delta\, p_j \geq \delta_{ij}$. But according to the linear GUP approach, the area and entropy, respectively,  can be re-written as
\bea
A_{GUP} &=& A - 4 \alpha\, \sqrt{\pi}\, \sqrt{A}+ 8\, \pi\, \alpha^2\, \ln\left(\sqrt{\frac{A}{\pi}}+2\,\alpha\right),\\
S_{GUP} &=& S - 2\, \alpha\, \sqrt{\pi}\, \sqrt{S} + \alpha^2\, \pi\, \ln S + C, \label{eq:sgupmschwrz}
\eea 
where $\alpha\ll \sqrt{A/\pi}$ and $C$ is an arbitrary constant. It is worthwhile to note that the coefficient of $\ln S$ is also positive, while the entropy gets an additional term,  $2\, \alpha\, \sqrt{\pi}\, \sqrt{S}$.

\subsubsection{Linear GUP approach and energy density of Schwarzshild black hole}

As given in sections \ref{sec:2GUPFRIED} and \ref{sec:1GUPFRIED}, the Friedmann equation (first law of thermodynamics) is $\left(\dot{H} - k/a^2\right)\, S^{\prime}_{GUP} = - 4\, \pi\, G\; (\rho + p)$, where the energy density is $\rho = - 3/(8\, G) \int S^{\prime}_{GUP}(A)\; \left(A/4\right)^{-2}\; dA$. By using Eq. (\ref{eq:sgupmschwrz}), 
\bea
\left(\dot{H} - \frac{k}{a^2}\right)\, \left[1-2\, \alpha \left(\frac{\pi}{A}\right)^{1/2} +4\, \alpha^2 \; \left(\frac{\pi}{A}\right) \right] &=& - 16\, \pi\, G\; (\rho + p).
\eea
Then, the modified energy density reads
\bea 
\rho_{GUP} &=& \frac{3}{8\, \pi\, G} \left[\left(\frac{\pi}{A}\right) - \frac{4}{3}\alpha \left(\frac{\pi}{A}\right)^{3/2} - 2 \alpha^2 \left(\frac{\pi}{A}\right)^2\right] 
 = \rho \left[1 - \frac{4}{3} \alpha \left(\frac{2}{3} \pi \rho\right)^{1/2} + \frac{4}{3} \pi \, \alpha^2\, \rho \right].
\eea

\subsubsection{Conclusions}

It was shown that the quantum correction of the geometric entropy of charged black hole has one great advantage. One can avoid being biased in favor of a certain theory of QG. For example, the correction to Bekenstein-Hawking entropy, which relates the entropy to the cross-sectional area of the BH horizon, includes a series of terms, where the coefficient of the leading-order correction, the logarithmic term, is suggested as a discriminator of prospective fundamental theories for QG. It is essential to suggest a method that fixes this, but it should not depend on the utilized models for QG, for instance, holographic principle. 


When comparing black hole entropy with the one that counts for the microstates $\Omega$, one can simply relate $A/4$ to $\ln \Omega$. This is valid as long as the gravity is sufficiently strong so that the horizon radius is much larger than the Compton wavelength. In order to apply the GUP approach, we start with the modified momentum and statistically derive expressions for area and entropy. Then, we apply the holographic principle.  Based on the linear GUP approach, the black hole thermodynamics and entropy get substantial corrections. We found that the logarithmic divergence in the entropy-area relation turns to be positive. Furthermore we find that $S$ gets an additional terms, such as $2\, \alpha\, \sqrt{\pi}\, \sqrt{S}$.

\subsection{Compact stellar objects}

By studying the ground state properties of a Fermi gas composed of $N$ ultra-relativistic electrons, it is obvious that  the ground state energy $\epsilon=c\, p$, i.e. $T\gg m$. At low $T$, the vacuum effect of fermions are negligible, i.e. the total particle number is conserved. For an isolated macroscopic interstellar object consisting of $N$ non-interacting  and ultra-relativistic particles, the background of the particles motion is assumed to be flat. Due to GUP, the modified number of particle of Fermi gas can be given as \cite{Ali:2013ii}
\bea \label{eq:pFermi1}
N(p) &=& \frac{8 \pi}{h^3}V \int_0^{p_F} \frac{p^2 dp }{(1-\alpha p)^4},
\eea
where $p_F$ is the Fermi momentum. Therefore, Eq. (\ref{eq:pFermi1}) can be re-written in terms of  Fermi energy $\epsilon_F$, $N(\epsilon_F) =(8\, \pi/3) (h\, c)^{-3}\, V\, \epsilon_F^3 \left(1-\frac{\alpha}{c} \epsilon_F\right)^{-3}$. By introducing $\kappa= \epsilon_F/\epsilon_H$, which is equivalent to $\alpha \epsilon_F/c$, then
\bea \label{eq:nkappa1}
N(\kappa) &=& \frac{8\, \pi}{(h\, c)^3}\, V\, \epsilon_H^3\, f(\kappa),
\eea
where $\epsilon_H=c/\alpha$ being Hagedorn energy and therefore 
\bea \label{eq:fkappa1}
f(\kappa) &=& \frac{1}{3} \frac{\kappa^3}{(1-\kappa)^3}.
\eea

The ground state energy can be estimated from
\bea \label{eq:tU}
U_0(\epsilon) &=& \frac{8 \pi }{(h c)^3} V \int_0^{\epsilon_F} \frac{\epsilon^3 d\epsilon}{\left(1-\frac{\alpha}{c} \epsilon\right)^4}.
\eea
Then, in terms of $\kappa$, the ground state energy and pressure, respectively, \cite{Ali:2013ii}
\bea
U_0(\kappa) &=& \frac{8 \pi}{(h c)^3}\,  V\, \epsilon_H^4\, g(\kappa), \label{eq:U1}\\
P(\kappa) &=& \frac{N}{V} \epsilon_F - \frac{U_0}{V} =\frac{8 \pi}{(h\, c)^3} \, \epsilon_H^4\, h(k),
\label{eq:P1}
\eea
where
\bea
g(\kappa)&=&\ln(1-\kappa) + \frac{\kappa}{(1-\kappa)^3} - \frac{15}{6} \frac{\kappa^2}{(1-\kappa)^3} + \frac{11}{6} \frac{\kappa^3}{(1-\kappa)^3}, \label{gk} \\ 
h(\kappa)&=& \frac{1}{3} \frac{\kappa^4}{(1-\kappa)^3} -  \left[\ln(1-\kappa) + \frac{\kappa}{(1-\kappa)^3} - \frac{15}{6} \frac{\kappa^2}{(1-\kappa)^3} + \frac{11}{6} \frac{\kappa^3}{(1-\kappa)^3}\right].
\label{hk}
\eea

\begin{figure}[htb!]
\centering{
\includegraphics[width=12.cm,angle=0]{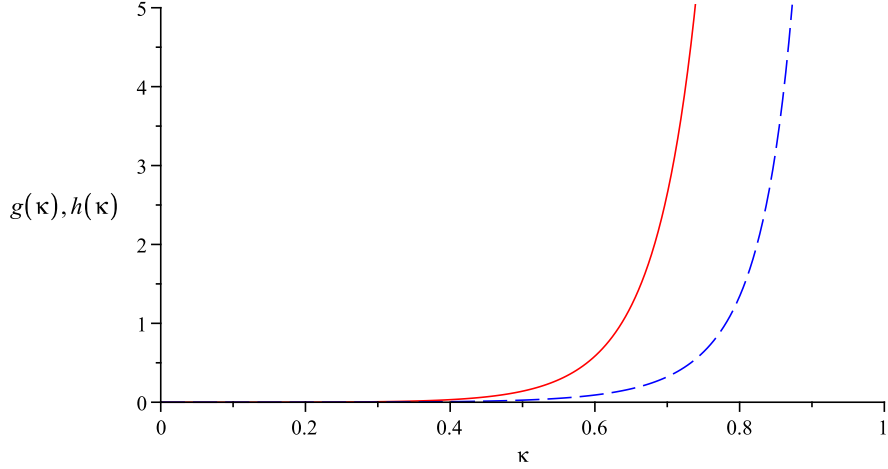}
\caption{The dimensionless quantities $g$ (solid curve), Eq. (\ref{gk}), and $h$ (dashed curve), Eq. (\ref{hk}), are presented as function of $\kappa$. The graph taken from Ref. \cite{Ali:2013ii}.  \label{cs:afig1} 
}}
\end{figure}

In Fig. \ref{cs:afig1}, it is obvious that both quantities diverge at $\kappa\rightarrow 1$. Also, $g(\kappa)$ seems to diverge much faster than $h(\kappa)$. So far, we conclude that the validity of this approach is strictly limited to the Fermi energy. By a maximum energy bound $c/ \alpha$, the given approach is bounded from above, which makes it  completely consistent with the predicted maximum measurable momentum $1/\alpha$ introduced in Ref. \cite{advplb,Das:2010zf}.

The Hagedorn energy (or equivalently temperature), which is defined as \hbox{$\epsilon_H = M_p c^2/\alpha_0$}  is a scale to set the limit of  applying the GUP approach. Accordingly, Eqs. (\ref{eq:U1}) and (\ref{eq:P1}) can be re-written  as \cite{Ali:2013ii}
\bea
U_0(\kappa) &=& \frac{8 \pi V}{(h c)^3} \epsilon_H^4 \left[\frac{\kappa^4}{4}+ \frac{4 \kappa^5}{5}\right], \label{eq:U2}\\
P(\kappa) &=& \frac{8 \pi}{(h c)^3} \epsilon_H^4 \left[ \frac{\kappa^4}{12}+\frac{\kappa^5}{5}\right]. \label{eq:P2}
\eea
In deriving these two expressions, the condition $\kappa\ll 1$ has been  implemented. Due to GUP. the corrections to $U_0(\kappa)$ and $P(\kappa)$ would be given in terms of $\delta$ \cite{Ali:2013ii}
\bea
\frac{U_0(\kappa)}{V} &=& \frac{3^{4/3}}{4} \frac{h c}{(8 \pi)^{\frac{1}{3}}} \left(\frac{N}{V}\right)^{4/3} \left(1+\frac{16}{5} (3 \pi^2)^{1/3}\delta\right), \label{eq:U3}\\
P(\kappa)  &=&  \frac{3^{1/3}}{4} \frac{h c}{(8 \pi)^{\frac{1}{3}}} \left(\frac{N}{V}\right)^{4/3} \left(1+ \frac{12}{5} (3 \pi^2)^{1/3}\delta\right). \label{eq:P3}
\eea
$\kappa$ can be given in terms of number density $N/V$,
\bea \label{eq:kappaDE}
\kappa^3 \left(\frac{1}{3} + \kappa\right) &=& \frac{(h c)^3}{8 \pi}\, \frac{1}{\epsilon_H^3}\, \frac{N}{V}.
\eea
Thus, we can relate $\kappa$ with $\delta$, $\kappa= (3 \pi^2)^{\frac{1}{3}}\, \delta\, \left[1 - (3 \pi^2)^{\frac{1}{3}}\, \delta\right] + {\cal O}(\delta^3)$. 

Few remarks are in order now. By modifying the dispersion relations, Eq. (\ref{eq:disp}), these results can be interpreted. In fact, a framework  based on GUP and modified measure of the momentum space, Eq. (\ref{eq:pFermi1}) was implemented, which is beyond the effective field theory  \cite{Maccione:2007yc}.

\subsubsection{Compact stars with non-relativistic cold nuclei}

The white dwarfs have two properties \cite{wd}:
\begin{itemize}
\item the electrons are described by relativistic dynamics and 
\item the electron gas is completely degenerate. 
\end{itemize}
The major contributions to the mass of white dwarfs are non-relativistic cold nuclei having mass $M=2 N/m_p$ \cite{Chandrasekhar:1931ih}. Therefore, the electron gas would be treated as a zero-temperature gas. Thus, $\epsilon_F=2  N c^2/\alpha_0 m_p$ indicating that $\kappa\ll 1$ and Eq. (\ref{eq:P2}) seems to reflect that the QG effects increase the degenerate pressure. Should this effects is confirmed, then the QG corrections to the mass of white dwarfs arise. The degeneracy pressure of electron is supposed to resist the gravitational collapse and keep the electron gas at a given density. At equilibrium, 
\bea \label{eq:pequil}
P_0(R) &=& \frac{\lambda}{4 \pi} G \left(\frac{M}{R^2}\right)^2,
\eea
where $R^3\equiv V$ and $\lambda$ is free parameter of the order of unity. Nevertheless, its value depends on how the matter is distributed inside the white dwarf. From Eqs. (\ref{eq:P3}) and (\ref{eq:pequil}) and by ignoring the constants (assign them to unity), the pressure can be expressed in terms of internal energy $\left(N/V\right)^{4/3} \left( 1 + \delta\right) = G M^2/R^4$. By substituting $M=2 N m_p$, the mass correction reads \cite{Ali:2013ii}
\bea
M = M_0 \left(1+ \left(\frac{N}{V}\right)^{\frac{1}{3}}\, \alpha\, \hbar\right),
\label{resulte}
\eea
where $M_0=\left(h c/G\right)^{\frac{3}{2}} (2 m_p)^{-2}$. For white dwarfs, the density number $N= 10^{36}$ and $M_0$ approximately approaches the Chandrasekhar limit (about $1.44\; M_{\odot}$).

Apparently, from Eq. (\ref{resulte}), we can conclude that the quantum gravity correction seems to be proportional to the density number of the star \cite{Ali:2013ii}.  For a white dwarf, in which the average distance $\bar{d}= 10^{-12}$, and the Fermi energy $\epsilon_F= 10^{5}$ eV  \cite{Ali:2013ii}.

At $\alpha_0\times l_p$ (an upper bound), $\alpha\leq 10^{-2}$ GeV$^{-1}$. Other bounds on $\alpha_0$ are discussed in Ref. \cite{afa2}, which have been derived by calculating the effect of QG with non-relativistic heavy meson systems like charmonium \cite{afa2}. The latter is a relevant example for the white dwarfs. Then, the QG correction to the mass of the white dwarf is given by  \cite{Ali:2013ii}
\bea \label{eq:mGUP}
M_{GUP}= M_0\left(1 + 10^{-5}\right).
\eea

Two remarks are now in order \cite{Ali:2013ii}.
\begin{itemize}
\item The mass correction seems to be more stringent than the one derived for compact stars with QG corrections \cite{wang}. To compare with, the correction given in Ref. \cite{wang} is $10^{-10}$. 
\item The QG corrections \cite{Ali:2013ii} is positive referring to resisting the collapse of the compact stars. It is obvious that this conclusion agrees with the result in \cite{wang}.
\end{itemize}

\subsubsection{Compact stars with ultra-relativistic nuclei}

There are other configurations for constituents of the compact stars, such as ultra-relativistic nuclei. In this case, the mass of the nuclei is compressed as $M=U_0/c^2$. These constituents are characterized by an ideal Fermi gas with mass $M=U_0/c^2$.  Again, the degeneracy pressure of electrons is assumed to resist the gravitational collapse. At equilibrium, the radius of the white dwarf  is given by \cite{Ali:2013ii}
\bea
R &=& \frac{\lambda}{8 \pi} R_S \frac{g(\kappa)}{h(\kappa)} =
\frac{\lambda}{8 \pi} R_S Q(\kappa), \label{eq:Rr}
\eea
where $Q(\kappa)=g(\kappa)/h(\kappa)$ and the parameter $\lambda$ approximately equals unity.  In the considered case, the Schwarzschild radius reads $R_S = 2 \,G\, M/c^2 = 2 \,G\, U_0/c^4$.

\begin{figure}[htb!]
\centering{
\includegraphics[width=12.cm,angle=0]{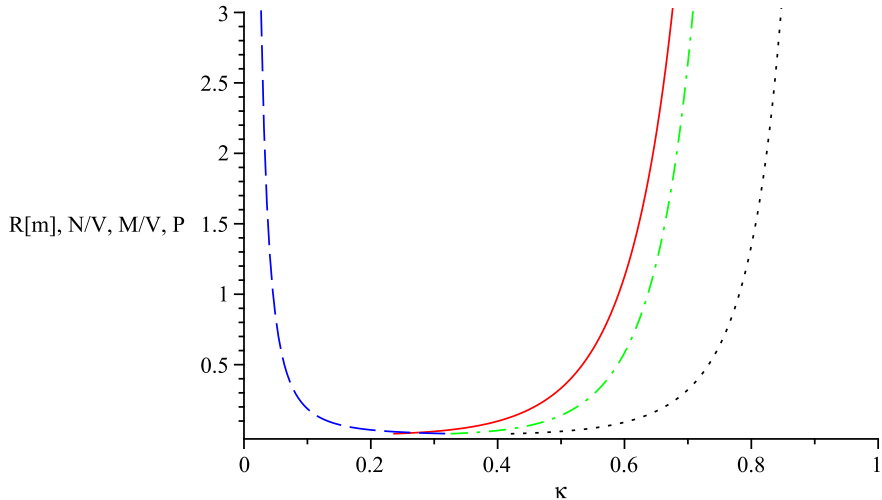}
\caption{The modified radius of white dwarf, Eq. (\ref{eq:Rr}), is given in dependence on $\kappa$ (dashed curve) at $\alpha_0 \approx 10^{17}$, i.e. $\alpha\approx10^{-2}~$GeV$^{-1}$ \cite{afa2}. At  vanishing $T$, the modified  normalized particle density in Fermi gas, $N(\kappa)/V$  is given as function of $\kappa$ (solid curve). The normalized mass density $M(\kappa)/V$ is given as dash-dotted curve. The normalized pressure is given as dotted curve. The graph taken from Ref. \cite{Ali:2013ii}. }
\label{cs:afig4}
}
\end{figure}

At t$\alpha\approx10^{-2}~$GeV$^{-1}$ and $\lambda\approx 1$, the results are presented in Fig. \ref{cs:afig4} \cite{afa2}. It is obvious that the radius approaches a minima as  $\kappa\rightarrow 1$.  We observe that the number density, the mass density and the pressure approach their minima as $\kappa\rightarrow 0$, but they reach their maximum values as $\kappa\rightarrow 1$.

Current observations seem to indicate that white dwarfs have smaller radii than expected \cite{Mathews:2006nq}. The behavior of $R$ vs $\kappa$ in Fig. \ref{cs:afig4} suggests that $R$ is decreasing as $\kappa\rightarrow 1$. This offers a possible explanation for the smaller radii observations. Similar analysis has been done in the context of DSR \cite{AmelinoCamelia:2009tv} and of the modified dispersion relations \cite{Camacho:2006qg,Gregg:2008jb}. 

\subsubsection{Conclusions}

Effects of the linear GUP approach on the thermodynamic properties of the compact stars are investigated.  Concretely, the impact on the Chandrasekhar limit and the  gravitational collapse is studied. It is concluded that the QG corrections would increase the Chandrasekhar limit and hence they resist  the gravitational collapse. Furthermore, it is found that the radius of the compact star is decreasing as the energy increasing, which might be  considered as a possible explanation for the smaller radii observations.

\subsection{Saleker-Wigner inequalities}

The proposal that fundamental limits can be utilized in governing mass and size of  physical systems in order to register time dates back to nearly six decades \cite{Tawfik:2013uza}. Salecker and Wigner proposed the use of a quantum clock \cite{wigner57,wigner58} in measuring {\it distances} between events in space-time \cite{wigner58},  where the quantum clock is given as constrains of the smallest accuracy and the maximum running time as function of mass and position uncertainties. The Salecker-Wigner constrains assume that repeating measurements should not disturb the clock. This makes them more severe than HUP, which requires that only one single simultaneous measurement of both energy and time, for instance, can be accurate. The quantum clock is supposed to be able to accurately register time over its total running period. Barrow applied them in describing the quantum constrains on black hole lifetime \cite{barrow96}. It was found that the BH running time should be correspondent to the Hawking lifetime while the Schwarzschild radius is correspondent to the constrains on the Salecker-Wigner size.

\subsubsection{Salecker-Wigner inequalities and black hole evaporation}
\label{sec:sw}

As anticipated in section \ref{sec:SW1}, the second Salecker-Wigner inequality is more severe than the standard Heisenberg energy-time uncertainty principle. This is simply because it requires that a quantum clock is able to show proper time even after the time was being read. In other words, the quantum uncertainty in its position does not produce a significant inaccuracy in its time measurement. This property is conjuncted to hold over long periods, i.e., coherent time intervals. The terminology {\it ''coherence"} has to do with the correlation properties of the signal used in the measurement. The {\it ''coherent time''} is defined as the time period within which the signal remains {\it ''coherent''} $\tau_{c} = (\Delta\, \nu_c)^{-1} \approx \lambda_c^2/(c\, \Delta\, \lambda_c$,
where the subscript $c$ refers to coherence.

From HUP, the momentum uncertainty in {\it single analogue} quantum  clock of mass $m$  is $\hbar/2\, \Delta\, x$, where $\Delta\, x$ is uncertainty in its quantum position. After time $t$, the clock position spread increases to 
\bea \label{eq:ts}
\Delta\, x^{\prime} &=& \Delta\, x + \frac{\hbar\, t}{m}\, \frac{1}{2\, \Delta\, x}. 
\eea
Assuming that the mass of quantum clock remains unchanged, then Eq. (\ref{eq:ts}) leads to a minimum time spread   
\bea \label{eq:ts2}
\Delta\, x &\geq& \sqrt{\hbar\, \frac{t_{max}}{2\, m}},
\eea
where $t_{max}$ is the total {\it ''coherent''} time.  Expression equation (\ref{eq:ts}) is known as Salecker-Wigner first inequality. If the mass depends on the position uncertainty, then the minimum time spread  reads \cite{Tawfik:2013uza}
\bea
\Delta\, x &\geq& \frac{\hbar\; t_{max}\, m^{\prime} - \sqrt{\hbar\; t_{max}\, \left[8\, m^2 + (m^{\prime})^2\, \hbar\, t_{max}\right]}}{4\, m^2},
\eea
where $m^{\prime}=d m/d \Delta\, x$. The positive sign is evaluated as non-physical.
 
If you insist in assuring reliability for the repeated time measurements, the position  uncertainty which in turn must be caused by the repeated measurements, should be smaller than the minimum wavelength of the reading signals, i.e. $\Delta\, x \leq c\, T_{min}$. For an unentangled, unsqueezed and Gaussian signal, the minimum size is defined by the minimum mass of the quantum clock. From Eq. (\ref{eq:ts2}),  the mass-time inequality (Salecker-Wigner second inequality) is given as \cite{Tawfik:2013uza}
\bea \label{eq:ts22}
m &\geq& \frac{\hbar}{2\, c^2}\; \frac{t_{max}}{t_{min}^2}.
\eea

\subsubsection{Salecker-Wigner inequalities and linear GUP approach}

The Salecker-Wigner first inequality, Eq. (\ref{eq:ts}), can be applied on a black hole with a size comparable to the Schwarzschild radius, $r_s=2\, G\, m /c^2$. From Eq. (\ref{eq:ts2}), the maximum running time (lifetime) of black hole reads
\bea
t_{max} & \leq & 8 \frac{G^2\, m^3}{\hbar \, c^4},  \label{eq:bhlt1}\\
 & \leq & 8\, \frac{G}{c^3}\left(\frac{m^3}{M_p^2}\right),  \label{eq:bhlt2}
\eea
where $M_p=\sqrt{c\, \hbar/G}$ is the Planck mass. Obviously, these expressions are compatible with the Hawking lifetime \cite{Inflation100q}. They answer the question, {\it ''how does the life of a black hole run out?''} \cite{Tawfik:2013uza}. As discussed in the previous sections, the mass of black hole quantum clock is the only parameter that describes a reliable mechanism and offers another alternative rather than black hole as a black body radiator \cite{Inflation100q}.

At Planck scale, the space-time fluctuation becomes significant. Therefore, it is natural to set a bound to the linear spread of the quantum clock, Eq. (\ref{eq:ts2}), which is the Planck distance. On the other hand, the GUP approach gives prediction for a minimal measurable length. Therefore, $\alpha_0\, \ell_{p}$ would be taken as the smallest linear spread of the quantum clock. At time $t$, the position uncertainty due to GUP \cite{Tawfik:2013uza}
\bea
\Delta\, x^{\prime} &=& \Delta\, x + \frac{2 \Delta\, x + \frac{4}{3}\, \alpha_0
~\ell_p~\sqrt{\mu}}{4~(1+\mu)~\alpha_0^2~\ell_{p}^2\, m} \, \hbar \, t \, \left[1-
\sqrt{1-\frac{8~(1+\mu)~\alpha_0^2\ell_{p}^2} {\left(2 \Delta\, x + \frac{4}{3}\, \alpha_0
\ell_p~\sqrt{\mu}~\right)^2}}~\right].
\eea
Then 
\bea
\Delta\, x_{GUP} &\geq & \frac{1}{2} \left[-A_1 + \frac{\sqrt{2}\, (m\, A_2 + 2\, \hbar\, t)^2}{\sqrt{m\,(m\, A_2 + 2\, \hbar\, t)^2\; (m\, A_2 + 4\, \hbar\, t)}}\right], \label{eq:Dx1}
\eea
where $A_1=\frac{4}{3} \alpha_0 \ell_p \sqrt{\mu}$ and  $A_2 = 4 (1+\mu)   \alpha_0^2 \ell_p^2$.  At $\alpha_0=0$, the Salecker-Wigner position uncertainty can be recovered
\bea
\Delta\, x_{SW} &\geq & \sqrt{\hbar \frac{t}{2\, m}}.  \label{eq:Dx2}
\eea
In Eqs. (\ref{eq:Dx1}) and  (\ref{eq:Dx2}), the negative solutions are evaluated as non-physical. It is apparent that Eq. (\ref{eq:Dx2}), in which GUP effects are excluded, is identical with the Salecker-Wigner first inequality, Eq. (\ref{eq:ts2}). The difference between Eq. (\ref{eq:Dx1}) and Eq. (\ref{eq:ts2}) simply reads \cite{Tawfik:2013uza}
\bea
\Delta\, x_{GUP}-\Delta\, x_{SW} &=& \frac{1}{2}\left[-A_1-\hbar^2\, t_{max}^2 \sqrt{\frac{2}{m\, \hbar^3\, t_{max}^3}} +  (m\, A_2 + 2 \hbar\, t_{max})^2 \sqrt{\frac{2}{m(m\, A_2 + 2 \hbar t_{max})^2 (m\, A_2 + 4 \hbar t)}}\right], \hspace*{10mm}
\eea
which obviously vanishes at vanishing $\alpha_0$.

\begin{figure}[tbp]
\includegraphics[width=.7\textwidth]{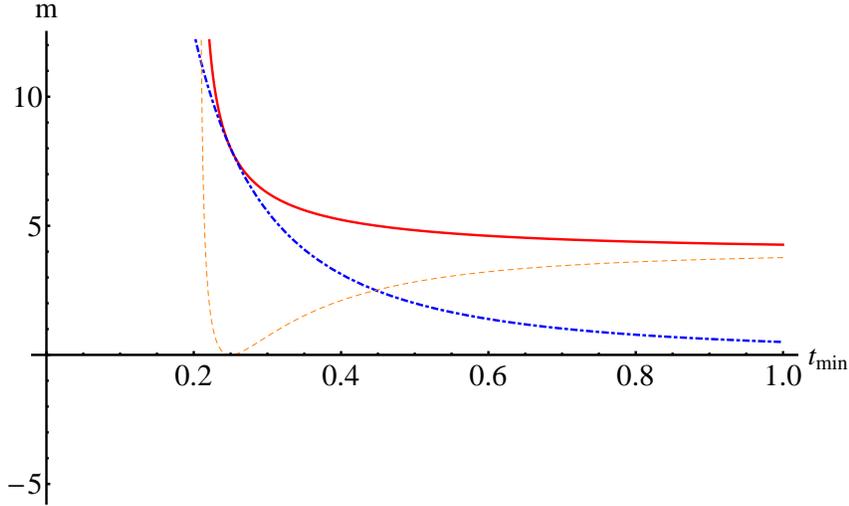}
\caption{The black hole mass is given in dependence on its lifetime with (solid line) and without (dash-dotted line) GUP and their difference (dashed line).  The values of the variables $A_1$, $A_2$, $\hbar$, and $c$ are taken unity. The graph is taken from Ref. \cite{Tawfik:2013uza}. }
\label{fig:tmax1}
\end{figure}

Assuming that the quantum position uncertainty should not be larger than the minimum wavelength of the measuring signal, so that in Eq. (\ref{eq:Dx1}), $\Delta\, x_{GUP}\leq c\, t_{min}$ \cite{Tawfik:2013uza},
\bea \label{eq:Wm22}
m_{GUP} &\geq & -\left[2\, \hbar \, t_{max} A_3 \pm 2 \hbar\, t_{max} \left(A_1+2 c\, t_{\min}\right) \sqrt{
A_3}\right]/(A_2\, A_3),
\eea  
where $A_3 = A_1^2-2 A_2+4 c\, t_{\min} \left(A_1+c \, t_{\min}\right)$. The positive sign defines a non-physical solution with $2 \hbar\, t_{max} \left(A_1+2 c\, t_{\min}\right) \sqrt{A_3} > 2\, \hbar \, t_{max} A_3$,
implies that $\sqrt{A_3} < A_1+2 c\, t_{\min}$.
At vanishing $\alpha_0$, Eq. (\ref{eq:Wm22}) goes back to the Salecker-Wigner second inequality, Eq. (\ref{eq:ts22}). The difference between Eq. (\ref{eq:Wm22}) and Eq. (\ref{eq:ts22}) results in \cite{Tawfik:2013uza}
\bea
m_{GUP}-m_{SW} &=& \frac{1}{2} \left(-\frac{4\, \hbar\, t_{max}}{A_2} - \frac{\hbar t_{max}}{c^2\, t^2_{min}}  -\frac{4\, \hbar^2\, t_{max}^2 (A_1 + 2\, c\, t_{min})^2}{A_2\,\sqrt{\hbar^2\, t_{max}^2\, (A_1 + 2\, c\, t_{min})^2\, A_3}} \right).
\eea

The modified black-hole lifetime can be estimated assuming that the spread of quantum clock has a minimum value, the Schwarzschild radius, $r_s$,
\bea \label{eq:tmbh}
t_{GUP} &=& \frac{1}{16\, \hbar^2}\left[-\hbar\, m\, A_4 - \hbar\, m\, A_4 \left(1-128\,  A_2\right)^{1/2}\right],
\eea
where $A_4=-4 A_1^2+8 A_2-16  r_s A_1-16 r_s^2$. The solution including negative sign is taken as physical. At $\alpha_0=0$, the modified black hole lifetime, Eq. (\ref{eq:tmbh}), goes back to Salecker-Wigner inequality, Eq. (\ref{eq:bhlt1}). The difference between the black-hole lifetime in the GUP approach and the Salecker-Wigner inequality reads \cite{Tawfik:2013uza}
\bea
t_{GUP}-t_{SW} &=& 2 \frac{m\, r_s^2}{\hbar}=8 \frac{G\, m}{c^3}\, \left(\frac{m}{M_p}\right)^2,
\eea
and depicted in Fig. \ref{fig:tmax1}.

%

\subsubsection{Conclusions}

Based on the assumption that the black hole is a perfect radiator, a reliable estimation of its lifetime is introduced. To this end, another approach based on the Salecker-Wigner inequalities was utilized \cite{Tawfik:2013uza}, which are assumed to be more severe than the Heisenberg energy-time uncertainty principle. The proposed quantum clock is conjectured to show proper time even after measuring the time while the quantum uncertainty in position does not produce a significant inaccuracy in the time measurement. This property is conjuncted to hold over long {\it ''coherent''} time intervals.

At Planck scale, the smallest linear spread of the quantum clock is set at $\alpha_0\, \ell_{p}$. Assuming that the mass remains unchanged, the Salecker-Wigner first inequality is reproduced. When applying GUP approach, the resulting position uncertainty seems not to match with the Salecker-Wigner first inequality. The discrepancy depends on the maximum lifetime. From the Salecker-Wigner second inequality, the latter can be related to the minimum lifetime. 

Assuming that the quantum position uncertainty is limited to the minimum wavelength of the measuring signal, the Salecker-Wigner second inequality can be reproduced, as well. The difference between black-hole mass with and without GUP is not negligible. The modified black-hole lifetime can be deduced if the spread of quantum clock is limited to a minimum value. The natural one is the Schwarzschild radius. Based on GUP, the resulting lifetime difference depends on black-hole mass and the bounds on $\alpha_0$.

\subsection{Minimal time measurement}
\label{sec:timem}

Towards measuring minimal time interval, we can highlight three milestones:
\begin{itemize}
\item In 1927, a hypothetical  indivisible interval of time taken as a ratio between the diameter of the electron and the velocity of light, being equivalent to approximately $\sim10^{-24}~$s, was proposed by Robert Levi  \cite{Levi}.
\item About fifty years ago, Shapiro pointed out that the possible time delay resulting from the observation that light slows down as it passes through a gravitational potential, could be measured in the solar system \cite{td1,td2,Tawfik:2012hz}. This was the first proposal about possible observation of time delay. As given in section \ref{sec:sw}, utilizing the fundamental limits governing mass and size, Salecker and Wigner \cite{wigner57,wigner58} suggested that a minimum time interval can be even registered by a quantum clock. 
\item Itzhaki considered the uncertainty principle and utilized the Schwarzchild solution in large scale in order to estimate the minimal measurable time interval \cite{itzhaki} and found that the uncertainty in time measurement depends on the distance separating the observer from the event, the clock accuracy and size, and the time taken by photon to reach the observer. 
\end{itemize}

Assuming distances, in which GR offers a good approach for QG, then the shortest distance $x_c=\beta\, (G \hbar /c^3)^{1/2}$, where $\beta$ is an arbitrary parameter. The minimum error in the time measurement is estimated as
\bea
\Delta\, t &=& \sqrt{\frac{8 \, G\, \hbar}{c^5}\, \ln\left(\frac{x}{x_c}\right)}=2\,  \sqrt{2\, \ln\left(\frac{x}{x_c}\right)}\; t_{pl},
\eea
where $t_{pl}=\sqrt{G\, \hbar/c^5}$ is the Planck time. This expression is valid at distance $x>x_c\, \exp(2/\beta^2)$, where  $x_c=\beta\, (G\, \hbar/c^3)^{1/2}$ is the shortest distance for which it is assumed that GR is a good approximation to QG. The corresponding minimum error in the energy is given by
\bea
\Delta\, E &=& \sqrt{\frac{\hbar \, c^5}{2\, G \, \ln\left(\frac{x}{x_c}\right)}} = \sqrt{\frac{1}{2 \,  \ln\left(\frac{x}{x_c}\right)}}\;  \frac{\hbar}{t_{pl}}.
\eea
At $x_c<x<x_c\, \exp(2/\beta^2)$, the minimal time and maximal energy,  $\Delta \, t_{min} = \frac{x_c}{c}\left[\frac{2}{\beta^2}+\ln\left(\frac{x}{x_c}\right) \right]$ and $\Delta \, E_{max} = \frac{c^4}{2 \, G}\, x_c = \frac{\hbar \, x_c}{2 \, c}\; \frac{1}{t_{pl}^2}$, respectively.

\subsubsection{Linear GUP approach: uncertainty in time and minimum measurable time}

In the linear GUP approach, the time uncertainty  reads
\bea
\Delta t & \geq & \frac{1}{2}\, \frac{\hbar}{\Delta E} \left[1 - 2\, \frac{\alpha}{c}\, \Delta E \right]  = \frac{\hbar}{2\, \Delta E} - \frac{\alpha}{c}\, \hbar,
\eea
implying that the physical limits require $2 \alpha \Delta E < c$. The minimum measurable time-interval $\Delta t_m$ is to be deduced under the condition that $d\, \Delta \, t / d\, \Delta E$ vanishes. Then, $-\hbar/(2\, (\Delta E)^2) = 0$,
which leads to $\Delta\, E_{max} = \infty$ and $\Delta\, t_{min} = -\alpha/c \, \hbar= -\alpha_0/(M_{pl}\, c^2) \, \hbar$, where $\alpha$ is replaced by $\alpha_0/M_{pl} c$. It is obvious that the measurable maximal energy gets infinite, while the measurable minimal time interval has a negative value. Both results are obviously non-physical. While $\Delta\, E$ apparently violates the conservation of energy, $\Delta\, t$ shows that the direction of the arrow of time becomes opposite.

\subsubsection{Uncertainty in time and minimum measurable time at the shortest distance $x_c$}

The time uncertainty as estimated from the Schwarzchild solution at distance $x_c$ at which GR is a good approximation to QG is given as \cite{itzhaki} $\Delta\, t \geq \hbar/(2\, \Delta E) + G\, \Delta E/c^5$. 
Then, the maximal measurable energy and minimal measurable time interval, respectively, read
\bea
\Delta\, E_{max} &=& c^2\, \sqrt{\frac{\hbar \, c}{2 \, G}}=\frac{\hbar}{\sqrt{2}}\;\frac{1}{t_{pl}}, \\
\Delta\, t_{min} &=& \frac{1}{c^2}\, \sqrt{\frac{2\, G \, \hbar}{c}}=\sqrt{2}\; t_{pl}.
\eea 
Both quantities are positive. It is obvious that both of them depend on the Schwarzshild radius, which is related to the black hole mass, $r_s=(2 G/c^2)\, m$. It is worthwhile to note that both quantities are related to the Planck time $t_{pl}$. Accordingly, they are bounded.

\subsubsection{Time uncertainty and minimum measurable time at distance larger than $x_c$ }

When the photon travels a distance $x$ larger than $x_c$, then the total time uncertainty is estimated as
\bea
\Delta \, t_{total} & \geq & \frac{\hbar}{2 \Delta E} + G \frac{\Delta E}{c^5} + 2 G  \frac{\Delta E}{c^5}  \ln\left(\frac{x}{x_c}\right).
\eea
The maximal measurable energy and corresponding minimal measurable time interval,  respectively, are given as
\bea
\Delta E_{max} &=& \sqrt{\frac{c^5 \hbar}{2 G [1+2 \ln(\frac{x}{x_c})] }} =  \sqrt{\frac{1}{2 [1+2 \ln(\frac{x}{x_c})] }}\; \hbar\; t_{pl}, \\
\Delta t_{min} &=& \sqrt{\frac{2 \hbar G}{c^5} \left[1+2 \ln\left(\frac{x}{x_c}\right)\right]} = \sqrt{2 \left[1+2 \ln\left(\frac{x}{x_c}\right)\right]}\; t_{pl}.
\eea
The resulting $\Delta E_{min}$ and $\Delta t_{min}$ are finite and positive. Both quantities are related to $t_{pl}$.

\subsubsection{Conclusions}

The maximal measurable energy $\Delta E$ and minimal measurable time $\Delta T$ are related to $t_{pl}$ and therefore both are accordingly bounded. The Itzhaki model used the most simple time measurement process. It was concluded that any particles that will be added must necessarily increase the uncertainty of the metric without decreasing the minimal measurable time. Furthermore, Itzhaki summarized that the measured uncertainty would represent a basic property of the Nature. 

The possibility of finding measurable maximal energy and minimal time are estimated in different quantum aspects. First, we find that the linear GUP approach gives non-physical results. The resulting maximal energy $\Delta\, E$ violates the conservation of energy. The minimal time interval $\Delta\, t$ shows that the direction of the arrow of time is backwards. So far, we conclude that the applicability of the linear GUP approach is accordingly limited or even altered. 

%
\subsection{Entropic nature of gravitational laws and Friedmann equations}

In 1D Ising model \cite{ising1925}, it is assumed that every single spin is positioned at a distance $d$ apart from the two neighbourhoods. Then, the macroscopic state of such a chain can be defined by $d$. Depending on $d$, the entire chain would have various configurations so that  if $d\rightarrow l$, the chain has much less configurations than if $d \ll l$, where $l$ is the chain's length. From statistical point-of-view, the entropy is given by the number of microscopic states $S=k_B\ln \Omega$. Due to second law of thermodynamics, such a system tends to approach a state of a maximal entropy. Accordingly, the chain in the macroscopic state $d$ tends to go to a another state with a much higher entropy. The force that causes such a statistical tendency is defined as entropic force. In light of this, the entropic force is a phenomenological mechanism deriving a system to approach maximum entropy, i.e. increasing the number of microscopic states which will be inhered in the phase-space. There are various examples on the entropic forces, for example, polymer molecules and even the elasticity of rubber bands.

Verlinde has proposed  that the gravity might not be a fundamental force and therefore be considered as an entropic force \cite{everlind1}. In light of this, we recall that the earliest idea about gravity as a non-fundamental interaction has been introduced by Sakharov \cite{sakharoov67}, where the space-time background was assumed to emerge as a mean field approximation of the underlying microscopic degrees of freedom. Similar behavior was observed  in the hydrodynamics \cite{hydro1}. As discussed in earlier sections, the BH entropy is to be related to the area of the BH horizon, while the temperature to the surface gravity. But both entropy  and temperature are assumed to be related to the BH mass \cite{entr3,entr1}. Thus, from the relations connecting heat, entropy and temperature \cite{jack}, the connection between thermodynamics and geometry leads to the Einstein's equations of the gravitational field. Also, the Einstein's equations themselves connect the energy-momentum tensor with the space geometry. Advocating the gravity as non-fundamental interaction leads to assuming that the gravity would be explained as an entropic force caused by changes in the information associated with the positions of material bodies \cite{everlind1}. When combining the entropic force with the Unruh temperature \cite{unruh2001}, the second law of Newton is obtained. But when combining it with the holographic principle and using the equipartition law of energy, the Newton's law of gravitation is obtained. The modification on the entropic force due to corrections to the area law of entropy, which is derived from quantum effects and extra dimensions, has been investigated  \cite{Zhang:2010hi}.

\subsubsection{Entropy and lack hole horizon area}

For a black hole absorbing a quantum particle with energy $E$ and size $R$, the area of the black hole is supposed to increase by $\Delta A \geq 8 \pi\, \ell_p^2\, E\, R$ \cite{entr1}. The quantum particle itself implies the existence of a finite bound given by $\Delta A_{min} \geq 8 \pi\, \ell_p^2\, E\, \Delta\, x$. 
Thus, we obtain
\be
\Delta A_{min} \geq 8 \pi \ell_p^2\left[ 1- \frac{2}{3}\alpha_0 \ell_p \sqrt{\mu} \frac{1}{\Delta x}\right]. \label{Area}
\ee
According to the argument given in Refs. \cite{Eliasentropy,gup-entropy1q}, the length scale is chosen to be
the inverse surface gravity $\Delta x= 2\, r_s$, where $r_s$ is the Schwarzschild radius. This argument implies that
 $(\Delta x)^2 \sim A/\pi$ and  \cite{Ali:2013ma}
\be
\Delta A_{min}= \lambda  \ell_{p}^2 \left[1- \frac{2}{3}\, \alpha_0\, \ell_p\, \sqrt{\frac{ \mu \, \pi}{A}}\right],
\ee
where parameter $\lambda$  will be fixed later. According to Refs. \cite{entr3,entr1}, the BH entropy is conjectured to depend on the horizon area. From the information theory \cite{Adami:2004mx}, it was found that the minimal increase of entropy should be independent on the area. It is just one bit of information, which is $b=\ln(2)$ \cite{Ali:2013ma}
\be
\frac{dS}{dA} = \frac{\Delta S_{min}}{\Delta A_{min}} = \frac{b}{\lambda  \ell_{p}^2 \left[1- \frac{2}{3}\, \alpha_0\, \ell_p\,  \sqrt{\mu\, \pi \, A}\right]},
\ee
where $b$ is a free parameter. By expanding the last expression in orders of $\alpha$ and then integrating, we get the entropy
\be
S = \frac{b}{\lambda \ell_p^2}\, \left[ A+ \frac{4}{3}\, \alpha_0\, \ell_p\, \sqrt{\mu\, \pi\, A}\right].
\ee
By using Hawking-Bekenstein assumption, $b/ \lambda= 1/4$, so that 
\be
S=\frac{A}{4\, \ell_p^2} + \frac{2}{3}\, \alpha_0\,  \sqrt{\pi\, \mu\, \frac{A}{4\, \ell_p^2}}. \label{correctENTROPY}
\ee
The entropy, which is directly related to the area, gets a correction due to the linear GUP approach. Although the ruling out of the power-law corrections to Bekenstein-Hawking entropy because of the Boltzmann-Einstein formula, it was found that these corrections may explain the observed cosmic acceleration today \cite{Wang:2005bi}. Furthermore, the black hole temperature can be given as \cite{Ali:2013ma}
\bea
T &=& \frac{\kappa}{8 \pi} \frac{dA}{dS} = \frac{\kappa}{8 \pi} \left[1- \frac{2}{3}\, \alpha_0\, \ell_p\, \sqrt{\mu\, \frac{\pi}{A}} \right].
\eea
Then, the temperature is not only proportional to the surface gravity but
also depends on the black hole's area.

\subsubsection{Linear GUP approach and entropic Newtonian laws}

Using the corrected entropy given in Eq. (\ref{correctENTROPY}), we find that  
\bea
N^{\prime} &=& \frac{4 S}{k_B}= \frac{A}{\ell_{p}^2}+ \frac{4}{3}\, \alpha_0\, \sqrt{\mu\, \pi\, \frac{A}{\ell_{p}^2}}. \label{bits} \\
E &= & F\, c^2\, \left(\frac{r^2}{m\, G}+ \frac{\alpha\, \sqrt{\mu}\, r}{3\, m\, G}\right).
\eea
This implies modifications in the Newtonian law
\bea 
F &=& G\, \frac{M\, m}{r^2} \left(1- \frac{\alpha\, \sqrt{\mu}}{3\, r}\right). \label{result}
\eea
From the Newtonian second law,
\bea
m\, \ddot{r} &=& - G\, \frac{M\, m}{r^2} \left(1 - \frac{\alpha \sqrt{\mu}}{3 r} \right). \label{eq:2ndLaw}
\eea
where $r$ is the apparent horizon radius
\bea
\ddot{r} &=& - \frac{4 \pi G}{3}\, \alpha\, \rho \left(1 - \frac{\alpha \sqrt{\mu}}{3 r} \right). \label{eq:2ndLaw2} 
\eea

\subsubsection{Entropic Newtonian laws and modifications in Friedmann equations}

By multiplying both sides of Eq. (\ref{eq:2ndLaw2}) by $a \dot{a}$, then \cite{Ali:2013ma} 
\bea
\dot{a}\, \ddot{a} &=& - \frac{4\, \pi\, G}{3} a\, \dot{a}\;  \rho \left(1 - \frac{\alpha \sqrt{\mu}}{3 r} \right), \label{eq:dota}
\eea
With $p = \frac{1}{3}\, \rho$, $\rho_0 = - 3 \, H\, (\rho + p) = -4 \, H\, \rho$, 
$\frac{d}{d t}\,  \dot{a}^2 = 2\, \dot{a}\, \ddot{a}$ and
$\frac{d}{d t}\,  (\rho\, a^2) = \rho_0\, a^2 + 2\, a\, \dot{a}\; \rho$. 
The integral of Eq. (\ref{eq:dota}) leads to
\bea
\dot{a}^2 + C &=& \frac{8\, \pi\, G}{3}\; \rho\, a^2\, \left(1 - \frac{\alpha\, \sqrt{\mu}}{3\, r} \int \frac{d\, (\rho\, a^2)}{\rho\, a^3} \right),
\eea
where $C$ is the integral constant, which, as it will explained below, is nothing but the curvature constant, $k$.  The energy density reads  $\rho = \rho_0\, a^{-3(1-\omega)}$, 
where $\omega$ is the speed of sound, $\omega=p/\rho\equiv c_s^2$.
\bea
d (\rho_0 \, a^{-3(1-\omega)}) &=& -3(1-\omega) \rho_0 \, a^{-3\omega-2}) \, d a, \\
\rho\, a^3 &=& \rho_0\, a^{-3\omega}.
\eea
Accordingly,
\bea
\dot{a}^2 + C &=& \frac{8\, \pi\, G}{3}\; \rho\, a^2\, \left(1 -  \frac{\alpha \, \sqrt{\mu} (3 \omega +1)}{3\, r\, a} \right),
\eea
which can be rewritten as \cite{Ali:2013ma}
\bea
\left[H^2 + \frac{C}{a^2}\right] +  \frac{\alpha \, \sqrt{\mu}}{3} (3 \omega +1) \left[H^2+\frac{C}{a^2}\right]^{3/2} &=& \frac{8\, \pi\, G}{3}\; \rho, \label{eq:modFried}
\eea
where $r\, a$ represents the apparent horizon radius, $(H^2+C/a^2)^{-1/2}$. Expression equation (\ref{eq:modFried}) is the modified Friedmann equation, where $C$ is equivalent to the curvature constant. A detailed solution of $H$ with respect to $\rho$ is presented in Appendix \ref{sec:appdx}.

\subsubsection{Conclusions}

Expression (\ref{result}) obviously states that the modification in the Newton's law of gravity agrees with the predictions of  Randall-Sundrum II model \cite{Randall:1999vf} which contains one uncompactified extra dimension and length scale $\Lambda_R$ (the sign is the only difference). accordingly, the modification in the Newton's  gravitational potential on brane  is given as \cite{potential}
\bea
V_{RS} =\begin{cases} -G\frac{m M}{r} \left(1+\frac{4 \Lambda_R}{3 \pi r}\right), &  r \ll \Lambda_R\\
& \\
-G\frac{m\, M}{r} \left(1+\frac{2 \Lambda_R}{3 r^2}\right), & r \gg \Lambda_R
\end{cases}, \label{VRS}
\eea
where $r$ and $\Lambda_R$ are radius and the characteristic length scale, respectively. It is clear that the gravitational potential  gets modified at short distance. When  $r\ll\Lambda_R$, Eq. (\ref{result}) agrees with Eq. (\ref{VRS}). Again, the sign is the only difference. This leads to $\alpha \sim \Lambda_R$, which helps in setting a new upper bound on the value of the GUP parameter $\alpha$. 

Apparently, there is a similarity between GUP and extra dimension length scale $\Lambda_R$. The proposed GUP approach \cite{advplb,Das:2010zf} is apparently able to predict the same physics as Randall-Sundrum II model. The latter assumes the existence of one extra dimension compactified on a circle whose upper and lower halves are identified. If the extra dimensions are accessible only to gravity and not to the standard model field, the bound on their size can be fixed by an experimental test of the Newton's law of gravity, which has only been led down to $\sim 4~$mm \cite{gt1}. This was the result, about ten years ago. In recent gravitational experiments, it is found that the Newtonian gravitational force, the $1/r^2$-law, seems to be maintained up to $\sim0.13-0.16~$mm \cite{gt2}. However, it is unknown whether this law is violated or not at sub-$\mu$m range. Further applications of this modifications have been discussed in Ref. \cite{Buisseret:2007qd} which could be the same for the GUP modification. 

Furthermore, the modification in Eq. (\ref{eq:2ndLaw2}) has various consequences, for example the Friedmann equations, Eq. (\ref{eq:modFried}). It is apparent that the entire modification is placed in the second term in lhs, which obviously depends on $H$, as well. The solution of $H$ with respect to $\rho$ is presented in Appendix \ref{sec:appdx}. The dependence of $H$ on $\rho$ is not monotonic. Reducing $\rho$, or increasing the cosmic time $t$, is accompanied with reduction in $H$. Another behavior is characterized by certain value of $\rho$ (or at concrete $t$). The Hubble parameter $H$ increases with the further decrease in $\rho$. The rate strongly depends on geometry of the Universe, $k$.

\subsection{Thermodynamics of high-energy collisions}

As discussed, the GUP approach apparently causes modifications in the fundamental commutator bracket between position and momentum operators. Then, it seems natural that this would result in considerable modifications in the Hamiltonian. For a particle of mass $M$ having a distant origin and an energy  comparable to the Planck scale, the momentum would be a subject of a tiny modification and so that the dispersion relation can be expressed as in Eq. (\ref{eq:disp}).  Modified dispersion relations have been observed in DSR \cite{Camacho:2006qg,Gregg:2008jb}. Calculations based on these have been presented \cite{AmelinoCamelia:2009tv}.

The phase space integral can be expressed as follows \cite{Elmashad:2012mq}.
\bea
\sum_i\, \dfrac{V}{(2\pi)^{3}}\int_{0}^{\infty}d^{3} p \rightarrow \sum_i\, \dfrac{V}{2\pi^{2}}\int_{0}^{\infty}\dfrac{p^{2}dp}{(1-\alpha {p})^4}.
\eea
The partition function of an ensemble of $N$ ideal (collision-free) constituents at vanishing chemical potential reads
\bea
\ln z(T,V,\,\alpha) &=& \sum_i^N \pm \frac{V\, g_i}{2 \pi^2} \int_0^{\infty}\,  \frac{p^2}{(1 - \alpha\, p)^{4}}\,  \nn \\
&&          \ln 
         \left\{1 \pm 
       \exp\left[-\frac{p\, \sqrt{(1 - 2\, \alpha\, p) + \left(\frac{m_i}{p}\right)^2}}{T} \right] 
         \right\} \, d p,  \label{eq:lnzalf1}
\eea
where $\pm$ stand for bosons and fermions, respectively. Equation (\ref{eq:lnzalf1}) can be decomposed into
\bea
\ln z(T,V,\alpha) &=& \sum_i^N \pm \frac{V\, g_i}{2\, \pi^2} \int_0^{\infty}\,  p^2 \;
         \ln \left\{1 \pm  \exp\left[-\frac{p\, \sqrt{(1 - 2\, \alpha\, p) + \left(\frac{m_i}{p}\right)^2 }}{T}\right]\right\} ~d p,  \label{eq:lnzalf2a} \\
  &+&   \sum_i^N \pm \frac{V\, g_i}{2\, \pi^2} \int_0^{\infty}\,  p^2 \;
             F(\alpha\, p)\; \ln \left\{1 \pm  \exp\left[-\frac{p\, \sqrt{(1 - 2\, \alpha\, p) + \left(\frac{m_i}{p}\right)^2 }}{T}\right]\right\} ~d p, \hspace*{1cm} \label{eq:lnzalf2b}
\eea
where $F(\alpha\, p)$ is a series function.

The pressure is directly related to free energy of the system of interest, $p(T,V,\alpha)=T \partial \ln z(T,V,\alpha)/\partial V$. The number density reads
\bea
n(T,V,\alpha) &=& \sum_i^N \pm \frac{g_i}{2\, \pi^2} \int_0^{\infty}\,  p^2 \;
         \frac{\exp\left[-\frac{p\, \sqrt{ (1 - 2\, \alpha\, p) + \left(\frac{m_i}{p}\right)^2 }}{T}\right]}{1 \pm \exp\left[-\frac{p\, \sqrt{ (1 - 2\, \alpha\, p) + \left(\frac{m_i}{p}\right)^2}}{T}\right]} ~d p,  \label{eq:n1a} \\
  &+&   \sum_i^N \pm \frac{g_i}{2\, \pi^2} \int_0^{\infty}\,  p^2 \;
             F(\alpha\, p)\; \frac{\exp\left[-\frac{p\, \sqrt{ (1 - 2\, \alpha\, p) + \left(\frac{m_i}{p}\right)^2 }}{T}\right]}{1\pm \exp\left[-\frac{p\, \sqrt{(1 - 2\, \alpha\, p) + \left(\frac{m_i}{p}\right)^2 }}{T}\right]} ~d p.  \label{eq:n1b}
\eea
The simplest way to calculate the energy density is to multiply the number of quantum states $n(T,V,\alpha)$ by the energy of each state. Equations (\ref{eq:lnzalf2b}) and (\ref{eq:n1b}) take into account possible modifications in the phase space \cite{Tawfik:2010aq,Tawfik:2010uh,Tawfik:2010kz}. In equations (\ref{eq:lnzalf2a}) and (\ref{eq:n1a}), the phase space is apparently not a subject of modification, while the dispersion relation is.

\subsubsection{Linear GUP approach at QCD scale}

The central question is whether the GUP approach is applicable at the level of QCD scale, $\sim1~$GeV. If this would raise drastic {\it havoc} in high-energy phenomena and probably would show up in the high-precision measurements at low energy before they showed up in the QGP phase. The experimental inferences of the QGP are barely better candidates to text this.  The phenomenology should be well thought out, as it seems that if the dispersion relation, Eq. (\ref{eq:disp}), were sufficiently modified to affect QGP observations, it would seem to alter other measurements in a more easily way. 

Instead of modifying the dispersion relation, we may allow the phase space to be modified. In doing this, we start with the single-particle equilibrium distribution function \cite{Tawfik:2010pt,Tawfik:2010kz,Tawfik:2010aq}. The maximum number of {\it micro}-states is given by solving
\bea
\frac{\partial}{\partial n_j}\left(S-\alpha N -\beta E \right)=\frac{\partial}{\partial n_j}\left(\ln N! + \ln \Pi_i^n g_i^{n_i} - \sum_i^n \ln n_i! - \alpha N -\beta E \right) &=&0,
\eea
which means that only the terms having same subscript $j$ remain finite. The coefficients $\alpha$ and $\beta$ are Lagrange multipliers in entropy maximization. Each of these multipliers basically adds some unknown amount of each independent constraint to the function being optimized and ensures that the constraints are satisfied. 
\bea \label{eq:nj1}
\frac{\partial}{\partial n_j} \ln \Pi_j g_
j^{n_j}- \frac{\partial}{\partial n_j}\ln n_j! - \alpha -\beta \epsilon_j  &=&0.
\eea
Utilizing the Stirling approximation, then the occupation number, $n_j= g_j\,\exp\left(-\alpha-\beta \epsilon_j\right)$, which apparently falls off exponentially with increasing $\epsilon$, since, as will be shown below, $\gamma=\exp(-\alpha)$ is constant. 

Then, the grand-canonical partition reads
\bea \label{eq:zTr-Lagrange}
Z_{gc}(T,V,\mu) &=& \Tr \, \left[\exp^{\frac{\mu \hat{b}-\hat{H}}{T}-\alpha}\right],\\
f_{gc}(T,V,\mu) &=& \frac{\exp\left(\frac{-\hat{H}}{T}-\alpha\right)}{Z_{gc}(T,V,\mu)}.
\eea
With these assumptions, the dynamics of the partition function can be calculated as sum over single-particle partition functions $Z_{gc}^i$ 
\bea
\ln Z_{gc}(T,V,\mu)&=&\sum_i \ln Z_{gc}^i(T,V,\mu) \nn \\
&=& \sum_i\pm \frac{g_i}{2\pi^2}\,V\int_0^{\infty} k^2 dk \ln\left(1\pm \gamma\,\lambda_i\, e^{-\epsilon_i(k)/T}\right),
\eea
where $\lambda_i=\exp(\mu_i/T)$ is the $i$-th particle fugacity and $\gamma=\exp(-\alpha)$ is the quark phase-space occupation factor.

With this regard, the constraints on the Lorentz invariance violation are very essential. If Lorentz invariance is instead deformed and the quantities in Eq. (\ref{eq:lnzalf1}) transform under this deformed transformation, this necessarily leads to a modification in the addition law of momenta, as well. The definition of the parameter $\beta$ given in Eq. (\ref{eq:nj1}), involves summing over energies. Then, it becomes non-trivial, i.e. $\beta=1/T$ would be also a subject of modification.

\subsubsection{Conclusions}

It is worthwhile to highlight that the works on this topics are still ongoing \cite{Elmashad:2012mq}. At the QCD scale, which is accessible by means of high-energy experiments and even the lattice QCD simulations at finite temperature, the GUP approach would be applicable. In this limit, modifications on the phase space, Lorentz invariance and even temperature have been obtained. The validity of the GUP approach  at the level of QCD scale, $\sim1~$GeV, is questionable, as one expects that it would raise drastic {\it havoc} in high-energy phenomena and probably would show up in the high-precision measurements at low energy.

\newpage

\section{Alternative approaches to GUP} 
\label{other}

In this section, we introduce other GUP approaches, which propose higher-order modifications and/or solve some some of the physical constrains/problems appeared when applying either linear or quadratic GUP approaches. One alternative approach gives predictions for the minimal length uncertainty, section \ref{sec:1HO}.  Second one foresees maximum momentum besides the minimal length uncertainty, section \ref{sec:2HO}. An extensive comparison between three GUP approaches is elaborated in section \ref{sec:cpmprsn}.

\subsection{Higher order GUP with minimal length uncertainty}
\label{sec:1HO}

Nouicer suggested a higher-order GUP approach \cite{Nouicer}, which agrees with the GUP given in Eq. (\ref{eg}) to the leading order and predicts a minimal length uncertainty, as well. The Heisenberg algebra of the new GUP approach can be given by $\left[x,\, p\right] = i\, \hbar\, \exp\left(\beta\, p^{2}\right)$.
Apparently, this algebraic basis can be fulfilled from the representation of position and momentum operators $X\, \psi(p) = i\, \hbar\, \exp\left(\beta\, p^{2}\right) \de_{p}\, \psi(p)$ and $P\, \psi(p) = p\, \psi(p)$, which are symmetric and imply modified completeness relation 
\bea
\langle \phi | \psi \rangle &=& \int_{-\infty}^{\infty} dp\, \exp \left(- \beta\, p^{2}\right)\, \phi ^{*}(p)\,   \psi (p).
\eea
The scalar product of the momentum eigenstates changes to $\langle p | p^{'} \rangle = \exp\left(\beta\, p^{2}\right)\, \delta (p - p^{'})$.
Also, the absolutely smallest position uncertainty is given as
\bea
(\Delta\, x)_{min} &=& \sqrt{\frac{e}{2}}\, \hbar\, \sqrt{\beta}.
\eea

\subsection{Higher-order GUP with minimal length  and maximal momentum uncertainty}
\label{sec:2HO}

Another  higher-order GUP* approach was proposed in Ref. \cite{pedram}, assuming $n$-dimensions and implying both minimal length uncertainty and maximal observable momentum, $[X_{i},\, P_{j}] = i\, \hbar\, \delta _{i j}/(1 - \beta\, p^{2})$, where $p^{2} = \sum_{j}^3\, p_{j}\, p_{j}$. If the components of the momentum operator are assumed to commutate, $[P_{i},\, P_{j}] = 0$.
The Jacobi identity determines the commutation relations between the components of the position operator 
\bea
[X_{i},\, X_{j}] &=& \frac{2\, i\, \hbar\, \beta}{(1- \beta\, p^{2})^{2}}\, (P_{i}\, X_{j} - P_{j}\, X_{i}),
\eea
which apparently results in a non-commutative geometric generalization of the position space. In order to fulfil  these commutation relations, the position and momentum operators in the momentum space representation should  be written as
\bea 
X_{i}\, \phi (p) &=& \frac{i\, \hbar }{1- \beta\, p^{2}} \de_{p_{i}} \phi (p),\\
P_{j}\, \phi (p) &=& p\, \phi.
\eea
In $1$D, the symmetricity condition of the position operator implies modified completeness relation with a domain varying from $-1/\sqrt{\beta}$  to $+1/\sqrt{\beta}$ \cite{pedram} $\langle \phi | \psi \rangle = \int_{-1/\sqrt{\beta}}^{+1/\sqrt{\beta}} dp (1-\beta p^{2}) \phi ^{*} (p)  \psi (p)$.
Apparently, this result differs from KMM \cite{16}.

Furthermore, the scalar product of the momentum eigenstates will be changed to $\langle p | p^{'} \rangle = \delta (p - p^{'})/(1-\beta p^{2})$. Also, the particle's momentum is bounded from above, $P_{max} = 1/\sqrt{\beta}$.
The presence of an upper bound agrees with DSR \cite{12,13}. As we shall see, the physical observables such as energy and momentum are not only non-singular, but they are  also bounded from above, as well. The absolutely smallest uncertainty in position reads
\bea
(\Delta X)_{min} &=& \frac{3\, \sqrt{3}}{4}\;	\hbar\, \sqrt{\beta}.
\eea

\begin{itemize}
\item This new GUP* approach \cite{pedram} estimates the minimal length uncertainty and the maximal observable momentum, simultaneously. 
\item It includes a quadratic term of the momentum and apparently assures non-commutative geometry. 
\item The maximal observable momentum agrees with the one estimated in DSR \cite{12,13}. 
\item If the binomial theorem is applied on this GUP* approach, the GUP approach which was predicted in string theory \cite{guppapers,2}, black hole physics \cite{7,3} can be reproduced.  
\end{itemize}

On the other hand, it is worthwhile to notice that this new GUP* approach \cite{pedram} does not agree with the commutators relation which was predicted in DSR \cite{12,13}. The latter contains a linear term of momentum that is responsible for the existence of maximal observable momentum. 

\subsection{Comparison between three GUP approaches}
\label{sec:cpmprsn}

\begin{table}[htb]
\label{tab:1}
\begin{center}
\begin{tabular}{|c||c|c|c|}
\hline 
Comparison  & KMM \cite{16} & ADV \cite{advplb,Das:2010zf} &Pedram \cite{pedram} \\ 
\hline  \hline
Algebra $\left[x,\, p\right]$ & $i \hbar \left(1+\beta p^{2}\right)$ & $i \hbar \left(1-\alpha p +2 \alpha ^{2} p^{2} \right)$& $\frac{i \hbar}{1-\beta p^{2}}$ \\ 
\hline 
$ (\Delta x)_{min} $ & $ \hbar \sqrt{\beta} $ & $ \hbar \alpha $ & $\frac{3\sqrt{3}}{4} \hbar \beta$   \\ 
\hline 
${(\Delta p)_{max}}$ & - & ${\frac{M_{pl} c}{\alpha _{0}}}$ & -  \\ 
\hline 
$P_{max}$ & Divergence  & $(\frac{1}{4\alpha })$  &  $(\frac{1}{\sqrt{\beta}})$  \\  
\hline 
\begin{tabular}{c}
$ \textbf{P} . \phi(p)$ \\ 
$\textbf{X} . \phi(p)$ \\ 
\end{tabular}  & \begin{tabular}{c}
$p \phi(p)$ \\ 
$i \hbar \left(1+\beta p^{2}\right) \partial_{p} \phi(p)$ \\ 
\end{tabular}  &\begin{tabular}{c}
$ p \phi(p)$ \\ 
$i \hbar \left(1-\alpha p +2 \alpha ^{2} p^{2} \right) \partial_{p} \phi(p)$ \\ 
\end{tabular} &\begin{tabular}{c}
$p \phi(p)$ \\ 
$\frac{i \hbar}{1- \beta p^{2}} \partial_{p} \phi(p)$\\  
\end{tabular}  \\
\hline 
Geometry & $\left[x_{i} ,x_{j}\right] \ne 0$ &  $\left[x_{i} ,x_{j}\right] \ne 0$  &  $\left[x_{i} ,x_{j}\right] \ne 0 $  \\ 
\hline 
 $\langle \frac{p^{2}}{2m} \rangle _{ max-localize-state}$ &$\frac{1}{2m \beta} $& $\frac{1}{32 m \alpha ^{2}}$ & $\frac{3}{2m \beta}$ \\ 
\hline 
 $(E(\lambda )$ or $\lambda (E))_{quasi-position}$  &$\frac{1}{2 m \beta} \left(\tan \frac{2 \pi \hbar \sqrt{\beta} }{\lambda} \right)^{2} $ & $\frac{2}{m \alpha ^{2}} \left( \frac{\tan(\frac{\hbar \alpha \sqrt{7} }{\lambda })\pi}{\tan(\frac{\hbar \alpha \sqrt{7} }{\lambda })\pi + \sqrt{7}} \right)^{2}$ & $\frac{2 \pi \hbar}{(1-\frac{2}{3} m \beta E ) \sqrt{2 m E}}$ \\ 
\hline 
$\lambda_{0}\;$ of~wavefuntion & $4 \hbar \sqrt{\beta}$ & $ \frac{\pi \alpha \hbar \sqrt{7}}{\left(\tan ^{-1} \frac{\eta}{3} + \tan ^{-1} \frac{4 \alpha p_{pl} - 1 }{\sqrt{7}} \right) }$&$3 \pi \hbar \sqrt{\beta} $  \\
\hline 
\end{tabular}
\caption{A comparison between the main features of the GUP approaches that were proposed by  KMM \cite{16}, ADV \cite{advplb,Das:2010zf} and Pedram \cite{pedram}.  }
\end{center}
\end{table}

Tab.~\ref{tab:1} summarizes an comprehensive comparison between the GUP approaches  of KMM \cite{16}, Ali, Das, Vagenas (ADV) \cite{advplb,Das:2010zf} and  Pedram \cite{pedram}. The minimum position uncertainty varies from $\hbar\,\alpha$ or $\hbar\,\sqrt{\beta}$ (both are equivalent) and $\sqrt{27}\, \hbar\, \alpha/4$, respectively. There is a maximum momentum uncertainty in ADV, although, it is wrongly called maximum momentum. The maximum momentum diverges in KMM, while it remains finite, $1/4\alpha$ and $1/\sqrt{\beta}$, respectively, in ADV and Pedram. The momentum operator and resulting geometry remain unchanged in all approaches. The position operator characterizes the different approaches. The maximum localised state slightly varies. The resulting energy (wavelength) related to quasiposition and wavefuction are very characteristic.

\newpage

\section{Argumentation against GUP}

In this section, we review argumentation against the GUP approaches. It would be entirely denied that some dogmatic concepts would convince other scientists to stand against the implementation of the GUP. But, the scientific discussion should be limited to the abstract argumentation.

The GUP effects on the equivalence principles shall be studied in section \ref{sec:GUPequv}. The universality of the gravitational redshif shall be discussed in section \ref{sec:unvredshift}.  The law of reciprocal action shall be given in section \ref{sec:RecLaw}. This leads to study of the universality of the free fall, section \ref{sec:UnFreeFall}. Finally, section \ref{sec:Kenrg} is devoted to the kinetic energy of composite system.

\subsection{Equivalence principles and kinetic energy} 
\label{sec:ev}

The {\it ''equivalence principle''} belongs to the five principles forming the basis of GR, where the motion of a gravitational test-particle in a gravitational field should be independent on the mass and composition of the test particle \cite{Ray}. On the other hand, when taking into consideration the Strong (SEP) \cite{7} and the Weak Equivalence Principle (WEP) \cite{7}, the gravitational field  is coupled to almost every system\cite{Ray}.

\subsubsection{GUP effects on equivalence principles}
\label{sec:GUPequv}

The Newtonian mechanics in a gravitational field is conjectured to fulfil the WEP effects \cite{7,wep2012}. This is nothing but the equivalence of inertial and the gravitational mass effects. That the QM does not violate the equivalence principles, can be shown from studying the Heisenberg equations of motion. For simplicity, let us consider $1$D motion with the Hamiltonian of a test particle. A macroscopic body considered as a point-like particle with mass $m$ embedded a uniform gravitational field. The Hamiltonian reads $H= p^{2}/2m - m\, g\, x$. The gravitational field is characterized by the acceleration $g$, which is directed along the $x$ axis. We note that the inertial mass $m$ (in the first term) is equal to the gravitational mass $m$ (in the second one). In the classical limit and by using the correspondence between the commutator in QM and the Poisson bracket in classical mechanics $\{ A,\, B\} = \frac{1}{i \hbar}[A,\, B]$, the Heisenberg equations of motion reads  
\bea
\dot{x} &=& \{x,\, H\} = \{x,\, p\} \frac{\de\, H}{\de\, p} = \frac{p}{m},\\
\dot{p} &=& -\, \{p,\, H\} =- \{x,\, p\} \frac{\de\, H}{\de\, p} = m\, g. 
\eea
These two equations ensure that the momentum at the quantum level is given as $p = m\, \dot{x}$ and the acceleration $\ddot{x}$ is mass-independent as the case in the classical physics. Therefore, the equivalence principle is preserved at the quantum level, where $\{x,\, p\}$ is unity.

According to KMM algebra, Eq. (\ref{KMM1}), the modified Heisenberg equations of motion leads to
\bea
\label{2666}
\dot{x} &=& \{x,\, H\}= \{x,\, p\} \frac{\de H}{\de p} = \frac{p}{m} \left(1+\beta\, p^{2}\right) ,\\
\dot{p} &=& - \{p,\, H\}=- \{x,\, p\} \frac{\de H}{\de p} = m g \left(1+\beta\, p^{2}\right).
\eea
In deformed space, the trajectory of point-like mass in the gravitational field depends on the mass of the test-particle \cite{Ray}. If an isotropic deformation is assumed, then the equivalence principle is violated. On the other hand, the acceleration $\dot{x}$ is not mass-independent, because of the mass-dependence through the momentum $p$. Therefore, the equivalence principle is dynamically violated, because of the GUP approaches \cite{Deformed,Subir}. In other words, any momentum term added to the Heisenberg relation leads to violating the equivalence principle  \cite{Deformed,Subir}. The predicted violations  are compared with experimental observations for the universality \cite{Claus} of the gravitational redshift, law of reciprocal action and universality of free fall. 

The bounds assigned to the  GUP parameter $\beta$ as in KMM are tighter than those obtained from QM predictions \cite{Das1}. Keeping the same level of approximation,  the modified geodesic equation is given as \cite{Subir}
\bea 
\label{255}
\frac{d^{2}\, x^{i}}{dt^{2}} &\approx & \frac{1}{2}\, (1+5\, \beta_{m})\, \de^{i}\, h_{0 0},
\eea
where $\beta_{m} = \beta\, m^{2}/2$.

\subsubsection{Universality of gravitational redshift}
\label{sec:unvredshift}

At $\beta _{m}=0$, Eq. (\ref{255}) and from the Newtonian equation of the gravitational potential at a distance $r$ away from a mass $M$ \cite{Claus,Subir}, $d^{2} x/dt^{2} = - \nabla \phi$ and $\phi = - G \frac{M }{r}$.
As given in Ref. \cite{Wiley}, $h_{00} = - 2 \phi \Longrightarrow g_{00} =-(1+2 \phi )$. We have $(1+5\beta _{m}) h_{00} = - 2 \phi  \Longrightarrow h_{00} \approx - 2 \phi (1 - 5 \beta_{m}) \Longrightarrow g_{00} = - (1 + 2 \phi (1 - 5 \beta_{m}))$ \cite{Subir}. 
 
Furthermore, any check for the universal influence of the gravitational field on the clocks, which are based on different physical principles requires {\it a clock comparison} during the common transport of the clock through different gravitational potentials \cite{Claus,Subir}. There is a large variety of clocks which can be compared in \cite{Claus}: \begin{itemize}
\item light clocks (optical resonators), 
\item various atomic clocks, 
\item various molecular clocks,
\item gravitational clocks based on the revolution of planets or binary systems, 
\item the Earth rotation, 
\item pulsar clocks based on the spin of stars
\item clocks based on decay of particles. 
\end{itemize}
In order to measure the gravitational redshift effect \cite{Wiley}, one needs two observation points, say $x_{1} , x_{2}$ and consider a given atomic transition \cite{Claus,Subir}.

The ratio of frequencies $\nu_{2}$ and $\nu_{1}$, where $\nu_{2}$ refers to a light beam coming from $x_{2}$ and goes to $x_{1}$ and $\nu_{1}$ refers to the end position, i.e. the position of observation, is given as \cite{Claus,Subir,Subir2}
\bea
\frac{\nu_{2}(x_{2})}{\nu_{1}(x_{1})} &=& \left(\frac{g_{00}(x_{2})}{g_{00}(x_{1})}\right)^{1/2} = \left(\frac{1+2 (1-5\beta_{m})\phi (x_{2}) }{1+2 (1-5\beta_{m}) \phi (x_{1})} \right)^{1/2} \approx 1+(1-5\beta_{m}) \left(\phi (x_{2}) - \phi (x_{1})\right).  \hspace*{1cm}
\eea

On a phenomenological level, the comparison of two coallocated clocks $A$ and $B$ was given in Ref. \cite{Wiley}
\bea
\frac{\nu_{A}(x_{2})}{\nu_{A}(x_{1})} &=& 1+(1-5(\beta_{m})_{A} )\left(\phi (x_{2}) - \phi (x_{1})\right),\\
\frac{\nu_{B}(x_{2})}{\nu_{B}(x_{1})} &=& 1+(1-5(\beta_{m})_{B} )\left( \phi (x_{2}) - \phi (x_{1})\right). 
\eea
Then the frequency rate reads 
\bea
\frac{\nu_{A}(x_{2})}{\nu_{B}(x_{2})} &\approx &(1-5\left[(\beta_{m})_{A}-(\beta_{m})_{B}\right]) \left( \phi (x_{2}) - \phi (x_{1})\right)\frac{\nu_{A}(x_{1})}{\nu_{B}(x_{1})}.
\eea

Any mismatch of $\nu_{A}(x_{2})/\nu_{B}(x_{2})$ signals violation in the equivalence principle. According to recent observational result, $|\alpha_{Hg} - \alpha_{Cs} | \le 5\times 10^{-6}$, where $\alpha_{Hg}$ and $\alpha_{Cs}$ are clock-dependent parameters for the two elements $Mg$ and $Cs$ \cite{Claus,Subir} ($\alpha_{Hg} = 5 \beta m_{Hg}^{2}$ and $\alpha_{Cs} = 5 \beta m_{Cs}^{2}$). Conventionally, one considers $\beta = \beta_{0}/M_{pl}^{2}$ with $\beta_{0} = 1$  \cite{Das1}, in which the mismatch turns to be $(m_{Hg}^{2} - m_{Cs}^{2})/M_{pl}^{2} \approx 10^{-32}$. This signal is very small. 

Another interpretation \cite{Das1} is to consider an upper bound for $\beta_{0} \approx  10^{26}$; the GUP parameter, which is much below of $\beta _{0} \le 10^{34}$. On the other hand, this new upper bound seems to be compatible with the electroweak scale but much tighter than the bounds suggested in Ref. \cite{Das1} from Lamb shift and Landau level measurements \cite{Das,Das1}. Furthermore, it is weaker than $\beta_{0} \le 10^{21}$, which was derived from  scanning tunnelling microscope  \cite{Das1}.

\subsubsection{Law of reciprocal action}
\label{sec:RecLaw}

Another key model characterizing the violation in the reciprocal action law is based on the estimate of the difference between active and passive gravitational masses \cite{Claus,Subir}. Through the motion of active and passive masses and their possible non-equality, this difference can be estimated. The active mass $m_{A}$ is responsible for the gravitational field. The passive mass $m_{p}$ reacts with the gravitational potential $\Delta U = 4 \pi m_{a} \delta (x)$ through $m_{i} \ddot{x}=m_{p} \nabla U (x)$. Here, $m_{i}$ is the inertial mass and $x$ the position of the particle.  For a gravitationally bound of a two-body system the equations of motion reads \cite{Claus,Subir}
\bea
m_{1i}\, \ddot{x_{1}} &=& G\, m_{1p}\, m_{2a}\, \frac{x_{2}-x_{1}}{|x_{2}-x_{1}|^{3}},\\
m_{2i}\, \ddot{x_{2}} &=& G\, m_{2p}\, m_{1a}\, \frac{x_{1}-x_{2}}{|x_{1}-x_{2}|^{3}},
\eea
where $G$ is the gravitational constant and the indices $1$ and  $2$ refers to first and second particles. Accordingly, the motion of the center-of-mass coordinate   \cite{Claus,Subir}.
\bea
X &=& \frac{m_{1i}\,  x_{1} + m_{2i}\, x_{2}}{m_{1i} + m_{2i}},\\
\ddot{X} &=& G\, \frac{m_{1p}\, m_{2p}}{m_{1i} + m_{2i}}\, C_{21}\, \frac{x_{2} - x_{1}}{|x_{2} - x_{1} |^{3}},
\eea
where $C_{21} = m_{2a}/m_{2p} - m_{1a}/m_{1p}$. In the case that $C_{21} \ne 0$, the active and passive masses become different. Also, the center-of-mass shows a self-acceleration along the direction of $x$. This violates the Newtonian {\it actio-equals-reactio}-law \cite{Claus,Subir}. 

By Lunar Laser Ranging, a limit has been derived (LLR) \cite{Claus,Subir,1986}. Accordingly, no self-acceleration of the moon should be observed. This leads to the limit $|C_{Al - Fe }|\le 7 \times 10^{-13}$ \cite{Claus,Subir,1986}, which obviously provides a considerably tighter bound $\beta _{0} \le 10^{19}$  than the one provided by the gravitational redshift and even the other earlier bounds \cite{Das1}.

\subsubsection{Universality of free fall}
\label{sec:UnFreeFall}

According to GR, the neutral free particles follow geodesic. Hence, their motion does not depend of the nature of the particles themselves.  By measuring the so-called E$\ddot{o}$tv$\ddot{o}$s parameter  \cite{Claus,Subir}
\bea
\eta &=& \frac{(g_{A} - g_{B})}{\frac{1}{2}(g_{A}+g_{B})},
\eea
where $g_{A}$ and $g_{B}$ are accelerations of the two particles $A$ and $B$, respectively, on which {\it ''same''} gravitational field is acting. A non-zero $\eta$ signals violation in the universality of the free fall. But, the active mass gets various corrections to $A$ and $B$. This makes the gravitational field which is perceived by them is not the same. In the mass field $M$, the acceleration of $A$ and $B$ are $g_{A} = (1-5(\beta _{m})_{A})\, g$ and $g_{B} = (1-5(\beta _{m})_{B})\, g$, respectively. Thus, we find that
\bea
\eta &=& \frac{(1 - 5 (\beta_{m})_{A})-(1 - 5 (\beta_{m})_{B})}{\frac{1}{2} [(1 - 5 (\beta_{m})_{B}) + (1 - 5 (\beta_{m})_{B})]}\approx \beta _{0} \frac{5 m_{B}^{2} - 5 m_{A}^{2} }{M_{ p\ell}^{2}} .
\eea
The torsion pendulum leads to $\eta \le 2 \times 10^{-13}$, which leads to $\beta \le 10^{19}$ \cite{Das,Das1}. It worthwhile to highlight that these results will not hold for the macroscopic bodies, due to  $\beta _{m} \ll 1$ \cite{Claus,Subir}. 

So far, it can be concluded that the minimally extended point-particle-model satisfying GUP approach leads to a modified geodesic equation \cite{Subir}. At low energy and in the limit of weak gravity, this effect can be translated into a modified gravitational potential. Furthermore, the correction should depend on the energy of the test particle (or on its mass) \cite{Subir}. This leads to violating the equivalence principle, as well. These results predict violation in the gravitational redshift, the law of reciprocal action and the universality of free fall. Improved bounds for the GUP parameter cab be obtained from the direct comparison with the experimental results \cite{Subir}.

\subsection{Kinetic energy of composite system}
\label{sec:Kenrg}

 {\it The kinetic energy has a additivity property}. Thus, it does not depend on the composition of a body but merely on its mass. When considering $N$ particles with masses $m_{i}$ and deformation parameters $\gamma_{i}$, this is equivalent to the situation when the macroscopic body is divided into $N$ parts. Each part can be treated as point-like particles with the corresponding masses and deformation parameters \cite{Deformed}. When each part/particle moves with the same velocity as the whole system, then the kinetic energy can be given as a function of the velocity. In the first approximation over $\gamma$ and from the relation between velocity and momentum and kinetic energy, Eq. (\ref{2666}), we find \cite{Deformed}
\bea
P &=& m \dot{X}\left(1- \gamma  m^{2}\dot{X}^{2}\right), \\
K.E &=& \frac{1}{2} m \dot{X}^{2}-\gamma m^{3} \dot{X}^{4}.
\eea

At the quantum level, the motion of the center-of-mass of a composite system in a deformed space is governed by an effective parameter. In other words, the deformation parameter for a macroscopic body reads $\gamma = \sum_{j} \mu_{i}^{3}\gamma_{i}$, where $\mu_{j} =m_{i}/\sum_{j}m_{i}$ and $\gamma_{i}$ are the masses and deformation parameters of particles of composite system (body), respectively.

\subsection{Conclusions}

We conclude that
\begin{itemize}
\item The presence of GUP effects implying some noise in GR, where the equivalence principle should be postulates of it. 
\item GUP introduces a mass term to the geodesic equation which  violates the equivalence principle. 
\end{itemize}
In section \ref{sec:ev}, various observations in GR have been reviewed, which should be reestimated in the presence of the GUP effects and compared to the upper bounds on the GUP approach \cite{Das,Das1}. The scanning tunnelling microscope \cite{Das1} appears differently  \cite{Subir}. This means that the violation of the equivalence principle does not support the idea of modification of the Heisenberg principle. The presence of GUP effect corrects the kinetic energy, which as known is independent on the composition of the system.

\newpage

\section{Discussion and final remarks}
\label{dis}

The Heisenberg uncertainty principle expresses one of the fundamental properties of the quantum systems. Accordingly, there should be a fundamental limit of the accuracy with which certain pairs of physical observables, such as the position and momentum, time and energy, can be measured, simultaneously. In other words, the more precisely one observable is measured, the less precise the other one can be estimated. In QM, the physical observables are described by operators acting on the Hilbert space representation of the states. Thus, the Heisenberg uncertainty principle uses operators in describing the relation between various pairs of  uncertainties.

The quantum aspects of the gravitational fields can emerge in a limit, in which the different types of interactions, like strong, weak and electromagnetism can be distinguished from each other. In string theory, the particles are conjectured to stem from fundamental strings. This fundamental scale is nothing but the string length, which is also supposed to be in order of the Planck length. The string cannot probe distances smaller than its own length.  The current researches of the quantum problems in the presence of gravitational field at very high-energy near to the Planck scale implies new physical laws and even corrections to the space-time. The quantum field theory in curved background can be normalized by introducing a minimal observable length as an effective cutoff in the ultraviolet domain. 

We have reviewed different approaches for GUP, which predict an existence of a minimal length uncertainty. The non-zero length uncertainty expresses a non-zero state in the description of the Hilbert space representation and is able to fulfil the non-commutative geometry. These should have impacts on the discreteness and the quantization of space and on the aspects related to the quantum field theory. The elicitation of the minimal length from various experiments, such as string theory, black hole physics and loop quantum gravity, imitates the quantum gravity. All of them predict corrections to the quadratic momentum in the Heisenberg algebra. Many authors represent such algebra under modification in the position operator which fits with the Hilbert space representation and takes into consideration the states of space (eigenvectors) corresponding to the energy (eigenvalue). Others represent such modified algebra by modification in the linear momentum. This is motived by momentum modification at very high energy, which is supposed to fulfil the Hilbert space representation but also approves the idea of modified dispersion relation of the energy-momentum tensor.

The doubly special relativity is conjectured to provide a GUP approach with an additional term reflecting the possibility to deduce information about the maximum measurable momentum. This new term and the one, which is related to the  minimal uncertainty on position are - in modified Heisenberg algebra - of first order of momentum. Some authors suggest a combination of all previously-proposed GUP-approaches in one concept, as anticipated in DSR and the string theory, black hole physics and Loop quantum gravity. Others prefer to revise the GUP of a minimal length in order to overcome some constrains. Another suggestion for GUP-dependent on the Feynman propagator should display an exponential ultra-violet cutoff. All of these verify the predication of minimal length at very high energy, despite of the different physical expression or the algebraic representation of Heisenberg principle. In summary, we have different GUP-approaches with many of applications in various branches of physics.

An unambiguous experiment evidence to ensure these ideas is till missing. Some physicists prefer to deny due to their convention. Some have other objections. Here we review both points-of-views. The value of the GUP parameter remains another puzzle to be verified. For example, the principles of GR developed by Einstein are seen as solid obstacles against the interpretation of the GUP approaches, which are thought to violate the equivalence principle, for instance. In thermodynamics, the natural property of the kinetic energies is assumed to be violated under the consideration of these approaches. As a reason, the symmetries can be broken in quantum field theory. Furthermore, the value of the Keplerian orbit and the correction of the continuity equation for some fields are no longer correct. 

In the present review, we have summarized all these proposals and discussed their difficulties and applications. We aimed to elucidate some of these proposals. On the other hand, from various {\it gedanken} experiments, which have been designed to measure the area of the apparent black hole horizon in QG, the uncertainty relation seems to be preformed. The modified Heisenberg algebra, which was suggested in order to investigate GUP, introduces a relation between QG and Poincare algebra. Under the effect of GUP in an $n$-dimension space, it is found that even the gravitational constant $G$ and the Newtonian law of gravity are subject of modifications. The interpretation of QM through a quantization model formulated in $8$-dimensional manifold implies the existence of an upper limit in the accelerated particles. Nevertheless, the GUP approaches given in forms of quadratic and linear terms of momenta  assume that the momenta approach some maximum values at very high energy (Planck scale).

In supporting the phenomena that uncertainty principle would be affected by QG many examples can be mentioned. In context of polymer quantization, the commutation relations are given in terms of the polymer mass scale. The standard commutation relations are conjectured to be changed or - in a better expression - generalized  at Planck  length. Such modifications are supposed to play an essential role in the quantum gravitational corrections at very high energy. Accordingly, the standard uncertainty relation of QM should be replaced by a gravitational uncertainty relation having a minimal observable length of the order of the Planck length.
On the other hand, the detectability of quantum space-time foam with gravitational wave interferometers has been addressed. The limited measurability of the smallest quantum distances has been criticized. An operative definition for the quantum distances and the elimination of the contributions from the total quantum uncertainty were given. In describing the quantum constrains on the black hole lifetime, Wigner inequalities have been applied. It was found that the black hole running time should be correspondent to the Hawking lifetime, which is to be calculated under the assumption that the black hole is a black body. Therefore, the utilization of Stefan-Boltzmann law is eligible. It is found that the Schwarzschild radius of black hole is correspondent to the constrains on the Wigner size. Furthermore, the information processing power of a black hole is estimated by the emitted Hawking radiation. 

There are several observations supporting GUP approaches and offer a valuable possibility to study the influence of the minimal length on the properties of a wide range of physical systems, especially at quantum scale. The effects of linear GUP-approach have been studied on the compact stars, the Newtonian law of gravity, the inflationary parameters and thermodynamics of the early Universe, the Lorentz invariance violation and the measurable maximum energy and minimum time interval. 
It was observed that GUP can potentially explain the small observed violations of the weak equivalence principle in neutron interferometry experiments and also predicts a modified invariant phase-space which is relevant to Lorentz transformation. It is suggested that GUP can be measured directly in Quantum Optics Lab. 

The experimental tests for Lorentz invariance become more accurate. A tiny Lorentz-violating term can be added to the conventional Lagrangian, then the experiments should be able to test the Lorentz invariance by setting an upper bound to the coefficients of this term, where the velocity of light $c$ should differ from the maximum attainable velocity of a material body. This small adjustment of the speed of light leads to modification in the energy-momentum relation and adding $\delta \textit{v}$ to the vacuum dispersion relation which could be sensitive to the type of candidates for the quantum gravity effect that has been recently considered in the particle physics literature. In additional to that, the possibility that the relation connecting energy and momentum in the special relativity may be modified at the Planck scale, because of the threshold anomalies of ultra-high energy cosmic ray (UHECR) is conventionally named as {\it Modified Dispersion Relations (MDRs)}. This can provide new and many sensitive tests for the special relativity. Accordingly, successful researches would reveal a surprising connection between the particle physics and cosmology. The speed of light not limited to that, but do many searches for the modification of the energy-momentum conservations laws of interaction such as pion photo-production by the inelastic collisions of the cosmic-ray nucleons with the cosmic microwave background and higher energy photon propagating in the intergalactic medium which can suffer inelastic impacts with photons in the Infra-Red background resulting in the production of an electron-positron pair.

The systematic study of the black hole radiation and the correction due to entropy/area relation gain the attention of theoretical physicists. For instance, there are nowadays many methods to calculate the Hawking radiation. Nevertheless, all results show that the black hole radiation is very close to the black body spectrum. This conclusion raised a very difficult question whether the information is conserved in the black hole evaporation process? The black hole information paradox has been a puzzled problem. The study of the thermodynamic properties of black holes in space-times is therefore a very relevant and original task. For instance, based
on recent observation of supernova, the cosmological constant may be positive. The possible corrections can be calculated by means of approaches to the quantum gravity. Through the comparison of the corrected results obtained from this alternative approaches, it can be shown that a suitable choice of the expansion coefficients in the modified dispersion relations leads to the same results in the GUP approach. 

The existence of minimal length and maximum momentum accuracy is preferred by various physical observations. Thought experiments have been designed to illustrate influence of the GUP approaches on the fundamental laws of physics, especially at the Planck scale. The concern about the compatibility with the equivalence principles, the universality of gravitational redshift and the free fall and  reciprocal action law should be addressed. The value of the GUP parameters remains a puzzle to be verified. Furthermore, confronting GUP approaches to further applications would elaborate essential properties. The ultimate goal would be an empirical evidence that the same is indeed quantized and its fundamental is given by the minimal length accuracy. If the current technologies would not able to implement this proposal, we are left with the empirical prove that the modifications of various physical systems can be estimated, accurately. To this destination, we should try to verify the given approaches, themselves. We believe that the compatibility with MDR would play the role of the Rosetta stone translating GUP in energy-momentum relations. The latter would have cosmological and astrophysical observations.
  
 \section*{Acknowledgement}
The last phase of this work has been accomplished in Erice-Italy. AT would like to acknowledge the kind invitation and the  great hospitality of Prof. Antonino Zichichi. This work is financially supported by the World Laboratory for Cosmology And Particle Physics (WLCAPP), http://wlcapp.net/

\appendix

\section{Solution of Eq. (\ref{eq:modFried}) \label{sec:appdx}}

\begin{figure}[htb]
\includegraphics[width=12cm,angle=0]{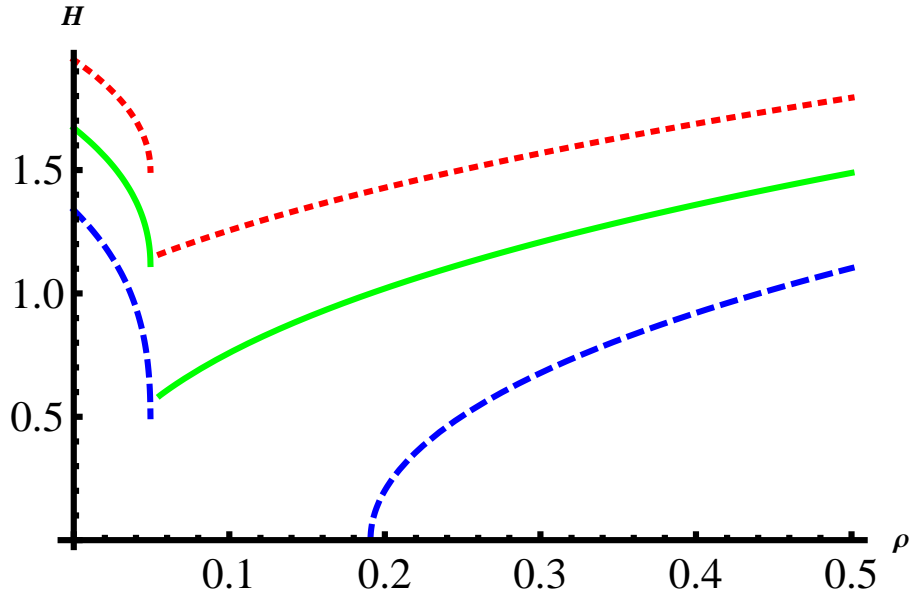}
\caption{The Hubble parameter $H$ is given as a function of the energy density $\rho$. The three curves represents the three values of curvature parameter $k$, $1$ (dotted), $0$ (solid) and $-1$ (dashed) from top to bottom. The discontinuity reflects the region, in which the square root gets imaginary.}
\label{fig:AppndxH}
\end{figure}

Equation (\ref{eq:modFried}) can be solved with respect to $H$. The first two real roots read 
\bea
H &=& \pm\left(-A+\frac{D}{3 \sqrt[3]{2} B^2}- \frac{2 \sqrt[3]{2} C \rho }{D}+\frac{\sqrt[3]{2}}{3 B^2 D}+\frac{1}{3 B^2}\right)^{1/2}, \label{eq:HsolutA}
\eea
where $A=k/a^2$, $B=\alpha \sqrt{\mu} (3 \omega +1)/3$ and $C=8 \pi G/3$. While 
\bea
D &=& \sqrt[3]{27 C^2 \rho ^2 B^4-18 C \rho  B^2+3 \sqrt{3} \sqrt{27 B^8 C^4 \rho ^4-4 B^6 C^3
\rho ^3}+2}
\eea
is real and strongly depends on $\rho$. $H$ remains real as long as
\bea
\frac{D}{3 \sqrt[3]{2} B^2}+\frac{\sqrt[3]{2}}{3 B^2 D}+\frac{1}{3 B^2} &>& A + \frac{2 \sqrt[3]{2} C \rho }{D},
\eea
which is apparently valid, because of the denominator $B$.  Fig. \ref{fig:AppndxH} shows the Hubble parameter $H$ as a function of energy density $\rho$, positive root in Eq. (\ref{eq:HsolutA}). The three curves represents the three values of the curvature parameter $k$, $1$ (dotted curve), $0$ (solid curve) and $-1$ (dashed curve). The region of discontinuity reflects $rho$-values, at which the square root, Eq. (\ref{eq:HsolutA}), gets imaginary. In calculating these curves, we use $a=G=\alpha=1$, $\omega=1/3$ and $\mu=(2.82/\pi)^2$. It is apparent, that the dependence of $H$ on $\rho$ is not monotonic. Reducing $\rho$, which is corresponding to increasing the cosmic time $t$, is accompanied with reducing $H$, as well. Then, starting from a certain value of $\rho$ (and indirectly of $t$), $H$ increases with the further decrease in $\rho$. In other words, the rate of expansion reduces. Then, then rate rapidly increases. The rate strongly depends on geometry of the Universe, $k$.

\end{document}